%% file: main.tex
\documentclass[3p]{article}
\usepackage[utf8]{inputenc}

\usepackage{amsmath,amssymb}
\newcommand{\dd}[1]{\mathrm{d}#1}
\newcommand{\vect}[1]{\mathbf{#1}}
\newcommand{\lra}[1]{\langle #1 \rangle }
\newcommand{\mbs}[1]{\mathbf{#1}}
\usepackage{relsize}
\usepackage{siunitx}
\usepackage{comment}
\usepackage{subcaption}
\usepackage{enumerate}

\usepackage[hidelinks]{hyperref}
\usepackage{algorithmicx}
\usepackage{algpseudocode}
\newcommand{\ALOOP}[1]{\ALC@it\algorithmicloop\ #1%
  \begin{ALC@loop}}
\newcommand{\ENDALOOP}{\end{ALC@loop}\ALC@it\algorithmicendloop}

\algdef{SE}[DOWHILE]{Do}{doWhile}{\algorithmicdo}[1]{\algorithmicwhile\ #1}%

\makeatletter
\newcounter {subsubsubsection}[subsubsection]
\renewcommand\thesubsubsubsection{\thesubsubsection .\@arabic\c@subsubsubsection}
\newcommand\subsubsubsection{\@startsection{subsubsubsection}{4}{\z@}%
     {12\p@ \@plus 6\p@ \@minus 4\p@}%
           {\p@}%
           {\normalfont\normalsize\itshape}}

\newcommand*\l@subsubsubsection{\@dottedtocline{4}{10.0em}{4.1em}}
\newcommand*{\subsubsubsectionmark}[1]{}
\makeatother

\usepackage{tikz}
\usetikzlibrary{math} 
\usepackage{pgffor} 
\usetikzlibrary{shapes.misc}
\usepackage{csvsimple} 
\usepackage{pgfplots}
\pgfplotsset{compat = newest}
\usetikzlibrary{shadows,patterns,shapes}
\usetikzlibrary{arrows,intersections}
\usetikzlibrary{positioning,backgrounds,fit,calc}
\usetikzlibrary{calc}
\usetikzlibrary{spy} 
\usetikzlibrary{math}
\usetikzlibrary{decorations}
\usetikzlibrary{decorations.text}
\usetikzlibrary{decorations.shapes}
\usetikzlibrary{decorations.pathmorphing}
\usetikzlibrary{decorations.pathreplacing}
\usetikzlibrary{shapes.geometric,decorations.fractals,shadows}
\usetikzlibrary{decorations.markings}

\tikzset{cross/.style={cross out, draw=black, minimum size=2*(#1-\pgflinewidth), inner sep=0pt, outer sep=0pt},
cross/.default={3pt}}

\definecolor{greenmf}{RGB}{0,128,0}
\definecolor{redmf}{RGB}{128,0,0}
\definecolor{blueedf}{RGB}{0,91,187}
\definecolor{orangeedf}{RGB}{240,160,47}
\definecolor{bluededf}{RGB}{9,53,122}
\definecolor{orangededf}{RGB}{250,88,21}
\definecolor{greenedf}{RGB}{80,158,47}
\definecolor{greendedf}{RGB}{196,214,47}
\definecolor{indigo}{RGB}{106,44,145}
\definecolor{gold}{RGB}{240,215,0}

\title{\textbf{A time-step-robust algorithm to compute particle trajectories in 3-D unstructured meshes for Lagrangian stochastic methods}}
\date{}

\begin{document}

\maketitle
\vspace{-1cm}
\vspace{1cm}
\begin{center}

\author{\large\textbf{G. Balvet$^{*,1,2}$, J.-P. Minier$^{1,2}$, C. Henry$^{3}$, Y. Roustan$^{2}$, M. Ferrand$^{1,2}$}
}~\\
\vspace{1cm}

$^*$ contact address: guilhem.balvet@edf.fr~\\
$^{1}$ EDF R\&D, Fluid Mechanics, Energy and Environment Dept., 6 Quai Watier,78400, Chatou, France~\\
$^{2}$ CEREA, \'Ecole des Ponts, Île-de-France, France  ~\\
$^3$  Université Côte d’Azur, Inria, CNRS, Sophia-Antipolis, France ~\\
\end{center}

DOI: 10.1515/mcma-2023-2002

\vspace{0.5cm}

\begin{abstract}
{The purpose of this paper is to propose a time-step-robust cell-to-cell integration of particle trajectories in 3-D unstructured meshes in particle/mesh Lagrangian stochastic methods.
The main idea is to dynamically update the mean fields used in the time integration by splitting, for each particle, the time step into sub-time steps, such that each of these sub-steps corresponds to particle cell residence times. This reduces the spatial discretization error. Given the stochastic nature of the models, a key aspect is to derive estimations of the residence times that do not anticipate on the future of the Wiener process. To that effect, the new algorithm relies on a virtual particle, attached to each stochastic one, whose mean conditional behavior provides free-of-statistical-bias predictions of residence times. After consistency checks, this new algorithm is validated on two representative test cases: particle dispersion in a statistically uniform flow and particle dynamics in a non-uniform flow.~ \\

MSC 2010 class: 76M35 ~\\ 

\textbf{keyword: Lagrangian stochastic modeling, particle-mesh PDF, temporal integration, trajectory in 3-D unstructured mesh, time-splitting methods, anticipation error}
}
\end{abstract}

 \newpage

\section{Introduction}
\label{sec: introduction}

As it transpires from their name, Lagrangian probability density function (PDF) methods consist in simulating the PDF of a number of variables, which have been selected to describe relevant information about single-phase or disperse two-phase flows. From the Lagrangian PDF, the Eulerian PDF is immediately deduced which gives access to the statistical moments of interest. In a weak formulation in which only statistics are of interest, Lagrangian PDF methods are equivalent to particle stochastic models. Consequently, these methods appear as stochastic equations written for a set of variables attached to each particle which form the particle state vector (often composed of the particle position, its velocity and other variables of interest). These methods present strong advantages for modeling and simulating turbulent reactive particle-laden flows. In particular, they can treat without approximation local source terms, however complex or non-linear, when these source terms are given as known expressions of the variables attached to each particle (provided that these variables are included in the particle state vector). Consequently, PDF methods can solve closure problems related to the average of non-linear local terms, such as chemical source terms or the drag force which enters the particle equation of motion. At the same time, the Lagrangian point of view ensures that convection is treated accurately. This explains that these methods are well-suited to simulate polydisperse non-homogeneous two-phase flows, as well as single-phase reactive flows, while still being interesting models for inert single-phase turbulent flows.

Lagrangian PDF models started being developed in the 1980s and 1990s, mostly by Pope and co-workers~\cite{haworth1986generalized, pope1990velocity, pope1991application, pope1994lagrangian, pope1994relationship}) and were later extended to the disperse two-phase flow situation by Minier and co-workers~\cite{minier1997derivation, pozorski1998lagrangian, minier2001pdf, minier2004pdf, minier2015lagrangian, minier2016statistical, minier2021methodology}. Over the years, there has been continuous effort to analyze stochastic models~\cite{sabelfeld2012random,minier2015lagrangian, minier2016statistical} as well as to come up with useful guidelines for their construction~\cite{minier2014guidelines}. This is particularly true in stochastic models for the velocity of the fluid seen in disperse turbulent two-phase flows, which is a subject of ongoing research \cite{innocenti2021lagrangian} (see also a recent proposal of a general methodology in~\cite{minier2021methodology}). Meanwhile, the situation seems to have reached a more mature state of affairs for single-phase turbulent flows with the formulation of general Langevin models (GLM) by Pope and various co-authors~\cite{haworth1986generalized, pope1994lagrangian, pope2000turbulent, haworth2010progress, innocenti2020lagrangian}. On the numerical side, there has been specific works dedicated to the analysis of the various issues related to the numerical implementation, which is often carried out in the context of particle/mesh methods~\cite{xu1999assessment, peirano2006mean}. At this stage, it is worth reminding that many of such developments have been made in research codes mostly with a view towards validating new stochastic models hence using structured meshes in relatively simple geometries. To the authors' knowledge, less attention has been devoted to the question of particle tracking methods within complex realistic domains ({e.g.}, for environmental purposes) used in combination with appropriate time integration schemes.

To integrate the stochastic differential equations (SDE) that model particle dynamics, a specific time-integration numerical scheme was introduced in~\cite{peirano2006mean, minier2003weak}. Building on the existence of relaxation terms in the particle velocity equations, this scheme was obtained, first by considering the integrated form involving exponential terms, and, second by discretizing these integrated expressions rather than the SDEs themselves. For this reason, it is referred to as an exponential numerical scheme. Note that this scheme is not applicable to any system of SDEs, since it assumes relaxation terms, but it is nevertheless of great interest for a wide range of modeled equations for particle dynamics in turbulent flows. This exponential scheme has interesting properties: by construction, it corresponds to the exact solution when all mean fields and timescales are constant (when the SDEs are linear), it remains explicit and unconditionally stable (thanks to the exponential factors in its formulation). Furthermore, careful discretization of the stochastic integrals allows to correctly reproduce the asymptotic regimes, when one or several of the timescales entering the relaxation terms become much smaller than the time step. As a typical example of key practical interest, this means that the diffusive regime is well captured without having to switch to a different numerical scheme. An in-depth description and analysis of this exponential scheme can be found in an extensive review on the subject and we refer interested readers to the available literature for further details~\cite{peirano2006mean}.

The unconditional stability of the time-integration exponential scheme, combined with its ability to capture the correct asymptotic limits, allow large time steps to be used in practical simulations. However, in that case, a particle can typically cross a distance over which the mean fields entering its modeled dynamical equation, such as the fluid mean velocity field, can vary significantly. Yet, in the spirit of Euler schemes, the first-order-in-time exponential scheme retains only the values of these mean fields at the initial particle location. Note that the same limitation exists even for the second-order scheme, which is based on an extension of prediction-correction steps and retains only two values of the mean fields (at the initial and at the predicted particle positions). In the case of strong gradients of these mean fields, this leads to a spatial discretization error that reduces the overall accuracy of the numerical scheme and of the statistics of interest extracted from the particle dynamics simulation. New methods are therefore required to couple particle tracking with the time-integration numerical scheme in complex geometries and non-uniform flows.

In this context, the present objective is to develop a new particle-tracking algorithm to reduce the spatial discretization error of the exponential scheme, especially for large time steps during which particles can cross several cells. The basic idea is to perform a cell-to-cell integration by introducing sub-time steps corresponding to the particle residence times in the cells that each particle crosses. This allows to update the mean fields and local timescales dynamically along particle trajectories during each time step. In essence, if a $P_0$ interpolation is retained (in which mean fields are considered as constant within a cell), this new time-integration method is meant to retrieve the exact solution when the exponential numerical scheme is applied. However, although this idea of using particle-attached local sub-time steps may seem rather straightforward and directly applicable to each particle trajectory, the estimation of the residence time of a particle within a given cell is actually a tricky question due to the stochastic nature of the dynamical models used to construct its trajectory. The estimations of the residence time have indeed to respect It\^{o}'s definition of stochastic integrals. This corresponds to the non-anticipating requirement, which is satisfied by the new algorithm detailed in this paper. Note that the idea of cell-to-cell integration was already proposed by \cite{popov2008accurate} but was only tested with prescribed velocity fields given on structured meshes (therefore without explicit velocity stochastic models). Hence, this study can be seen as an extension of this work in the context of stochastic models (i.e., addressing the non-anticipating requirement) and for unstructured meshes.

The paper is organized as follows. The theoretical background on Lagrangian PDF, or particle stochastic, models is first recalled in Section~\ref{sec:stochastic_model}, in which the limit case of fluid particles and the specific Langevin model retained are presented. In Section~\ref{sec:stochastic_impl}, an overview of the current numerical scheme is given to bring out the spatial discretization error. The new algorithm is then detailed in Section~\ref{sec:new_alg} and validated on two test cases in Section~\ref{sec:results}. Conclusions are then drawn in Section~\ref{sec:conclusion}.

\section{Theoretical background on stochastic models for turbulent flows}
\label{sec:stochastic_model}

\subsection{PDF models for turbulent flows}
\label{sec:turb_flow_model}

Disperse turbulent two-phase flows involve discrete elements, or `particles', transported by turbulent fluid flows. For the sake of simplicity, we focus here on particle dynamics while leaving out thermal effects, changes in the particle radius or mass, and particle collision effects. To model such a system, a suitable modeling framework consists in adopting a Lagrangian stochastic point of view. In this approach, a system of stochastic differential equations (SDEs) is written to describe the evolution in time of the particle-attached variables making up the particle state-vector $\mbs{Z}_p$. Here, the state vector is taken equal to $\mbs{Z}_p=(\mbs{X}_p,\mbs{U}_p,\mbs{U}_s)$, with $\mbs{X}_p$ the particle position, $\mbs{U}_p$ its velocity and $\mbs{U}_s$ the velocity of the fluid seen. The system of SDEs is
\begin{subequations}
\label{SDEs}
\begin{align}
\dd X_{p,i} & = U_{p,i}\, \dd t~, \label{SDE xp} \\
\dd U_{p,i} & = \frac{U_{s,i} - U_{p,i}}{\tau_p}\, \dd t + A_{p,i}\, \dd t~, \label{SDE Up} \\
\dd U_{s,i} & = (\text{stochastic model})~. \label{SDE Us}
\end{align}
\end{subequations}
In Eqs.~\eqref{SDEs}, the velocity of the fluid seen $\mbs{U}_s$ is defined as the instantaneous fluid velocity sampled at the particle location, $\mbs{U}_s(t)=\mbs{U}(t,\mbs{X}_p(t))$ where $\mbs{U}(t,\mbs{x})$ is the fluid velocity field. Note that, for discrete particles in non-fully-resolved turbulent flows, the velocity of the fluid seen cannot be obtained from the reduced information available on the fluid velocity field (typically, its first and second moments). This means that $\mbs{U}_s$ needs to be introduced as a separate particle-attached variable (see comprehensive discussions in~\cite{minier2001pdf, minier2015lagrangian, minier2016statistical}). On the right-hand side (RHS) of Eq.~\eqref{SDE Up}, it is seen that $\mbs{U}_s$ enters the expression of the drag force while the term $\mbs{A}_p$ contains possible additional forces acting on discrete particles (such as gravity). This drag force is expressed in terms of the particle relaxation timescale $\tau_p$, defined as
\begin{equation}
\label{definition taup}
\tau_{p}=\frac{\rho_p}{\rho}\frac{4\,d_p}{3\,C_D \vert \,\mbs{U}_{s}-\mbs{U}_{p}\, \vert}~,
\end{equation}
with $\rho$ and $\rho_p$ the fluid and particle densities, respectively, $d_p$ the particle diameter and $C_D$ the drag coefficient. This timescale represents the typical time over which particle velocities adjust to the local fluid velocity seen and is a measure of particle inertia. In the Stokes regime, valid when ${\rm Re}_p \leq 1$ (with ${\rm Re}_p$ the particle Reynolds number defined by ${\rm Re}_p=\vert \, \mbs{U}_r \, \vert\, d_p/\nu$, with $\mbs{U}_r=\mbs{U}_{s}-\mbs{U}_{p}$ and $\nu$ the fluid kinematic molecular viscosity), the drag coefficient is $C_D=24/{\rm Re}_p$. In that case, the particle relaxation timescale is given by the Stokes formula:
\begin{equation}
\label{definition taup_Stokes}
\tau_{p}=\frac{\rho_p}{\rho}\frac{d_p^2}{18\nu_f}.
\end{equation}
For general values of ${\rm Re}_p$, the drag coefficient is obtained through empirical correlations such as~\cite{clift2005bubbles,brennen2005fundamentals}
\begin{equation}
\label{definition correlation taup}
C_D=
 \begin{cases}
  \dfrac{24}{{\rm Re}_p}\left[\,1 + 0.15 {\rm Re}_p^{0.687}\, \right] & \text{if} \; {\rm Re}_p \leq 1000,  \\
  0.44 & \text{if} \; {\rm Re}_p \geq 1000.
 \end{cases}
\end{equation}

It is worth recalling that stochastic particles, modeled by equations such as Eqs.~\eqref{SDEs}, must be regarded as samples of a corresponding PDF and, consequently, the description can also be referred to as a two-phase PDF model~\cite{minier2001pdf, minier2016statistical, minier2014guidelines}. In the present PDF framework, we are basically handling a two-time one-particle Lagrangian PDF, from which a one-time one-point Eulerian PDF is derived, allowing moment equations to be extracted in a straightforward manner~\cite{minier2001pdf, minier2016statistical, pope2000turbulent}. This means that once a stochastic model, such as the one in Eqs.~\eqref{SDEs}, is written, we obtain directly the corresponding expressions of the discrete Lagrangian $F_{p,N}^L$ and resulting Eulerian $F_{p,N}^E$ mass density functions (MDF) through
\begin{align}
F_{p,N}^L(t\,;\,\mbs{Y}_p, \mbs{V}_p,\mbs{V}_s) &=\sum_{i=1}^N m_p^{(i)}\,\delta (\mbs{Y}_p-\mbs{X}_p^{(i)}(t))
               \,\delta (\mbs{V}_p-\mbs{U}_p^{(i)}(t)) \,\delta (\mbs{V}_s-\mbs{U}_s^{(i)}(t))~, \\
F_{p,N}^E(t,\mbs{x}\,; \,\mbs{V}_p,\mbs{V}_s) &= F_{p,N}^L(t\,;\, \mbs{Y}_p=\mbs{x}, \mbs{V}_p,\mbs{V}_s)~,
\end{align}
where $m_p$ is the particle mass, $N$ is the number of stochastic particles in the domain and $\mbs{Y}_p$, $\mbs{V}_p$ and $\mbs{V}_s$ the sample space values corresponding to the random variables $\mbs{X}_p(t)$, $\mbs{U}_p(t)$ and $\mbs{U}_s(t)$, respectively. It is then straightforward to derive the field values for the average $\lra{H_p}(t,\mbs{x})$ of any particle variable $H_p(t;\mbs{V}_p,\mbs{V}_s)$,
\begin{equation}
\alpha_p(t,\mbs{x}) \, \rho_p \lra{H_p}(t,\mbs{x})= \int H_p(t; \mbs{V}_p,\mbs{V}_s)
                           F_p^E(t,\mbs{x}; \mbs{V}_p,\mbs{V}_s)\,d\mbs{V}_p\,d\mbs{V}_s~,
\end{equation}
where $\alpha_p(t,\mbs{x})$ is the mean particle volumetric fraction defined through a proper probabilistic normalization constraint~\cite{minier2001pdf, peirano2002probabilistic}. In a numerical simulation, these theoretical expressions imply that, in a small volume around a given location $\mbs{x}$, mean values are estimated as the ensemble averages over the $N^p_{\mbs{x}}$ particles present in that volume, or as Favre averages (or mass-weighted averages)
\begin{equation}
  \lra{H_p}(t, \mbs{x}) \simeq \dfrac{
  \displaystyle
  \sum_{i=1}^{N^p_{\mbs{x}}} m_p^{(i)}
  H_p(t; \mbs{V}_p^{(i)}(t), \mbs{V}_s^{(i)}(t))}{
  \displaystyle
  \sum_{i=1}^{N^p_{\mbs{x}}} m_p^{(i)}}~.
\end{equation}

For reasons explained elsewhere~\cite{minier2001pdf}, the velocity of the fluid seen is generally represented by a Langevin model, which takes the following form (see extensive accounts in~\cite{minier2001pdf, minier2004pdf, minier2015lagrangian, minier2014guidelines})
\begin{equation}
\label{complete Langevin stochastic model}
\dd U_{s,i} = -\frac{1}{\rho}\frac{\partial \lra{P}}{\partial x_i}\, \dd t + \left( \lra{U_{p,j}} - \lra{U_{j}} \right)
\frac{\partial \lra{U_{i}}}{\partial x_j} \dd t  + G^{*}_{ij}\left( U_{s,j} - \lra{U_{j}} \right) \dd t + B^{*}_{ij}\, \dd W_j~.
\end{equation}
where $P$ is the fluid pressure and $\dd \mbs{W}$ the increments of a Wiener process (see more details in Section~\ref{sec:stochastic integrals}). In Eq.~\eqref{complete Langevin stochastic model}, the expressions retained to close the matrices $G^{*}_{ij}$ and $B^{*}_{ij}$ can be intricate and we refer to the relevant literature for their complete formulas~\cite{minier2001pdf, minier2015lagrangian}. As explained in these works, the stochastic model for $\mbs{U}_s$ is built as an extension of the ones used to simulate fluid particles. Starting from the general disperse two-phase model in Eqs.~\eqref{SDEs}, the fluid limit case (this is also referred to as the tracer-particle limit) is indeed retrieved by considering the limit of vanishing particle timescale $\tau_p$.

In the present work, we essentially focus on particle tracking for general non-homogeneous flows in complex geometries, and the emphasis is therefore on particle locations $\mbs{X}_p$. To bring out the essential elements of the new algorithm introduced in Section~\ref{sec:new_alg}, the stochastic model retained to model particle velocity is chosen as simple as possible to avoid handling cumbersome expressions in the developments to follow. For these reasons, from now on, we consider only the limit case of fluid particles in turbulent flows and we choose to select the simplest possible Langevin model for the velocity of these fluid particles. However, the general context recalled in this section is meant to indicate that the new algorithm is indeed applicable to a much larger class of stochastic formulations than the one retained in the rest of the paper.

\paragraph*{PDF models for single-phase turbulent flows}
As just indicated, the dynamic of fluid particles is based on the simplest Langevin model, the simplified Langevin model (SLM) \cite{pope2000turbulent, haworth1987pdf}. The particle state vector is now $\mbs{Z}=(\mbs{X},\mbs{U})$ and the evolution equations write
\begin{subequations}
\begin{align}
  \dd X_i & = U_i \dd t , \label{SLM_cont_X}\\
  \dd U_i & = - \dfrac{1}{\rho }\dfrac{\partial \lra{P}}{\partial x_i}\left(t,\mbs{X}(t)\right)\dd t - \dfrac{U_i - \lra{U_i}(t,\mbs{X}(t))}{T_L(t,\mbs{X}(t))}\dd t+ \sqrt{ C_0 \epsilon(t,\mbs{X}(t)) } \, \dd W_i~. \label{SLM_cont_U}
\end{align}
\label{SLM_cont}
\end{subequations}
The RHS of Eq.~\eqref{SLM_cont_U} involves several mean field values, including the turbulent dissipation rate $\epsilon$ , the fluid turbulent kinetic energy $k$ and the Lagrangian integral time scale $T_L$. All these mean fields are to be evaluated locally in the vicinity of the particle, hence the local dependence that is explicitly written. The Lagrangian integral time scale is defined as:
\begin{equation}
T_L=\left( \dfrac{1}{\frac{1}{2}+\frac{3}{4} C_0} \right) \frac{k}{\epsilon}~,
\end{equation}
with $C_0$ the so-called Kolmogorov constant. The PDF formalism, outlined above, can then be followed to reveal that, in terms of resulting statistics for the fluid velocity, the SLM model corresponds directly to a second-order turbulence model, namely the Rotta model \cite{pope1994relationship, pope2000turbulent}. Furthermore, we are considering only incompressible flows. Hence, since we associate the same mass $\delta m$ to each fluid particles, this means that the particle concentration must remain uniform to respect the incompressible condition.

\subsection{A reminder on stochastic integrals}
\label{sec:stochastic integrals}

The SLM for fluid particles is retained as the model of reference in this work, although, as indicated above, the numerical developments to follow in Section~\ref{sec:new_alg} can be extended to the case of dispersed turbulent two-phase flows by using the extended numerical integration scheme presented in~\cite{peirano2006mean}. Present ideas can also be applied to more general stochastic models where the particle state vector does include particle location and velocity as well as extra variables $\mbs{Z}=(\mbs{X},\mbs{U},\mbs{Z'})$, where $\mbs{Z'}$ stands for additional variables attached to each stochastic particle to address more complex physical issue (in the rest of the paper, we have $\mbs{Z'}=\varnothing$). Furthermore, other stochastic models than Langevin ones can be considered for the modeled particle velocity equation but we assume that it takes the form of a general stochastic diffusion process, implying white-noise terms (note that, in that case, the numerical scheme for time-integration has then to be adapted, see Section~\ref{sec:new_alg:description:overview}).

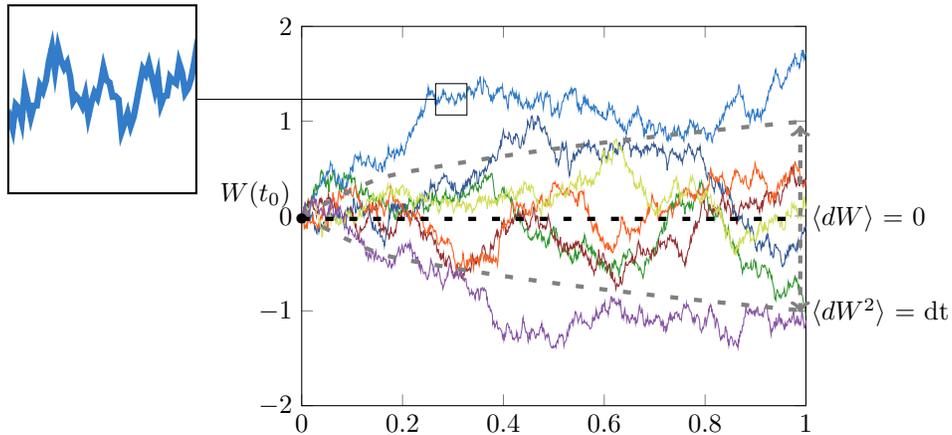
\begin{figure}[h]
    \centering
    \input{tikz/Gaussian_traj}
    \caption{Some realizations of a conditional Wiener process, starting from $0$ at time $t=0$.}
    \label{Gaussian_traj}
\end{figure}

However, a few words are in order to recall the definition of stochastic integrals and the important issue of non-anticipating formulations. To that effect, we consider a general stochastic diffusion process for a particle state vector $\mbs{Z}=(Z_i)$ with the form
\begin{equation}
    \dd Z_i = \underbrace{A_i \left (t,\mbs{Z}, \lra{ \mathcal{F}(t,\mbs{Z})} \right) \dd t}_{\text{deterministic Part}}+ \underbrace{B_{ij} \left(t,\mbs{Z} , \lra{ \mathcal{G}(t,\mbs{Z}) } \right) \dd W_j~. }_{\text{random Part}}
\label{diffusion_process}
\end{equation}
In Eq.~\eqref{diffusion_process}, $\mbs{A}=(A_i)$ and $\mbs{B}=(B_{ij})$ stand for the drift vector and diffusion matrix, respectively. They are written with the most general possible form, involving a possible dependence on mean values represented by the two functional terms, namely $\mathcal{F}$ and $\mathcal{G}$, which makes Eq.~\eqref{diffusion_process} a stochastic diffusion process of the McKean type (see~\cite{minier2015lagrangian, minier2016statistical}). The Wiener process $\mbs{W}$ is a continuous time Markov process whose increments follow independent Gaussian centered distributions of variance $\dd t$~\cite{ottinger1996stochastic, sabelfeld2012random}. Although continuous, its trajectories are highly erratic and even of infinite total variance in any interval (see the illustration in \figurename{}~\ref{Gaussian_traj}), implying that classical integration rules cannot be applied.

The formulation of the SDE in Eq.~\eqref{diffusion_process} is actually a short-hand notation for the proper mathematical expression, which is its integrated version
\begin{equation}
Z_i(t)= Z_i(t_0) + \int_{t_0}^{t} A_i(s, \mbs{Z}(s), \mathcal{F}(s,\mbs{Z}(s))\, \dd s + \underbrace{ \int_{t_0}^t B_{ij}(s,\mbs{Z}(s),\mathcal{G}(s,\mbs{Z}(s)))\, \dd W_j(s) }_{\text{stochastic integral}}~,
\end{equation}
where the stochastic integral must be carefully defined. This is done by applying the It\^{o} definition for non-anticipating processes. Using a partition $[t_k \, ; \, t_{k+1}]$, $k=1,\ldots, N$, of the interval $[t_0\,;\, t]$, the stochastic integral in the It\^{o} sense is then obtained as
\begin{equation}
\int_{t0}^t B_{ij}(s,\mbs{Z}(s),\mathcal{F}(s,\mbs{Z}(s)))\, \dd W_s = \text{ms-}\lim_{N \to \infty} \sum_{k=1}^N B_{ij}(t_k,\mbs{Z}({t_k}), \mathcal{F}(t_k,\mbs{Z}(t_k)) )\left( W_j (t_{k+1}) - W_j(t_k) \right)~,
\end{equation}
where the limit must be understood as a limit in the mean-square sense (since a convergence trajectory by trajectory is not possible). The choice of this definition allows two fundamental properties of stochastic integrals to be always respected for any smooth-enough functions $\Psi_1$ and $\Psi_2$ (vector indexes and functional dependencies are left out here, for the sake of keeping simple notations):
\begin{itemize}
\item $\bigg\langle \displaystyle \int_{t_0}^t \Psi_1(\mbs{Z}(s)) \dd W(s) \bigg\rangle =0~,$
\item $\bigg\langle \displaystyle \left( \int_{t_0}^{t_2 } \Psi_1(\mbs{Z}(u)) \dd W(u) \right)  \left( \int_{t_1}^{t_3 } \Psi_2(\mbs{Z}(v)) \dd W(v) \right) \bigg\rangle = \int_{t_1}^{t2 } \lra{ \Psi_1(\mbs{Z}(s)) \Psi_2(\mbs{Z}(s)) } \dd s~, \; t_0\leq t_1 \leq t_2 \leq t_3~.$
\end{itemize}
A key consequence of the non-anticipation nature of the definition of the stochastic integrals is that the mean conditional increment of the stochastic process $\mbs{Z}$ in Eq.~\eqref{diffusion_process} over a small time increment $\Delta t$ is governed only by the drift term, which means that we have
\begin{equation}
\langle \Delta Z_i \, | \, \mbs{Z}(t_0)=\mbs{z}_0 \rangle = A_i (t_0,\mbs{z}_0,\mathcal{F}(t_0,\mbs{z}_0) )\, \Delta t~.
\end{equation}
These relations play a central role in the development of the new algorithm in Section~\ref{sec:new_alg}.

\section{Current numerical algorithm}
\label{sec:stochastic_impl}

To obtain numerical solutions of the SDEs in Eqs.~\eqref{SLM_cont}, three issues need to be addressed:
\begin{enumerate}[1)]
\item How do we calculate the mean fields at particle positions?
\item How do we integrate in time the SDEs?
\item How do we extract statistics from the ensemble of particles?
\end{enumerate}
These three separate points are related to three sources of error: the first one corresponds to the spatial discretization error due to an approximate expression of mean fields at particle locations; the second one corresponds to the time error due to the integration scheme selected to update particle variables at discrete times; while the third one is related to the statistical error which is inherent in Monte Carlo methods. In a complete numerical formulation, these errors can impact each other and can even induce a fourth one referred to as the bias error. However, for reasons set forth below, this bias error is not present in the present context and we can safely concentrate on the three ones mentioned above.

Although relatively few studies have been devoted to their numerical analysis, these questions have been analyzed in depth in at least two detailed investigations~\cite{xu1999assessment,peirano2006mean}, to which readers are referred to for further details. In this section, we limit ourselves to recalling the key aspects of the current algorithm putting the emphasis on the spatial discretization error so as to pave the way for the developments described in Section~\ref{sec:new_alg}.

\subsection{Hybrid FV/PDF approach and particle-mesh implementation}\label{sec:spatial_disc}

When dealing with fluid particles, a first possibility is to extract the mean fields entering Eqs.~\eqref{SLM_cont} directly from the particle simulation which are then fed back into the computation. This can be done using either an auxiliary grid to compute such statistics from the ensemble of particles or by resorting to a local kernel estimation centered around each particle position. Such numerical formulations are called PDF stand-alone codes and one interest is that they are consistent by construction while, on the other hand, the inherent statistical noise due to the use of a finite number of particles can induce a deterministic bias error~\cite{xu1999assessment,peirano2006mean}. In the present study, we have, however, chosen to rely on a hybrid formulation in which the particle simulation is coupled to a classical FV (finite volume) solver, thereby leading to so-called hybrid FV/PDF methods. One interest of such formulations is that the mean fields provided to the particle solver are now free of statistical noise, thereby avoiding bias errors. On the other hand, this hybrid formulation raises immediately a consistency issue for fluid particles since, for example, the mean fluid velocity field is simulated twice (in the FV solver but also as the mean value of fluid particles located in a given small volume). As discussed in several works~\cite{minier2015lagrangian, chibbaro2011note}, it is then important to couple FV and particle solvers that correspond to the same turbulence model. For instance, with the present simple Langevin model retained in Eqs.~\eqref{SLM_cont}, the turbulence model within the FV solver must be the Rotta model with a constant consistent with the value of $C_0$ in Eq.~\eqref{SLM_cont_U}. These points have been discussed in detail in several papers~\cite{pope1994relationship, minier2015lagrangian, pope2000turbulent, chibbaro2011note} and are not repeated here. A further reason behind the choice of such hybrid formulations is that, as pointed out in Section~\ref{sec:stochastic_model}, we are interested in extending present developments to the case of discrete particles where, for example, the fluid mean velocity field cannot be extracted from the particle simulation (the statistics of the velocity of the fluid seen being different from the fluid ones~\cite{minier2001pdf,minier2015lagrangian,minier2016statistical}) and, in which, such hybrid formulations are of specific interest. Note that, although not addressed in the present work, the introduction of an underlying mesh in the particle solver is also useful to simulate particle collision effects \cite{schmidt2000new, sigurgeirsson2001algorithms}.

\begin{figure}[h]
 \begin{minipage}{0.6 \linewidth}
    \input{tikz/tikz_hybrid_scheme}

 \end{minipage}
 \begin{minipage}{0.4 \linewidth}
  \begin{algorithmic}
   \For{time interval $t$ to $t+\Delta t$}
    \State \hrulefill
    \For{each $\text{Cell 'k'}$ in the grid}
     \State Compute the mean fields at $t+\Delta t$
     \\ \hspace{3.2em} ({e.g.}  $\langle\mbs{U}\rangle_{\text{Cell k}}$, $\langle P\rangle_{\text{Cell k}}$)
    \EndFor
    \State \hrulefill
    \For{each particle $p$}
     \State Interpolate mean fields at particle
     \\ \hspace{3.5em} position ({e.g.}  $\langle\mbs{U}\rangle(t,\mbs{X}_{p}(t))$)
     \State Compute its dynamics $\mbs{X}_{p}(t+\Delta t)$
    \EndFor   \EndFor
  \end{algorithmic}
 \end{minipage}
 \caption{Sketch illustrating the principle of hybrid FV/PDF approaches: for each time step, the FV solver is first run to provide the needed mean fields on a grid and, then, the particle dynamics is solved with a PDF approach which implies to have an interpolation method to determine the mean field values at the particle position.}
 \label{fig:hybrid_scheme}
\end{figure}
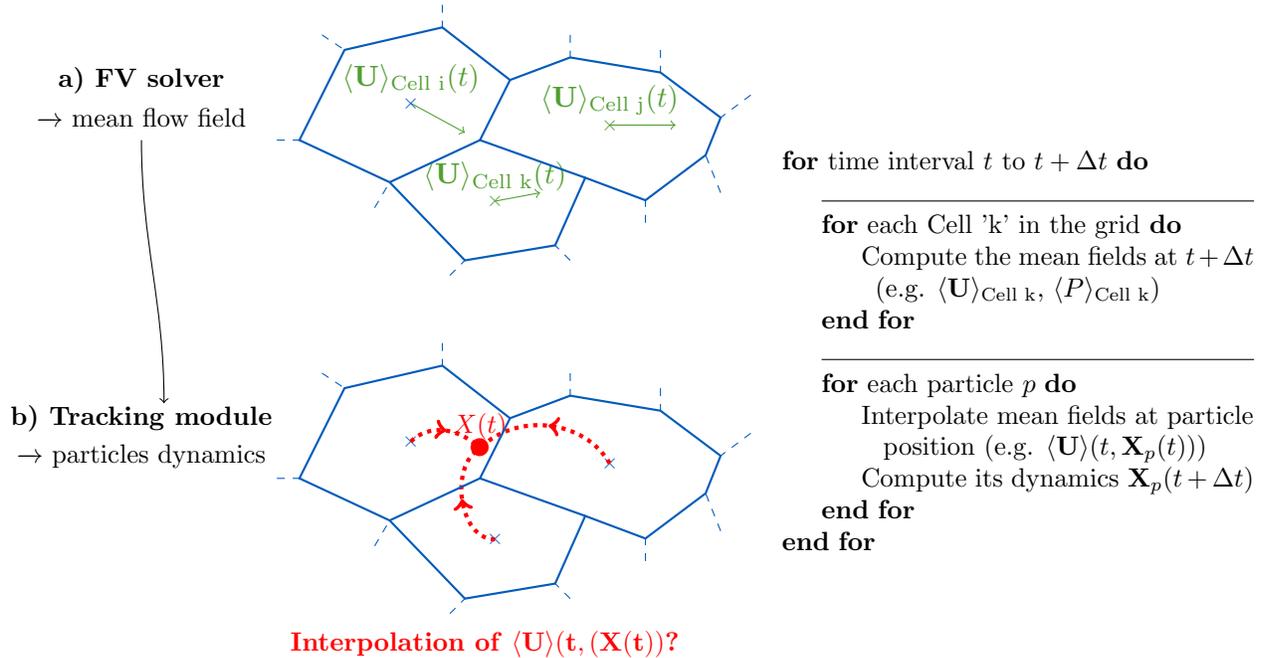

In practice, as displayed in \figurename{}~\ref{fig:hybrid_scheme}, the hybrid FV/PDF coupled formulation is translated into a particle/mesh simulation in which, during each iteration, the FV solver is first run and then provides the needed mean fields on a grid to the particle solver. The difficulty in such hybrid formulations is that the mean fields in the FV solver are known at determined spatial locations (typically at the center of each cell of the mesh for collocated discretizations). However, fluid tracers in the particle solver are continuously distributed within the domain and very rarely correspond to the cell center locations. We are then faced with an additional issue which is to determine how mean-field values are interpolated from the mesh to the particles. In the present study, we have retained the simplest possibility, which assigns the same cell-centered mean values to all particles located within that cell: this corresponds to a $P_0$ interpolation at particle positions. While this implies discontinuous profiles of mean quantities when a particle moves across an interface between two adjacent cells, this drawback is offset by the simplicity of the numerical implementation and, in particular, by the fact that it remains valid and easy to apply even in the case of unstructured meshes (sometimes made up by several overlapping meshes) in complex geometries for which more advanced methods, such a linear interpolation between neighboring cell centers quickly become very intricate and nearly intractable.

Based on this hybrid formulation and using the $P_0$ interpolation assumption, the particle equations, cf Eqs.~\eqref{SLM_cont}, can be expressed as:
\begin{subequations}
\begin{align}
\dd X_i = & U_i \, \dd t , \label{SLM_disc_X}\\
\dd U_i = & - \dfrac{1}{\rho }
\left[
\dfrac{\partial \lra{P} }{\partial x_i} \right]_0(t,\mbs{X}(t))\dd t - \dfrac{U_i -  \left[\lra{U_i}\right]_0(t, \mbs{X}(t)) }{\left[T_L\right]_0(t,\mbs{X}(t)) } \dd t+ \sqrt{ C_0 \left[\epsilon\right]_0(t,\mbs{X}(t)) }  \dd W_i \label{SLM_disc_U}~,
\end{align}
\label{SLM_disc}
\end{subequations}
where $[.]_0$
stands for the $P_0$ interpolation and will be omitted in the following for the sake of clarity. In this hybrid FV/PDF formulation, the same grid is used to extract particle statistics from the subset of particles present in a given cell at a given time. This corresponds to the nearest-grid-point (NGP) method in which all the statistical weight associated to each particle is associated to the cell it belongs to, regardless of its position in that cell. This NGP method is actually consistent with the $P_0$ interpolation used to provide information from the grid to the particles and appears as its reverse operation, that is from the particles to the grid (see discussions in~\cite{xu1999assessment,peirano2006mean}).

At this stage, a few comments are in order with respect to the issue of interpolating mean fields at particle positions. In stand-alone PDF models implemented as particle-mesh methods, the interpolation scheme and how particle statistics are extracted appear as two adjoint operators \cite{xu1999assessment, peirano2006mean}. This means that if, for example, a linear interpolation scheme is used to obtain interpolated mean-field values, then a cloud-in-cell (CIC) method must be selected to derive statistics from the particle set. In structured meshes, this corresponds to a classical approach \cite{hockney1988computer}. However, the implementation of such a consistent method becomes quickly intricate in unstructured meshes. In present hybrid FV/PDF formulations for fluid particles, these two operators are decoupled since particle statistics are not injected in the stochastic model for particle dynamics. It is then tempting to resort to more precise interpolation schemes than $P_0$, which are likely to contribute to reducing the overall spatial discretization error.  Nevertheless, such schemes remain delicate to implement in unstructured meshes. Furthermore, considering higher-order interpolations would essentially be beneficial if they can be applied in the different cells crossed by particles during one time step. In that sense, the interpolation issue appears as complementary to the development of cell-to-cell integration schemes. In the following, we want to concentrate on building such a cell-to-cell integration method. Therefore, we have chosen to retain the simplest case, namely the $P_0$ interpolation scheme for the sake of simplicity.

Therefore, questions (1) and (3) have been addressed and we can turn our attention to the question (2) and how time integration of the SDEs is performed.

\subsection{Current numerical scheme for particle transport}\label{sec:integration_scheme}

Once mean fields at particle location have been determined, the next step consists in devising a numerical scheme to predict the fluid particle properties making up the particle state vector $\mbs{Z}=(\mbs{X},\mbs{U})$. This typically introduces the notion of a time step $\Delta t$ for the integration of the particle equations of motion and discrete approximations of $\mbs{Z}$ at times $t^n=n\Delta t$, noted $\mbs{Z}^n$. These discrete approximations $\mbs{Z}^n$ are obtained by successive updates ({i.e.}, computing $\mbs{Z}^{n+1}$ from $\mbs{Z}^n$), using an integrated form of the RHS of Eqs.~\eqref{SLM_disc}. In the frame of present hybrid methods, the current algorithm involves a two-step process:
\begin{itemize}
 \item A time-integration scheme, which allows to obtain $\mbs{Z}^{n+1}$ from $\mbs{Z}^n$ using known values of the mean fields at particle location entering the SDEs in Eqs.~\eqref{SLM_disc};
 \item A trajectory algorithm, which determines the location of each particle in the spatial domain. This second step is needed in hybrid FV/PDF approaches since we must know in which cell each particle is with respect to the mesh defined for the computation of the fluid phase. This information is necessary in order to update the values of the mean fields at each particle location for the next step of the time-integration process.
\end{itemize}
In the following, we briefly recall the details of the current algorithm used for each step.

\subsubsection{Time-integration scheme to predict the particle state vector}\label{sec:intregration_scheme:position}

The numerical scheme used to integrate the SDEs, cf. Eqs.~\eqref{SLM_disc}, has been described in previous papers~\cite{peirano2006mean, minier2003weak}, which provide comprehensive information on its derivation and main characteristics (see, in particular, the complete description of the different steps leading to its construction in~\cite{peirano2006mean}). Therefore, only the salient aspects are recalled here along with the resulting formulation.

\smallskip

\paragraph*{Requirements} The time-integration scheme has been developed according to the following guidelines (more details can be found in \cite{peirano2006mean, minier2003weak}):
\begin{itemize}
\item The numerical scheme is explicit (for simplicity reasons);
\item The numerical scheme is unconditionally stable (this is of key interest when the time step cannot be reduced, as in hybrid methods, to respect potential stability criteria);
\item The numerical scheme corresponds to the exact solution when the mean fields and timescales entering the modeled equations are constant;
\item The numerical scheme should capture the correct physical behaviors in the limit cases when the time step becomes much larger than the physical timescales \cite{peirano2006mean}.
\end{itemize}
The main reason behind these requirements is that the time step used in hybrid FV-PDF formulations is generally the same for the fluid solver and for the particle solver. This is mandatory for unsteady flows, where fluid and particle properties must be obtained at the same time. Consequently, since the time step is usually imposed by the fluid solver (to properly compute the flow field on a given mesh), it cannot always be reduced/adapted for the particle solver (as could be done in PDF stand-alone approaches). In complex flows in which mean fluid velocities can drastically vary from one area to another one, this means that we do not control the number of cells that are crossed by particles during a time step. Furthermore, when the fluid timescales are also widely different, it is important to properly capture the diffusive limit rather than limiting the time step for the whole particle set \cite{peirano2006mean}.

\paragraph*{Chosen scheme} The guiding principle underlying the present time-integration scheme is to express first the integrated form of the SDEs (using an assumption of constant properties) which leads directly to an exponential analytical form. In the general case, when mean fields and timescales are not constant but vary in space and time, the idea is to use Euler-like schemes by freezing their values at the beginning of the time step in the integrated form rather than in the SDEs (see a detailed description in~\cite{peirano2006mean}). Thanks to the formulation with exponential factors, unconditional stability is then automatically guaranteed. Although second-order scheme are quite possible using the same approach (see~\cite{peirano2006mean}), we limit ourselves to a first-order formulation in the following since our main concern is about the prediction of relevant mean fields in such schemes. Consequently, the resulting discrete numerical scheme for fluid particles writes:
\begin{subequations}
\begin{align}
X_{i}^{n+1} &= X_{i}^n +U_{i}^n T_L^n \Bigg( 1- \exp\left(-\frac{\Delta t}{T_L^n}\right) \Bigg) + C_i^n T_L^n \Bigg(   \Delta t - T_L^n \bigg( 1- \exp\left(-\frac{\Delta t}{T_L^n}\right) \bigg)  \Bigg) \nonumber \\
 & + \underbrace{ \sqrt{\frac{C_0 \epsilon^n (T_L^n )^3}{2}} \frac{\Bigg(1-\exp\left(-\frac{\Delta t}{T_L^n}\right)\Bigg)^2}{\sqrt{1-\exp\left( - 2\frac{\Delta t}{T_L^n}\right)}}\; \zeta_{i}^U  +\sqrt{ C_0 \epsilon^n  (T_L^n)^2 \Bigg( \Delta t - 2 T_L^n \frac{1-\exp\left(-\frac{\Delta t}{T_L^n}\right)}{1+\exp\left(-\frac{\Delta t}{T_L^n}\right)} \Bigg)}\; \zeta_{i}^X}_{I_{i}^X(\Delta t)}, \label{expo_scheme_simple_X} \\
U_{i}^{n+1} &= U_{i}^n \exp\left(-\frac{\Delta t}{T_L^n}\right) +  C_i^n T_L^n \Bigg( 1- \exp\left(-\frac{\Delta t}{T_L^n}\right)\Bigg) + \underbrace{\sqrt{C_0 \epsilon^n \frac{T_L^n}{2}\Bigg(1-\exp\left(-2\frac{\Delta t}{T_L^n}\right)\Bigg)} \; \zeta_{i}^U}_{I_{i}^U(\Delta t)}. \label{expo_scheme_simple_U}
\end{align}
\label{expo_scheme_simple}
\end{subequations}
The two terms $\zeta_i^X$ and $\zeta_i^U$ correspond to independent random numbers sampled in a normalized Gaussian distribution $\mathcal{N}(0,1)$. As indicated in Eqs.~\eqref{expo_scheme_simple}, these random numbers intervene in the Choleski decomposition of the two correlated stochastic integrals, namely $I_{i}^U(\Delta t)$ and $I_{i}^X(\Delta t)$ (see descriptions in~\cite{peirano2006mean}).

\begin{figure}
 \centering
  \pgfmathsetseed{1}
 \begin{tikzpicture}[ultra thick, scale=0.8, transform shape]
 \coordinate(P) at (0,0);
  \draw[dotted, ->, orangeedf,line width =3pt] (P)++(0.5*rnd,0.5*rnd-0.5*0.5) coordinate(P) node[black!50,scale=1.75]{$\bullet$} node[below left,black!50, scale =1.5]{$\{\vect{X}^n, \, \vect{U}^n\}$}
 \foreach \i in {0,1,...,20}
{
   (P) -- ++(0.6*rnd,0.7*rnd-0.7*0.5) coordinate(P)
}
node[black,scale=1.75]{$\bullet$} node[above, right ,black, scale =1.5]{$\{\vect{X}^{n+1}, \,\vect{U}^{n+1}\}$};
 \draw (3,-1.5) node[greenedf, scale =1.5]{$\boldsymbol{+}$ \textbf{Knowledge of }$(T_L^n, \epsilon^n, \lra{P}^n,\lra{\vect{U}}^n)$};

 \end{tikzpicture}
 \caption{Sketch illustrating the time-integration with the current numerical scheme: the particle position $X_{i}^{n+1}$ and velocity $U_{i}^{n+1}$ at the next iteration are computed using information on the particle position and velocity at the previous iteration $n$ as well as values of the fluid characteristics at the position $X_{i}^{n}$.}
 \label{fig:numerical_scheme:time_integration}
\end{figure}
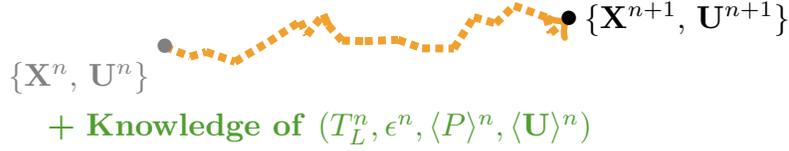

As displayed in \figurename{}~\ref{fig:numerical_scheme:time_integration}, the time-integration scheme predicts the particle state vector at the next time step $(\mbs{X}^{n+1},\mbs{U}^{n+1})$ using information on its values at the beginning of the time step $(\mbs{X}^n,\mbs{U}^n)$ and local fluid characteristics at the position $\mbs{X}^n$.

\subsubsection{Trajectory algorithm for spatial location}\label{sec:intregration_scheme:trajecto}

The time-integration scheme requires knowledge on the fluid mean fields at the particle position at a given time ({e.g.} , fluid velocity, turbulent dissipation rate), which are known on a given mesh. As a result, the time-integration scheme has to be supplemented by a trajectory algorithm to locate each particle and assign the corresponding cell.

\paragraph*{Assumptions} The trajectory algorithm has to satisfy the following requirements:
\begin{itemize}
 \item First, since we resort to a NGP approach, the trajectory algorithm is expected to provide the cell in which each particle is located.
 \item Second, the trajectory algorithm has to be applicable in the case of generic 3-D unstructured meshes, with cells being star-shaped around their center. This choice is motivated by the typical meshes used in the computation of the average flow-field.
 \item Third, the trajectory algorithm assumes a free-flight motion of particles during each time step. This means that, during each time step, particles are moving in a straight line from their initial position $X_i^n$ to their final position $X_i^{n+1}$.
\end{itemize}

\begin{figure}[h!]

    \begin{minipage}{0.5 \linewidth}
        \begin{tikzpicture}[ultra thick]
       \draw[blueedf,dashed]   (0,0)  rectangle (8,2);
       \draw[blueedf,dashed]   (4,2)  -- (4 ,0);
       \draw[orangeedf,dotted,->] (2.4,0.6)node[black!50,scale=1.75]{$\bullet$} node[above left ,black!50, scale =1.2]{$\{\vect{X}^n, \, \vect{U}^n\}$} -- (6.3,1.4) node[black,scale=1.75]{$\bullet$} node[below = 5pt ,black, scale =1.2]{$\{\vect{X}^{n+1}, \,\vect{U}^{n+1}\}$};
 \draw (2,-0.5) node[greenedf]{ \textcolor{blueedf}{ \textbf{Cell i}}$ \boldsymbol{\rightarrow } $
 {\small
 $\begin{pmatrix}
  \ \  T_L^n \ \  , \ \epsilon^n\quad \ \ \\
   \lra{P}^n , \lra{\vect{U}}^n
\end{pmatrix}$}};
\draw (6,-0.5) node[blueedf]{\textbf{Cell j}};
       \end{tikzpicture}
    \end{minipage}
    \begin{minipage}{0.5 \linewidth}
      \begin{algorithmic}
       \For{time interval $n$ to $n+1$}
        \State Integrate $\mbs{X}^{n+1}$
        \State Draw vector $\mbs{\delta X}= \mbs{X}^{n+1} - \mbs{X}^{n}$
        \Do
         \State Determine if $\mbs{\delta X}$ leaves the cell through a face
         \State Update the current cell
         \State Increment the sub-iteration $m$
        \doWhile{A face is crossed}
        \State Track the final integrated position
       \EndFor
      \end{algorithmic}
    \end{minipage}
  \caption{Current trajectory algorithm used after a single integration for the whole time step: the cell in which the particle resides at the end of the time step is tracked using a successive neighbor search algorithm.}
  \label{fig:numerical_scheme:tracking}
\end{figure}
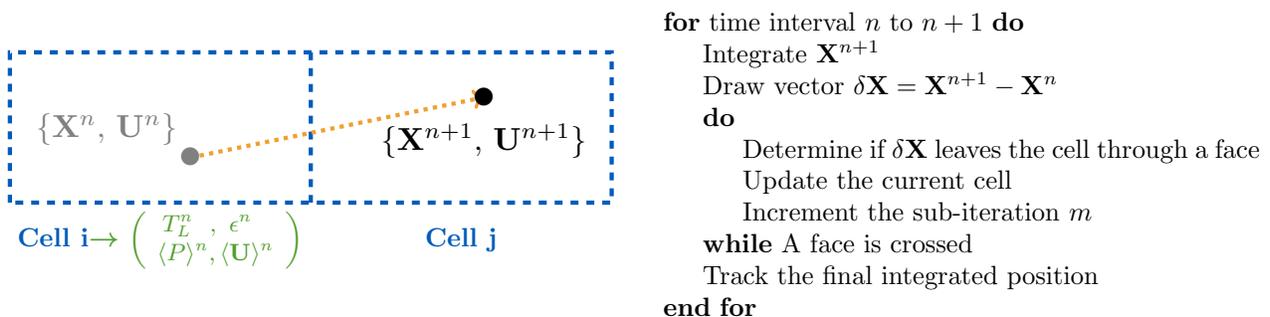

\newpage
\paragraph*{Chosen algorithm} With respect to the previous requirements, we resort to an algorithm based on the successive neighbor searches. Such tracking algorithms determine the new cell inside which a particle is by using information only on its initial and final positions (see also \figurename{}~\ref{fig:numerical_scheme:tracking}). The new cell is then determined by searching intersections between the trajectory vector (joining the initial and final position) and faces of the current cell. To do so, we rely on the Möller-Trumblemore algorithm \cite{moller1997fast} which has been developed to determine the intersections between a vector and triangular sub-faces. This process is then repeated to check if the final position is in the new cell or if another face is crossed by the trajectory vector. The details and validation of this tracking algorithm are provided in Appendix~\ref{sec:app:neighbor_searches}.

Such tracking algorithms have been retained since they have been shown to be efficient for unstructured meshes ({e.g.} , for Finite Difference approaches in \cite{lohner1990vectorized} or for Finite Volume approaches in \cite{muradoglu2002local, subramaniam2000probability}). Moreover, in the case of unstructured meshes, these tracking algorithms are more adequate than simple locating algorithms (which determine the location only using the information on the final position).

At this point, it is worth mentioning that additional conditions can be taken into account for each of the faces that are crossed during the tracking algorithm. In particular, when the crossed face corresponds to a physical boundary, specific boundary conditions can be added: for instance, particles can be removed from the simulation when reaching outlet boundaries while boundary conditions can be applied to wall surfaces (interested readers are referred to \cite{dreeben1997wall, minier1999wall,bahlali2020well}). Similarly, specific conditions can be added to properly treat periodicity by translation and/or periodicity by rotation.

\section{A new algorithm based on cell-to-cell integration}
\label{sec:new_alg}

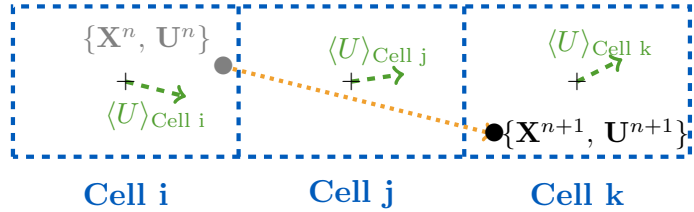
\begin{figure}[h]
 \centering
  \begin{tikzpicture}[ultra thick]
       \draw[blueedf,dashed]   (0,0)  rectangle (9,2);
       \draw[blueedf,dashed]   (3,2)  -- (3 ,0);
       \draw[blueedf,dashed]   (6,2)  -- (6 ,0);

		\draw[greenedf,->,dashed] (1.5,1) node[black]{$+$} -- (2.3,0.8) node[midway,below,scale=1.1]{$\lra{U}_{\text{Cell i}}$};
		\draw[greenedf,->,dashed] (4.5,1) node[black]{$+$} -- (5.2,1.1) node[midway,above,scale=1.1]{$\lra{U}_{\text{Cell j}}$};
		\draw[ greenedf,->,dashed] (7.5,1) node[black]{$+$} -- (8.1,1.3) node[midway,above,scale=1.1]{$\lra{U}_{\text{Cell k}}$};

       \draw[orangeedf,dotted,->] (2.8,1.2)node[black!50,scale=1.75]{$\bullet$} node[ above  left,black!50,scale=1.1]{$\{\vect{X}^n, \, \vect{U}^n\}$} -- (6.4,0.3) node[black,scale=1.75]{$\bullet$} node[right =-2pt ,black,scale=1.1]{$\{\vect{X}^{n+1}, \,\vect{U}^{n+1}\}$};

\draw (1.5,-0.5) node[blueedf,scale=1.2]{\textbf{Cell i}};
\draw (4.5,-0.5) node[blueedf,scale=1.2]{\textbf{Cell j}};
\draw (7.5,-0.5) node[blueedf,scale=1.2]{\textbf{Cell k}};

       \end{tikzpicture}

 \caption{Sketch illustrating the issue with the current numerical scheme over large time steps in non-uniform flows: the particle can cross many cells in a single time step, meaning that the fluid characteristics (including velocities) in the intermediate cells are not taken into account in its trajectory.}
 \label{fig:numerical_scheme:limits}
\end{figure}

In the context of present hybrid FV/PDF approaches, the $P_0$-interpolation assumption for mean-field values means that the exponential scheme described in Section~\ref{sec:integration_scheme} is not only stable but provides an exact solution for the particle state vector as long as particles remain in the same cell they started from during the whole time step. However, in such hybrid formulations, the time step is often imposed by the mean fluid flow computation and cannot always be reduced so as to respect this criterion. This is especially encountered in disperse two-phase flow simulations with discrete particles having inertia that can range over several orders of magnitude, as indicated above. In the present work, we limit ourselves to fluid particles but this general picture is an indication that we are faced with a situation where, during a given time step, a particle can cross several cells. The inherent assumption made in Euler schemes, and therefore also in the present exponential one, retains only mean-field values and timescales evaluated at the beginning of the time step, that is at the initial particle location. In non-uniform flows, with potential variations of mean flow field quantities, this implies that a spatial discretization error is introduced, as illustrated in \figurename{}~\ref{fig:numerical_scheme:limits}.

\newpage
\paragraph*{Objectives} The aim of this paper is to develop a new algorithm that remains accurate even for non-uniform flows in the case of large time steps when particles can cross several cells. For that purpose, we extend the numerical scheme described in Section~\ref{sec:integration_scheme} while keeping similar assumptions/requirements:
\begin{enumerate}[a)]
 \item it should be explicit;
 \item it should be unconditionally stable with respect to the time step;
 \item it should be exact for constant flow mean fields (and thus within a given cell in line with the $P_0$ interpolation);
 \item it should take into account spatial variations of flow mean fields encountered over large time steps.
\end{enumerate}
\subsection{Leading Principle: a cell-to-cell integration for large time steps}\label{sec:new_alg:leading_principle}

\begin{figure}[h]
  \begin{minipage}{0.58\linewidth}
    \input{tikz/new_scheme_cell_to_cell}
  \end{minipage}
  \begin{minipage}{0.42\linewidth}
    \begin{algorithmic}
     \For{time interval $n$ to $n+1$}
      \Do
       \State Get relative intersection time $\theta_{[m+1]}$
       \State Integrate until the face
       \State Update cell and remaining time
       \State Increment sub-iteration $m$
      \doWhile{A face is crossed}
     \EndFor
    \end{algorithmic}
  \end{minipage}
 \caption{Sketch illustrating the principle of a cell-to-cell integration for large time steps and the corresponding algorithm.}
 \label{fig:new_scheme:cell_to_cell}
\end{figure}
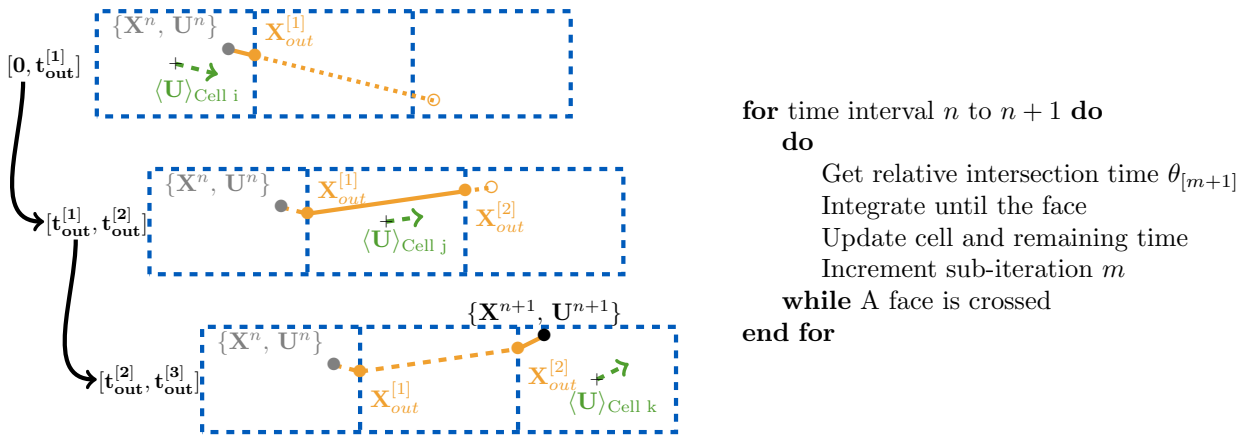

\paragraph*{Principle} To improve the numerical accuracy in non-uniform flows and in the case of large time steps, the idea is to extend the present exponential scheme with a cell-to-cell integration. As illustrated in \figurename{}~\ref{fig:new_scheme:cell_to_cell}, the leading principle consists in splitting the time step in several sub-iterations: the integration is stopped every time the particle exits a cell, such that each sub-iteration corresponds to the motion of the particle within one cell. By doing so, it is straightforward to account for the change of flow characteristics every time a particle enters a new cell. Since we need to detect whenever a particle crosses an interface between two adjacent cells, the following two steps are applied for each sub-iteration : (1) a time-integration scheme and (2) a trajectory algorithm. When the time step is small enough for a particle to remain inside the same cell during a whole time step, the new scheme is the same as the one previously described.

\newpage
\paragraph*{Additional points} Compared to the two-step process described in Section~\ref{sec:integration_scheme}, the new cell-to-cell integration brings out two additional questions:
\begin{itemize}
 \item When does a particle exit a cell?
 \item Where does a particle exit a cell?
\end{itemize}
This means that the trajectory algorithm has to be modified since it only provides information on the new cell (simple tracking). The new trajectory algorithm should provide information on the exit time $t_{\text{out}}$ and exit location $\mbs{X}_{\text{out}}$. These two additional pieces of information are indeed required to compute the motion of the particle during the remaining part of the time step. In fact, as displayed in \figurename{}~\ref{fig:new_scheme:cell_to_cell}, the motion of the particle after crossing the face is given by a time-integration scheme starting at $\mbs{X}_{\text{out}}$ over the remaining time $\Delta t-t_\text{out}$. In that sense, the present algorithm draws on the cell-to-cell integration proposed by \cite{popov2008accurate} while extending it to respect the non-anticipating requirement for stochastic integrals in the It\^{o} sense (this is discussed in the next paragraph).

\paragraph*{The key issue of non-anticipating estimations}
Implementing the principle of the cell-to-cell algorithm may seem straightforward. A naive formulation would indeed consider that, starting from a particle position $\mbs{X}^n$ at time $t^n$, we apply the exponential scheme, cf. Eqs.~\eqref{expo_scheme_simple}, to predict $\mbs{X}^{n+1}$ at time $t^{n+1}$ and, then, use the trajectory algorithm to determine $\mbs{X}_{\text{out}}$ and the exit time of that cell $t_{\text{out}}$ (see the top figure in \figurename{}~\ref{fig:new_scheme:cell_to_cell}). However, whatever the geometrical method used to determine these quantities, it is clear that the resulting time $t_{\text{out}}$ will depend on the random numbers $\boldsymbol{\zeta}$ which, in Eqs.~\eqref{expo_scheme_simple}, represent the normalized increments of the Wiener process in the stochastic model, cf. Eqs.~\eqref{SLM_cont}. When combining the different diffusion coefficients obtained from each sub-iteration (for instance, the three sub-iterations depicted in \figurename{}~\ref{fig:new_scheme:cell_to_cell}), the resulting diffusion coefficient will then be a function of the future of the Wiener process since the random numbers $\boldsymbol{\zeta}$ stand for the normalized values of $\mbs{W}^{n+1}-\mbs{W}^n$. In other words, the misleadingly simple approach will yield diffusion coefficients that anticipate on the future of the Wiener process. This is in direct violation of the very definition of the stochastic integrals in the It\^{o} sense, thereby inducing spurious overall drift and diffusion numerical values (see a similar analysis in \cite{minier2003comment}).

\paragraph*{Assumptions} Drawing on the aforementioned analysis, the new numerical scheme is designed so as to respect the following additional assumptions/constraints:
\begin{enumerate}[i)]
 \item It should respect the non-anticipation rule with respect to the Wiener process;
 \item The trajectory algorithm is based on a neighbor search;
 \item The trajectory algorithm considers free-flight motion of particles within each sub-iteration.
\end{enumerate}

\subsection{Non-anticipating estimations of intermediate time steps using virtual partners}
\label{sec:new_alg:description}

To obtain an estimate of the fraction of time spent by a particle in a given cell that remains independent of the future of the Wiener process entering the stochastic terms, a virtual partner is associated to each particle within each time step of the computation. This means that at the beginning of each time step $t=t^n$, this virtual partner starts from the same location as the real particle which is considered but moves in a deterministic manner (to be precised below) based only on the particle variables at time $t^n$ and local mean-field values. The fraction of time spent by this virtual partner in the corresponding cell is therefore, by construction, free of any statistical dependence with the Wiener process driving the random terms and provides an estimate for the time spent by the real particle within that cell.

In the following, details about the algorithm are first provided in Section~\ref{sec:new_alg:description:overview}. The consistence of this algorithm with the current integration scheme is then demonstrated in Section~\ref{sec:new_alg:description:verif}.
\newpage
\subsubsection{Overview of the new algorithm}
\label{sec:new_alg:description:overview}

As indicated, the new algorithm is based on a cell-to-cell integration for large time steps. This implies that each time iteration is split in a series of sub-iterations, each one corresponding to the motion of a particle in a given cell. As a result, the algorithm combines a number of successive time-integration and trajectory steps. During each time step, the virtual partners are used in the trajectory steps to determine the cell in which the particle is as well as to estimate the exit time and location.

In practice, the algorithm is composed of the following steps, which are depicted in \figurename{}~\ref{fig:new_scheme_details}:
\begin{itemize}
 \item[step-1 ] Based on the particle initial location $\mbs{X}^{[0]}$ and the knowledge of the fluid mean fields in that cell, a deterministic estimation of the particle position at the end of the time step $\Delta t$, noted $\widehat{\mbs{X}^{[1]}}$, is made. Note that the superscript $[m]$ means that it corresponds to the sub-iteration number $m$ within the cell-to-cell integration. This is achieved by forcing the stochastic integrals to zero in the exponential time-integration scheme, which gives the following equation for $\widehat{\mbs{X}^{[1]}}$:
 \begin{equation}
  \widehat{\mbs{X}^{[1]}} = \mbs{X}^{[0]} +\mbs{U}^{[0]} T_L^{[0]} \left\lbrace 1- \exp\left(-\frac{\Delta t}{T_L^{[0]}}\right) \right\rbrace + C_i^{[0]} T_L^{[0]} \left( \Delta t - T_L^{[0]} \left\lbrace 1- \exp\left(-\frac{\Delta t}{T_L^{[0]}} \right) \right\rbrace \right)~.
 \end{equation}
It is important to note that $\widehat{\mbs{X}^{[1]}}$ corresponds to the mean conditional particle location at time $\Delta t$ conditioned on the fact that its initial location is $\mbs{X}^{[0]}$, that is $\widehat{\mbs{X}^{[1]}}=\lra{\mbs{X}(\Delta t) \, | \, \mbs{X}(0)=\mbs{X}^{[0]} }$. Therefore, if we release a number of particles at $\mbs{X}^{[0]}$ at time $t=0$, $\widehat{\mbs{X}^{[1]}}$ represents the position of the center of mass of the released cloud at time $t=\Delta t$.

 \item[step-2 ] Then, the trajectory algorithm based on a free-flight assumption is applied to determine if the particle leaves/remains in the cell. To avoid any anticipation issue, the trajectory algorithm is applied on the virtual partner and not on the real particle. The motion of the virtual partner is assumed to follow a straight line between the first location of the virtual partner (initialized at the particle position $\widetilde{\mbs{X}^{[0]}} = \mbs{X}^{[0]}$) and $\widehat{\mbs{X}^{[1]}}$. The trajectory algorithm is detailed in Appendix~\ref{sec:app:neighbor_searches}. Its outcome allows to distinguish between two cases:
 \begin{itemize}
  \item[step-2a ] The virtual partner leaves the cell. In that case, the trajectory algorithm provides information on the face through which the virtual particle exits the cell but also on the exit time $\theta^{[1]} \Delta t$ (see the extension of the trajectory algorithm in Appendix~\ref{sec:app:neighbor_searches:extended}). Thanks to the free-flight assumption, the exit location of this virtual partner, $\widetilde{\mbs{X}^{[1]}}$, is derived directly from the exit time. This case will activate a next sub-iteration to compute the motion of the particle during the remaining time (see Step-3a).
  \item[step-2b ] The virtual partner remains in the cell. In that case, $\theta^{[1]} = 1$, and the position of the virtual partner at the end of the time step is equal to the first estimated position $\widetilde{\mbs{X}^{[1]}}=\widehat{\mbs{X}^{[1]}}$. This case will not activate a new sub-iteration and the cell-to-cell integration will be stopped (see Step-3b).
 \end{itemize}

 \item[step-3 ] The position $\mbs{X}^{[1]}$ and velocity $\mbs{U}^{[1]}$ of the particle at the end of this sub-iteration are then computed using the full time-integration scheme, {i.e.}, including the stochastic integrals (see Eq.~\eqref{expo_scheme_simple}). The only issue is to determine the amount of time that the particle has spent in the cell. From step-2, there are two possibilities:
 \begin{itemize}
  \item[step-3a ] If the virtual partner exits the cell (step-2a), the elapsed time retained for the prediction of ($\mbs{X}^{[1]}$, $\mbs{U}^{[1]}$) is taken as being equal to the fraction of time computed for the virtual partner $\theta^{[1]} \Delta t$.

\smallskip
  Then, we have to pursue the computation of the particle motion during the remaining time $\Delta t^{[1]} = (1 - \theta^{[1]})\Delta t$. In the spirit of this cell-to-cell algorithm, this is achieved by repeating the previous three steps and updating the information required at each step (see sub-iterations 2 and 3 in \figurename{}~\ref{fig:new_scheme_details}). In step-1 of the second sub-iteration, the mean conditional particle position $\widehat{\mbs{X}^{[2]}}$ at the end of the new time step $\Delta t^{[1]}$ is estimated starting from the particle position $\mbs{X}^{[1]}$ and particle velocity $\mbs{U}^{[1]}$. The mean fields required to estimate $\boldsymbol{\widehat{X}}^{[2]}$ are now ($T_L(t,\boldsymbol{\widetilde{X}}^{[1]})$, $C_i(t,\boldsymbol{\widetilde{X}}^{[1]})$) taken in the cell in which the virtual partner is. Afterward, step-2 is applied to compute the trajectory of the virtual partner, assuming that it starts at the last exit location  $\widetilde{\mbs{X}^{[1]}}$ and moves towards the estimated mean conditional position $\widehat{\mbs{X}^{[2]}}$. It is worth noting that the starting position is taken as the last exit location so that the trajectory of the virtual partner remains continuous. Subsequently, from the estimation of $\theta_2$ obtained in step-2, the position $\mbs{X}^{[2]}$ and velocity $\mbs{U}^{[2]}$ are computed in step-3 of this second sub-iteration.

\smallskip
  This three-step process is actually repeated until the virtual partner reaches a cell in which it remains for the whole remaining time step $\Delta t^{[m]}$ (recall that the superscript $[m]$ corresponds to the sub-iteration number $m$).

  \item[step-3b ] If the virtual particle remains in the cell (step-2b), the elapsed time is actually equal to the remaining time step $\Delta t^{[m]}$ ({i.e.} $\theta^{[m]}=1$). The position of the virtual partner is equal to the last estimated position made in step-1 $\widetilde{\mbs{X}^{[m]}}=\widehat{\mbs{X}^{[m]}}$. In that case, the cell-to-cell iterative integration is stopped.

\item[Final step ] When the virtual partner finally reaches a cell in which it remains for the remaining time, we obtain the particle position at the end of the whole time step $\mbs{X}(t+\Delta t)$. Yet, as can be seen in \figurename{}~\ref{fig:new_scheme_details}, the particle might end up in a different cell than the virtual partner due to the stochastic terms. For that reason, we apply a last trajectory algorithm: we track the change of cell from the virtual partner last position $\widetilde{\mbs{X}^{[m]}}$ to the particle position $\mbs{X}(t+\Delta t)$. This last trajectory step ensures that the particle is associated to the correct cell inside which it resides at the end of the time step. This ensures that no error on the location is introduced for the next time step.

 \end{itemize}
\end{itemize}
\begin{figure}[h!]
 \begin{minipage}{1.0 \linewidth}
 \centering
  \input{tikz/summary_new_alg}
 \end{minipage}
\begin{minipage}{1.0 \linewidth}
 \centering
 \begin{algorithmic}
 \For{time interval $n$ to $n+1$}
  \Do
   \State (a) Integrate {\color{orangededf}$\widehat{\mbs{X}^{[m+1]}}$} from {\color{red} $\mbs{X}^{[m]}$} with $I^Z = 0$
   \State (b) Compute the exit point {\color{greenmf}$\widetilde{\mbs{X}^{[m+1]}}$} using the vector $\mbs{{\color{greenmf}\widetilde{\mbs{X}^{[m]}}}{\color{orangededf}\widehat{\mbs{X}^{[m+1]}}}}$
    \\ \hspace{4.2em} Compute the exit time {\color{violet}$\theta^{[m+1]} dt^{[m]}$ } with {\color{violet}$\theta^{[m+1]}$} $ = \frac{|\mbs{{\color{greenmf}\widetilde{\mbs{X}^{[m]}} \widetilde{\mbs{X}^{[m+1]}}}}|}{|\mbs{{\color{greenmf}\widetilde{\mbs{X}^{[m]}}}{\color{orangededf}\widehat{\mbs{X}^{[m+1]}}}}|}$
   \State (c) Integrate {\color{red} $\mbs{X}^{[m+1]}$} from {\color{red} $\mbs{X}^{[m]}$} using the exit time {\color{violet}$\theta^{[m+1]} \dd  t^{[m]}$} and random numbers $(\zeta^{Z})^{[m+1]}$
   \State (d) Update cell, flow fields and remaining time  {\color{violet}$\dd t^{[m+1]} = (1-\theta^{[m+1]}) \dd t^{[m]}$}
    \State Increment the sub-iteration $m$
  \doWhile{(A face is crossed)}
  \State Track the final integrated position
 \EndFor
 \end{algorithmic}
\end{minipage}
\caption{Summary of the proposed algorithm.}
\label{fig:new_scheme_details}
\end{figure}

A few comments are in order. At first sight, it could be argued that the difference between the position of the virtual partner $\widetilde{\mbs{X}^{[m]}}$ and the position of the particle $\mbs{X}(t+\Delta t)$ at the last sub-iteration points to an error in the algorithm. However, it should be recalled that this difference is due to the stochastic terms that are taken into account when computing the particle position but not in the position of the virtual partner (which is fully deterministic). At this stage, it is worth emphasizing that, within each time step of the computation, the role of the virtual partner is essentially to provide information on the time spent in each cell and to determine the neighboring cells that a particle can cross during the time step, while ensuring that the anticipation issue is avoided. This further supports the choice of a tracking algorithm based on successive neighbor searches (see Section~\ref{sec:intregration_scheme:trajecto}), since it naturally provides information on the successive cells crossed by a particle during a time step (contrary to simple localization algorithms). At each sub-iteration of the algorithm, the increments of this virtual partner represent the mean conditional displacement and, in that sense, remain coherent with the underlying physical process modeled. Each virtual partner is therefore an auxiliary in the calculation of the (true) particle motion. Furthermore, we are basically interested in developing weak approximations, which means capturing particle dynamics in a statistical sense. As a result, what really matters is to obtain accurate estimations of statistics extracted from particle dynamics through Monte Carlo methods ({e.g.} , average concentration, mean velocity), that is from an ensemble of particles. In other words, improvements in the prediction of individual particle variables (location as well as velocity) should always be assessed in a statistical sense. For that purpose, we now turn our attention to the resulting behavior of such particle statistics in the rest of the paper.

\subsubsection{Consistency analysis of the time step decomposition}\label{sec:new_alg:description:verif}

Before considering proper validation test cases in Section~\ref{sec:results}, it is important to check that the new algorithm meets the criterion (c) set forth at the beginning of Section~\ref{sec:new_alg}.

\paragraph*{Verification procedure} To that effect, we consider a situation where all mean fields and timescales are constant. In that case, we already know that the current numerical scheme, cf. Eqs.~\eqref{expo_scheme_simple} in which $T_L$, $C_i$ and $\epsilon$ are now constant, provides the exact solution in the weak sense. We then consider the new algorithm and assume that one sub-iteration has taken place (the extension to several sub-iterations being immediate) at a time $\theta \Delta t$, implying that the particle state vector $\mbs{Z}^{n+1}$ is now obtained from $\mbs{Z}^n$ as the sum of two iterations (one over $\theta \Delta t$ and one over $(1-\theta) \Delta t$). The issue is then to check that, still in the weak sense, both predictions are identical.

In the following, we leave out the direction index $i$ since this is basically a 1D formulation. Starting from a given particle state vector value $\mbs{Z}^n$ at time $t^n$, predictions obtained in one iteration ({i.e.}, the current algorithm) are indicated by the index $[1]$, while those obtained with two sub-iterations ({i.e.}, the new algorithm) are indicated by the index $[2]$.

Predictions in one iteration are expressed directly by Eqs.~\eqref{expo_scheme_simple}, which are rewritten here as
\begin{subequations}
\begin{align}
(X^{n+1})^{[1]} &= X^n + U^n T_L \left( 1- \exp\left(-\frac{\Delta t}{T_L}\right) \right) + C T_L \left[ \Delta t - T_L \left( 1- \exp\left(-\frac{\Delta t}{T_L}\right) \right)  \right] + (I^X)^{[1]} (\Delta t)~, \label{verif_one-go_X} \\
(U^{n+1})^{[1]} &= U^n \exp\left(-\frac{\Delta t}{T_L}\right) +  C T_L \left[( 1- \exp\left(-\frac{\Delta t}{T_L}\right)\right] + (I^U)^{[1]}(\Delta t). \label{verif_one-go_U}
\end{align}
\label{verif_one-go}
\end{subequations}
The two stochastic integrals are represented by the two correlated centered Gaussian random variables $(I^X)^{[1]}(\Delta t)$ and $(I^U)^{[1]} (\Delta t)$ which are fully determined by the matrix of their second-order moments~\cite{peirano2006mean}
\begin{subequations}
\begin{align}
\bigg \langle\left( (I^U)^{[1]} (\Delta t) \right)^2\bigg \rangle & = C_0 \epsilon \frac{T_L}{2}\left(1-\exp\left(-2\frac{\Delta t}{T_L}\right)\right)~, \label{correl_IU2}\\
\bigg \langle (I^U)^{[1]} (\Delta t) \, (I^X)^{[1]} (\Delta t) \bigg \rangle & = C_0 \epsilon \frac{T_L^2}{2}\left(1-\exp\left(-\frac{\Delta t}{T_L}\right)\right)^2~, \label{correl_IUIX} \\
\bigg \langle \left( (I^X)^{[1]} (\Delta t) \right)^2 \bigg \rangle & = C_0 \epsilon T_L^2 \left\{ \Delta t - \frac{T_L}{2} \left[ 1-\exp\left(-\frac{\Delta t}{T_L} \right)\right] \left[ 3-\exp\left(-\frac{\Delta t}{T_L} \right)\right]  \right\}~. \label{correl_IX2}
\end{align}
\label{correl_integrals}
\end{subequations}

In the case of two iterations, a first estimation of $\mbs{Z}(\theta \Delta t)$ is made with Eqs.~\eqref{verif_one-go} using the two correlated Gaussian random variables noted $(I^U)^{[2]}(\theta \Delta t)$ and $(I^X)^{[2]}(\theta \Delta t)$ and a second iteration is performed over the remaining time interval $(1 - \theta)\Delta t$ to obtain the prediction of $\mbs{Z}^{n+1}$ using two other correlated Gaussian random variables noted $(I^U)^{[2]}((1-\theta) \Delta t)$ and $(I^X)^{[2]}((1-\theta) \Delta t)$. Note that $\left[(I^U)^{[2]}((1-\theta) \Delta t),(I^X)^{[2]}((1-\theta) \Delta t)\right]$ are independent of $\left[(I^U)^{[2]}(\theta \Delta t),(I^X)^{[2]}(\theta \Delta t)\right]$. Combining the two gives
\begin{subequations}
\begin{align}
(X^{n+1})^{[2]} & = X^n + U^n T_L  \left( 1 - \exp\left(-\frac{\Delta t}{T_L}\right)\right) + C T_L \left[ \Delta t - \exp \left(-\frac{\Delta t}{T_L}\right)\right] \nonumber \\
    &  \hspace{0.05 \linewidth} +\underbrace{(I^X)^{[2]}(\theta \Delta t) +  (I^X)^{[2]} ((1- \theta) \Delta t) + T_L (I^U)^{[2]}(\theta \Delta t) \left( 1 - \exp \left(-\frac{(1- \theta) \Delta t}{T_L}\right)\right) }_{(\widetilde{I}^X)^{[2]}( \theta, \Delta t) }, \label{decomposed_value_eq_x} \\
(U^{n+1})^{[2]} &= U^n \exp \left(-\frac{\Delta t}{T_L}\right) +  C T_L \left( 1- \exp \left(-\frac{\Delta t}{T_L}\right)\right)\nonumber \\
     & \hspace{0.2 \linewidth} + \underbrace{(I^U)^{[2]}(\theta \Delta t)  \exp \left(-\frac{(1 - \theta)\Delta t}{T_L} \right)+ (I^U)^{[2]}((1- \theta) \Delta t)}_{(\widetilde{I}^U)^{[2]}(\theta,\Delta t)}. \label{decomposed_value_eq_u}
\end{align}
\label{decomposed_value_eq}
\end{subequations}

The variance of $(\widetilde{I}^U)^{[2]}(\theta,\Delta t)$ is easily calculated and is:
\begin{subequations}
\begin{align}
\bigg\langle \left(  (\widetilde{I}^U)^{[2]}(\theta,\Delta t) \right)^2 \bigg\rangle & = \bigg\langle \left(  (I^U)^{[2]}(\theta \Delta t) \right)^2 \bigg\rangle \exp \left(- \frac{2(1-\theta) \Delta t}{T_L}\right)+ \bigg\langle \left( (I^U)^{[2]} ((1- \theta) \Delta t) \right)^2 \bigg\rangle \\
& = C_0 \epsilon \frac{T_L}{2} \left[ \left(1-\exp \left(- \frac{2\theta \Delta t}{T_L}\right) \right) \exp \left(- \frac{2(1-\theta) \Delta t}{T_L}\right) + 1 - \exp \left(- \frac{2(1-\theta) \Delta t}{T_L}\right) \right] \\
& = C_0 \epsilon \frac{T_L}{2} \left[ 1- \exp \left(- \frac{2 \Delta t}{T_L}\right) \right] \\
& = \bigg\langle \left( (I^U)^{[1]}(\Delta t) \right)^2 \bigg\rangle, \label{eq:I_U2}
\end{align}
\end{subequations}
where the last equality comes from Eq.~\eqref{correl_IU2}. Similarly, the variance of $(\widetilde{I}^X)^{[2]}( \theta, \Delta t)$ is obtained as:
\begin{eqnarray}
\label{eq:I_X2}
\bigg\langle \left( (\widetilde{I}^X)^{[2]}( \theta, \Delta t) \right)^2 \bigg\rangle & = T_L^2 \left[ 1 - \exp \left(-\dfrac{(1- \theta) \Delta t}{T_L}\right)\right]^2 \bigg\langle \left( (I^U)^{[2]}(\theta \Delta t) \right)^2 \bigg\rangle  \nonumber \\
      & + 2 T_L \left[ 1 - \exp \left(-\dfrac{(1- \theta) \Delta t}{T_L}\right)\right] \bigg\langle \left( (I^X)^{[2]}(\theta \Delta t) (I^U)^{[2]}(\theta \Delta t)  \right)\bigg\rangle \\ \nonumber
      & + \bigg\langle \left( (I^X)^{[2]}(\theta \Delta t) \right)^2 \bigg\rangle + \bigg\langle \left( (I^X)^{[2]} ((1- \theta) \Delta t) \right)^2 \bigg\rangle~,
\end{eqnarray}
while the covariance is
\begin{multline}
\bigg\langle \left( (\widetilde{I}^X)^{[2]}( \theta, \Delta t) (\widetilde{I}^U)^{[2]}(\theta,\Delta t) \right) \bigg\rangle = T_L \left[ 1 - \exp \left(-\dfrac{(1- \theta) \Delta t}{T_L}\right)\right] \exp \left(-\dfrac{(1- \theta) \Delta t}{T_L}\right) \bigg\langle \left( (I^U)^{[2]}(\theta \Delta t) \right)^2 \bigg\rangle\\
  +\exp \left(-\dfrac{(1- \theta) \Delta t}{T_L}\right) \bigg\langle \left( (I^X)^{[2]} (\theta \Delta t) (I^U)^{[2]} (\theta \Delta t)  \right)\bigg\rangle  + \bigg\langle \left( (I^X)^{[2]} ((1- \theta) \Delta t) (I^U)^{[2]} ((1- \theta) \Delta t)  \right) \bigg\rangle~.
\label{eq:I_UI_X}
\end{multline}

The formulas in Eqs.~\eqref{correl_integrals} can then be applied, using either $\theta \Delta t$ or $(1 - \theta)\Delta t$ as the proper time interval, and injected in Eqs.~\eqref{eq:I_X2} and~\eqref{eq:I_UI_X}. Tedious but straightforward calculations show then that we have
\begin{subequations}
\begin{align}
\bigg\langle \left( (\widetilde{I}^X)^{[2]}( \theta, \Delta t) \right)^2 \bigg\rangle & = \bigg\langle\left( (I^X)^{[1]} (\Delta t) \right)^2\bigg\rangle  \\
\bigg\langle \left( (\widetilde{I}^X)^{[2]}( \theta, \Delta t) (\widetilde{I}^U)^{[2]}(\theta,\Delta t) \right) \bigg\rangle & = \bigg\langle (I^U)^{[1]} (\Delta t) (I^X)^{[1]} (\Delta t)\bigg\rangle~,
\end{align}
\end{subequations}
which, with Eq.~\eqref{eq:I_U2}, proves that $\left[(\widetilde{I}^X)^{[2]}( \theta, \Delta t),(\widetilde{I}^U)^{[2]}( \theta, \Delta t)\right]$ is statistically equivalent to $\left[(I^X)^{[1]} (\Delta t), (I^U)^{[1]} (\Delta t)\right]$ and, consequently, that Eqs.~\eqref{decomposed_value_eq} are identical to Eqs.~\eqref{verif_one-go}.

\paragraph*{An important remark} These calculations remain valid even when $\theta$ is a function of $\mbs{Z}^n$ and of the mean fields and timescale (here $C$ and $T_L$). However, it is worth emphasizing that the fact that $\theta$ is independent of the random variables representing the stochastic integrals is a crucial point in the above verification and that these properties would break down otherwise.
Indeed in this spurious case we would have:
\begin{equation}
    \bigg\langle \exp \left(-\dfrac{(1- \theta) \Delta t}{T_L}\right)  (I^{Z_1})^{[2]}( \theta \Delta t)  (I^{Z_2})^{[2]}( \theta \Delta t) \bigg\rangle \neq \bigg\langle \exp \left(-\dfrac{(1- \theta) \Delta t}{T_L}\right)\bigg\rangle \bigg\langle (I^{Z_1})^{[2]}( \theta \Delta t)  (I^{Z_2})^{[2]}( \theta \Delta t) \bigg\rangle,
\end{equation}
with $Z_1$ and $Z_2$ either $X$ or $U$ and:
\begin{equation}
\bigg\langle \exp \left(-\dfrac{(1- \theta) \Delta t}{T_L}\right)\bigg\rangle \neq \exp \left(-\dfrac{(1- \theta) \Delta t}{T_L}\right).
\end{equation}

This reflects the non-anticipation issue brought out repeatedly throughout this section.

Having carried out this verification test case, we can now consider validation test cases. This is addressed in the next section.

\section{Numerical results}
\label{sec:results}

Having checked that all the criteria set forth at the beginning of Section~\ref{sec:new_alg} are met, the new algorithm is now validated. For that purpose, it has been implemented in the open source CFD solver code\_saturne \cite{archambeau2004code} and numerical results are compared to analytical ones in two situations:
\begin{itemize}
 \item The first case consists in checking that statistics obtained from particle-attached variables by Monte-Carlo methods are in line with analytical expressions when all mean fields and timescales are constant (i.e. a uniform mean flow). This is done in Section~\ref{sec:results:point_source} using a point source particle dispersion in a stationary homogeneous isotropic turbulent flow.
 \item The second case corresponds to a simple but challenging non-uniform flow involving curved streamlines, which allows to assess the accuracy of the new algorithm in complex situations where classical particle-tracking algorithms often encounter limitations (see Section~\ref{sec:results:Couette}).
\end{itemize}

\subsection{Validation in a uniform flow}
\label{sec:results:point_source}

Drawing on the verification test case carried out in Section~\ref{sec:new_alg:description:verif}, the aim is twofold: first, to check that, with the new proposed time-splitting method, numerical results are exact when flow properties are constant in time and space; second, to bring out discrepancies induced when using an anticipation method even in such a simple case. To that end, the dispersion of particles released from a point source is analyzed in a stationary homogeneous isotropic turbulent flow. This choice is motivated by the fact that, in such a context, analytical formulas can be derived \cite{minier2003weak}.

In the following, we start by describing the system considered, including physical aspects ({e.g.}, flow characteristics), as well as numerical ones ({e.g.}, spatial discretization, time step) and theoretical results. Then, numerical results are validated with respect to these theoretical expressions.

\subsubsection{System considered: point source dispersion in homogeneous isotropic turbulence}
\label{sec:results:point_source:case}

\paragraph*{Physical aspects} The case studied consists of a point-source (fluid) particle dispersion in a stationary homogeneous isotropic turbulent flow. This means that the flow is uniform and stationary within the domain considered and that it is periodic in all directions. Here, we have imposed a velocity based on the turbulent dispersion $ U_\alpha = \sqrt{\langle u^2_\alpha \rangle}= \SI{1}{m.s^{-1}}$ and a Lagrangian time scale $T_L =$ \SI{1}{s}.

\paragraph*{Numerical aspects} Simulations were carried out using the CFD software code\_saturne. The mean flow field is imposed taking into account an extra artificial forcing to maintain a constant level of energy within the system \cite{pope2000turbulent}. This leads to a flow field with a constant dissipation rate equal to $\epsilon=  \frac{2}{C_0} \frac{U_\alpha^2}{T_L}$ where the proper closure of $T_L$ for homogeneous stationary flow is $T_L=4k/(3C_0\epsilon)$ \cite{pope2000turbulent,minier2001pdf} with $k=3/2U_\alpha^2$, and $C_0$ the Kolomogorov constant in the diffusion coefficient of the Langevin model.

Once fluid mean fields are obtained, \num{100000} fluid particles are injected at the center of the box. Their trajectories are computed using the new algorithm described in Section~\ref{sec:new_alg}. The total size of the domain (box) is taken large enough so that the particles do not reach boundaries during each simulation.

\paragraph*{Theoretical aspects} As shown in Section~\ref{sec:new_alg:description:verif}, the algorithm is exact when all mean fields and timescales are constant. By resorting to It\^{o}'s calculus, it is then straightforward to derive analytical formulas for the variances $\lra{X^2}$ and $\lra{U^2}$ as well as for the covariance $\lra{XU}$ as a function of time, taking $t_0=0$ as the initial time when particles are released. From these analytical solutions, we can extract two limit cases: the ballistic ({i.e.}, when $t \ll T_L$) and the diffusive one ({i.e.}, when $t \gg T_L$). This was done in \cite{minier2003weak}, yielding the following expressions in these two limit cases:

\begin{minipage}{0.45 \linewidth}
\vspace{2ex}
 \center{ Ballistic limit case ($t \ll T_L$)}
 \begin{subequations}
  \begin{align}
   \langle X^2 \rangle \sim & \ C_0 \frac{\epsilon t^2 T_L}{2}  = U_\alpha^2\, t^2, \\
   \langle XU \rangle \sim & \, \, \, \, \, \   C_0 \epsilon\, t^2 \, \, \, \,  = 2 U_\alpha^2\frac{ t^2}{T_L},\\
   \langle U^2 \rangle \sim & \, \, \, \, \,  \ C_0 \epsilon\, t \, \, \, \, \, \,  \,= 2 U_\alpha^2\frac{t}{T_L} .
  \end{align}
 \label{ballistic_regime_eq}
 \end{subequations}
\end{minipage}
\begin{minipage}{0.45 \linewidth}
\vspace{2ex}
 \center{ Diffusive limit case ($t \gg  T_L$)}
 \vspace{0ex}
 \begin{subequations}
  \begin{align}
   \langle X^2 \rangle \sim & \ C_0 \epsilon  T_L^2 t = 2 U_\alpha^2 T_L t,  \phantom{ \frac{1}{1}} \\
   \langle XU \rangle \sim &  \ \frac{C_0 \epsilon T_L^2 }{2}= U_\alpha^2 T_L, \\
   \langle U^2 \rangle \sim &\  \frac{C_0 \epsilon T_L }{2}= U_\alpha^2.
  \end{align}
 \label{diffuse_regime_eq}
 \end{subequations}
\end{minipage}

It is worth mentioning that, in the diffusive regime, the only physically-relevant statistic is $\langle X^2 \rangle$ since the instantaneous particle velocity does not play a role anymore in the particle position evolution equation. This corresponds to the notion of fast-variable elimination (here $U$ becomes a fast variable that can be eliminated), which is addressed extensively in standard textbooks \cite{gardiner1985handbook} as well as in previous works both from the theoretical standpoint \cite{minier2001pdf,minier2016statistical} and the numerical one \cite{peirano2006mean}. The constant value of the velocity second-order moment still retains its typical kinetic energy meaning (showing that we are actually dealing with particles in contact with a kind of thermostat) but the correlation $\langle XU \rangle$, which is less physically-meaningful, is also considered since both results are useful to discuss statistical noise in Monte Carlo estimations.

\subsubsection{Validation}
\label{sec:results:point_source:validation}

In the following, we check that numerical results are in agreement with  analytical formulas regardless of the number of occurrences the time-splitting algorithm is called in two cases:

\begin{itemize}
 \item A first set of simulations was performed with a focus on the ballistic regime. For that purpose, the time step was taken equal to $\Delta t =0 .05 T_L$. Two spatial discretizations were considered: $\Delta x= U_\alpha \Delta t / 50 $ and $\Delta x= 10  U_\alpha \Delta t $. These two grid sizes were carefully chosen so that first one (resp. the second one) corresponds to the case where the average particle displacement during one time step is much smaller (resp. much larger) than the grid size $\Delta x$. The simulations were run for a total time equal to $6 T_L$.

Figures~\ref{fig:disp_short_time_im_xx_new}-\ref{fig:disp_short_time_im_ux_new} display the second-order moments as a function of the dimensionless time $t^* = t/T_L$, comparing results obtained with the new algorithm to  analytical formulas. It can be seen that the outcomes of the new algorithm remain exact for both cases. This validates the new algorithm (which has been activated on average $50$ times every time step in the case where $\Delta x = U_\alpha \Delta t/50$). Small differences between results obtained with the cell-to-cell algorithm and analytical values are simply related to statistical errors using Monte Carlo methods (due to a finite number of fluid particles), as discussed below.

As can be seen in Figs.~\ref{fig:disp_short_time_im_xx_naive}-\ref{fig:disp_short_time_im_ux_naive}, completely different results are obtained when using an anticipation method, such as the naive formulation outlined in Section~\ref{sec:new_alg:leading_principle}. This is especially revealed by the strong dependence on the cell size, which indeed governs how often the time-splitting algorithm is applied. Whereas results remain roughly acceptable in the case of a large-enough cell size (meaning that the new algorithm is seldom called), numerical predictions quickly deteriorate when smaller cell sizes are considered (i.e. using $\Delta x = U_\alpha \Delta t / 50 $), therefore inducing severe errors.

 \begin{figure}[h!]
    \begin{minipage}{0.5 \linewidth}
     \begin{subfigure}{0.8\linewidth}
     \centering
   \includegraphics[width = 1.1 \linewidth]{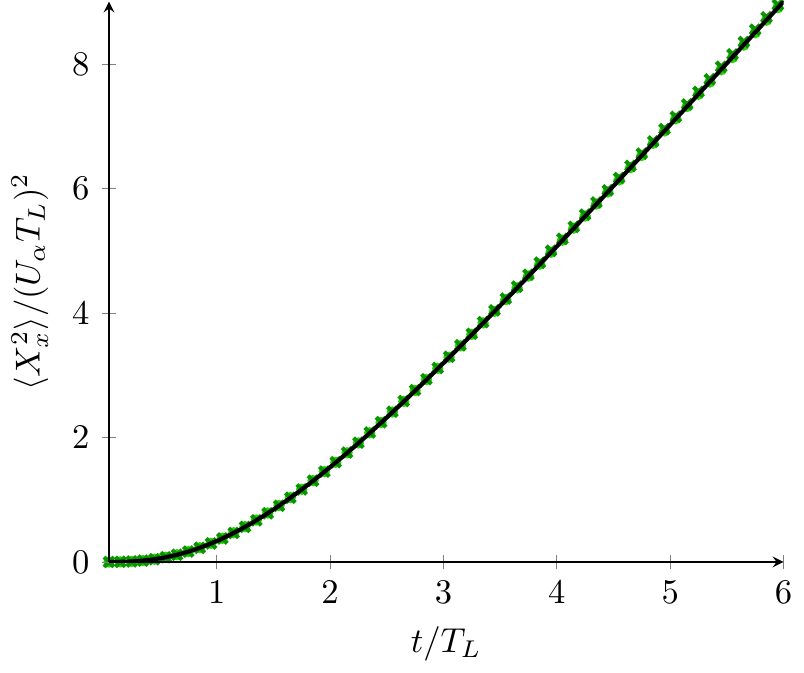}
   \caption{$\langle XX \rangle $ using the proposed method.}
    \label{fig:disp_short_time_im_xx_new}
    \end{subfigure}
  \begin{subfigure}{0.8 \linewidth}
   \includegraphics[width = 1.1 \linewidth]{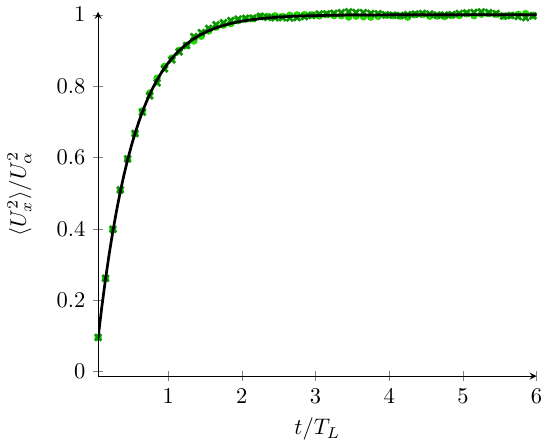}
   \caption{$\langle UU \rangle $ using the proposed method.}
    \label{fig:disp_short_time_im_uu_new}
  \end{subfigure}
  \begin{subfigure}{0.8 \linewidth}
   \includegraphics[width = 1.1 \linewidth]{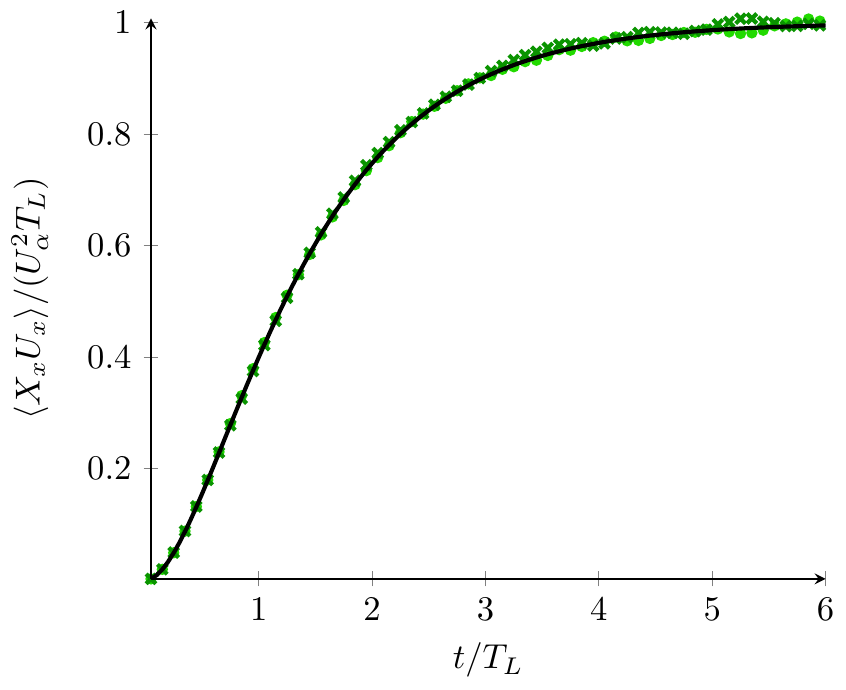}
   \caption{$\langle XU \rangle $ using the proposed method.}
    \label{fig:disp_short_time_im_ux_new}
  \end{subfigure}
         \end{minipage}
    \begin{minipage}{0.5 \linewidth}

     \begin{subfigure}{0.8\linewidth}
   \includegraphics[width = 1.1 \linewidth]{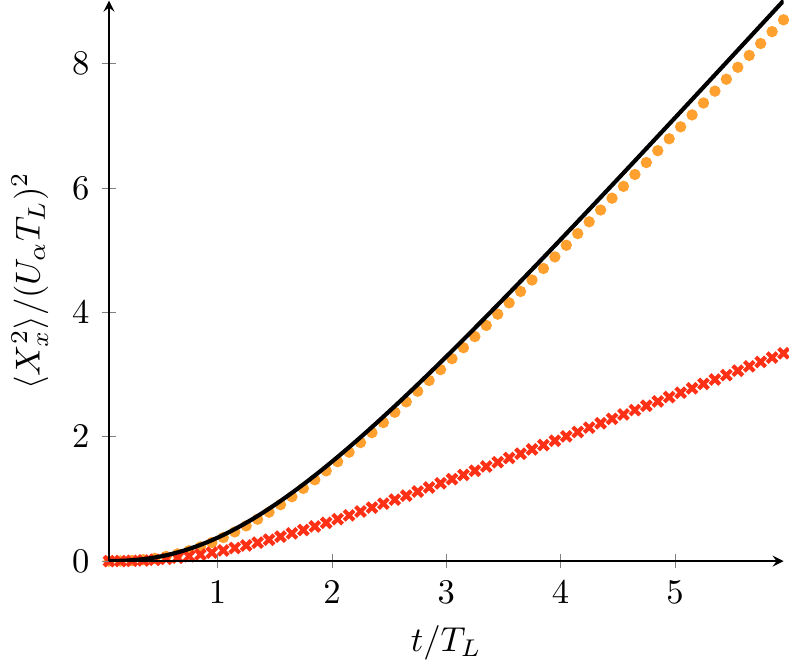}
   \caption{$\langle XX \rangle $ using an anticipating method.}
     \label{fig:disp_short_time_im_xx_naive}
    \end{subfigure}
  \begin{subfigure}{0.8 \linewidth}
   \includegraphics[width = 1.1 \linewidth]{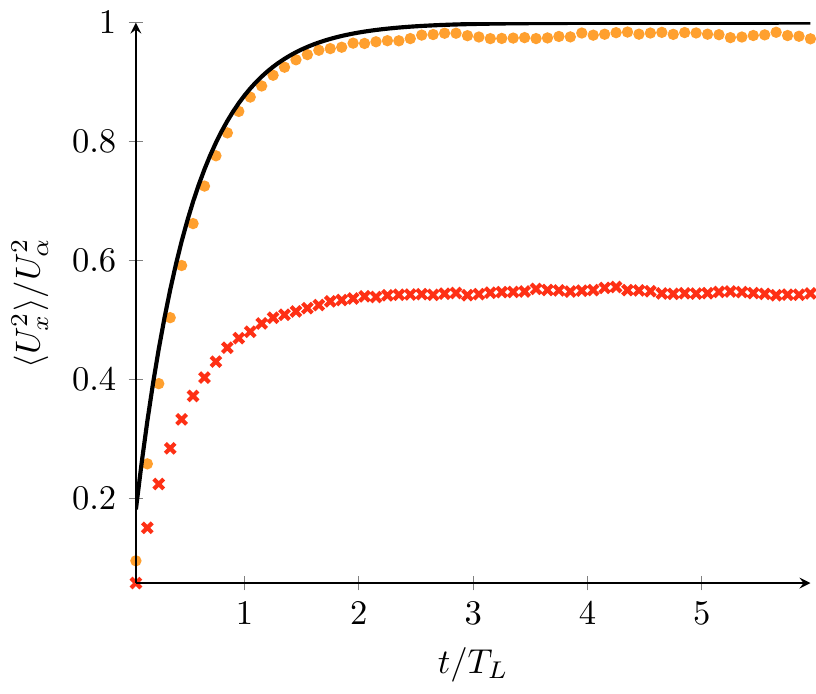}
   \caption{$\langle UU \rangle $ using an anticipating method.}
   \label{fig:disp_short_time_im_uu_naive}
  \end{subfigure}
  \begin{subfigure}{0.8 \linewidth}
   \includegraphics[width = 1.1 \linewidth]{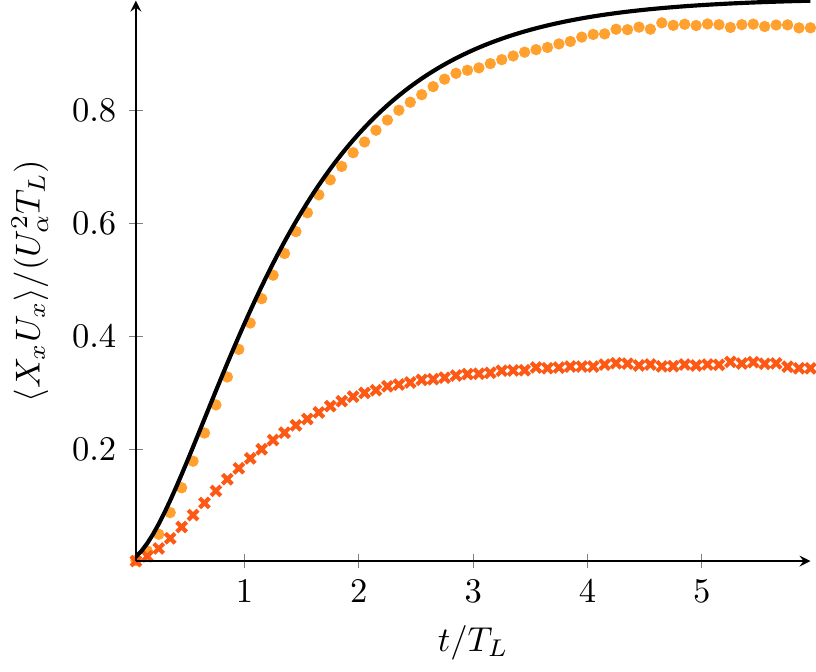}
   \caption{$\langle XU \rangle $ using an anticipating method.}
    \label{fig:disp_short_time_im_ux_naive}
  \end{subfigure}
         \end{minipage}

  \caption{Evolution of $\langle XX \rangle $,  $\langle UU \rangle $, $\langle UX \rangle $ as a function of the dimensionless time in the ballistic limit case with a time step $\Delta t = 0.05 T_L$. Two spatial refinements are considered: $\times$ (resp. $\bullet$) corresponds to simulations with a cell size $\Delta x = U_\alpha \Delta t / 50 $ (resp. $10\  U_\alpha \Delta t$). On the left part, Comparison between the analytical solution (black line) and numerical results obtained with an anticipating method. On the right part, errors obtained with the new algorithm. The dashed lines correspond to the envelop for the \num{99}\% confidence interval (analytical formula).}
  \label{fig:disp_short_time_im}
 \end{figure}

 \item A second set of simulations was carried out in the diffusive limit case. For that purpose, the time step was taken equal to $\Delta t = 200 T_L$. To assess that numerical results are independent of the number of times the new algorithm is applied, two spatial discretizations were also considered: $\Delta x= U_\alpha \Delta t / 20$ and $\Delta x= 2.5 \ U_\alpha \Delta t.$ The simulations were run for a total time equal to \num{24000} $T_L$.

Figures~\ref{fig:disp_long_time_im_xx_new}-\ref{fig:disp_long_time_im_ux_new} display the evolution of the second-order moments as a function of time, indicating that the new algorithm provides accurate results regardless of the number of occurrences the time-splitting algorithm is used within a time step. In \figurename{}~\ref{fig:disp_long_time_im_xx_new}, it is seen that $\langle X^2 \rangle$ follows a linear evolution right from the outset, as it should be in the diffusive regime, and that the slope is properly reproduced when using non anticipating methods. This demonstrates that the particle dispersion coefficient is well captured. As mentioned above, it is interesting to consider the numerical predictions of $\langle U^2 \rangle$ in Fig.~\ref{fig:disp_long_time_im_uu_new} and $\langle XU \rangle$ in Fig.~\ref{fig:disp_long_time_im_ux_new} to bring out the statistical noise inherent to Monte Carlo methods. When considering $\langle U^2 \rangle$, the variance of the Monte Carlo estimator is constant in time since $\text{Var}(U^2)=2\, \langle U^2 \rangle^2$ and the \num{99}\% confidence interval is indicated by the two horizontal lines shown in Figs.~\ref{fig:disp_long_time_im_uu_new} and~\ref{fig:disp_long_time_im_uu_naive}. However, for the correlation $\langle XU \rangle$, the variance of the estimator is a function of time since $\text{Var}(XU)(t)=(C_0\epsilon)^2T_L^4/2\, ( 1 + t/T_L)$ and the envelope lines limiting the \num{99}\% confidence interval are now increasing with time, as displayed  in Figs.~\ref{fig:disp_long_time_im_ux_new} and~\ref{fig:disp_long_time_im_ux_naive}. Note again that the resulting increasing level of noise for $\langle XU\rangle$ is a mere observable and has no feedback effect on the particle simulation.

On the other hand, as can be seen in Fig.~\ref{fig:disp_long_time_im_xx_naive}, an anticipating method, such as the naive formulation outlined in Section~\ref{sec:new_alg:leading_principle}, is unable to reproduce the correct dispersion coefficient, especially when the time-splitting algorithm is often called (in Fig.~\ref{fig:disp_long_time_im_xx_naive}, this corresponds to the case $\Delta x = U_\alpha \Delta t / 20 $ where 20 cells are crossed per iteration on average). For the sake of completeness, we also display $\langle U^2 \rangle$ in Fig.~\ref{fig:disp_long_time_im_uu_naive} and the correlation $\langle UX \rangle$ in Fig.~\ref{fig:disp_long_time_im_ux_naive}, which further confirms that an anticipation method yields results that fluctuate but around incorrect averages.

  \begin{figure}[h!]

    \begin{minipage}{0.5 \linewidth}

     \begin{subfigure}{0.8\linewidth}
   \includegraphics[width = 1.1 \linewidth]{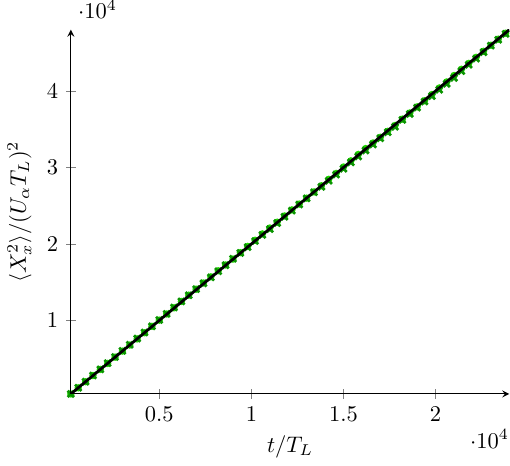}
   \caption{$\langle XX \rangle $ using the proposed method.}
    \label{fig:disp_long_time_im_xx_new}
    \end{subfigure}
  \begin{subfigure}{0.8 \linewidth}
   \includegraphics[width = 1.1 \linewidth]{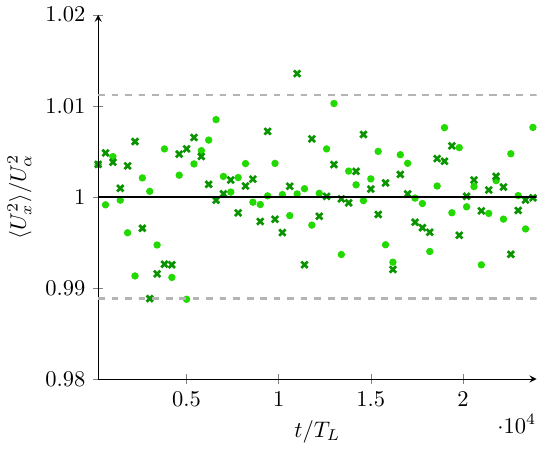}
   \caption{$\langle UU \rangle $ using the proposed method.}
    \label{fig:disp_long_time_im_uu_new}
  \end{subfigure}
  \begin{subfigure}{0.8 \linewidth}
   \includegraphics[width = 1.1 \linewidth]{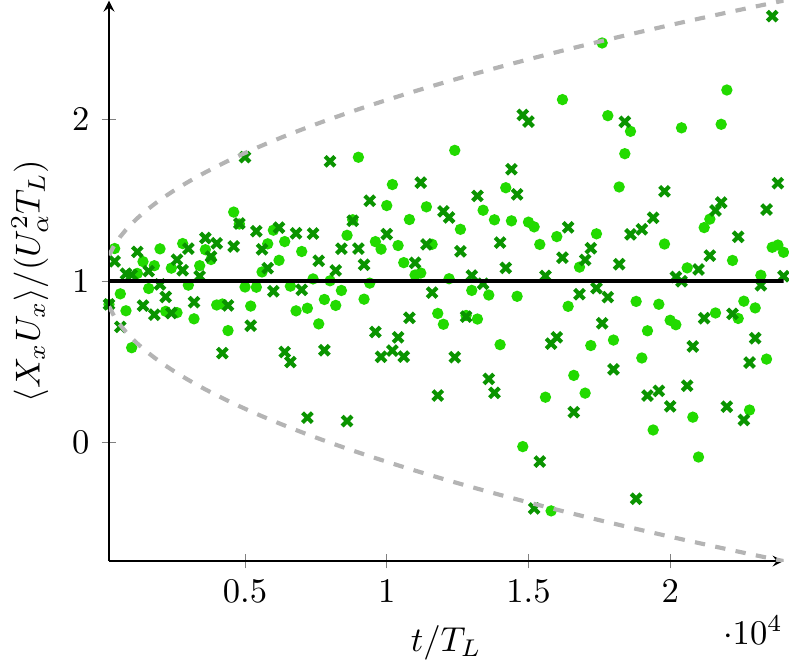}
   \caption{$\langle XU \rangle $ using the proposed method.}
    \label{fig:disp_long_time_im_ux_new}
  \end{subfigure}
         \end{minipage}
   \begin{minipage}{0.5 \linewidth}
     \begin{subfigure}{0.8\linewidth}
   \includegraphics[width = 1.1 \linewidth]{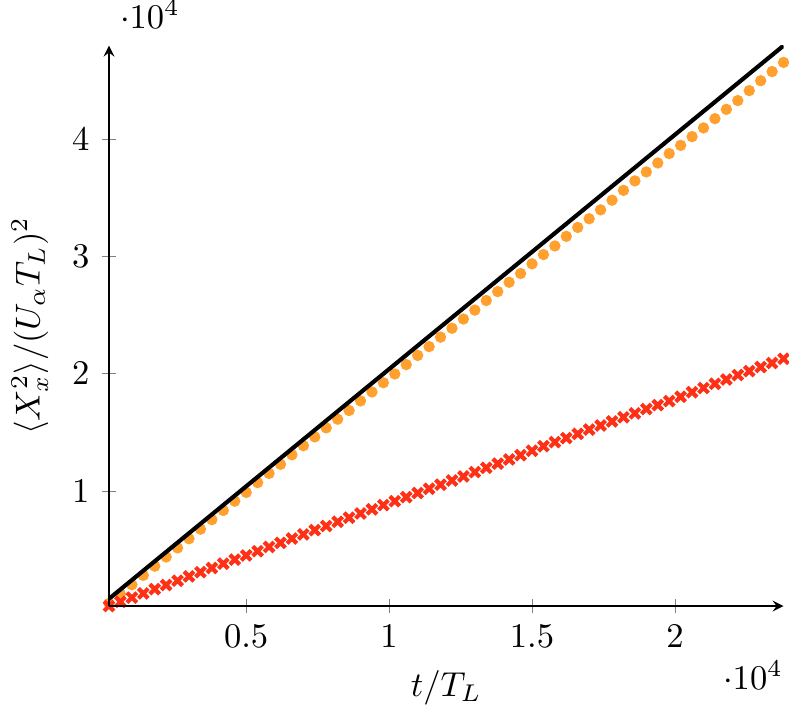}
   \caption{$\langle XX \rangle $ using an anticipating method.}
    \label{fig:disp_long_time_im_xx_naive}
    \end{subfigure}
  \begin{subfigure}{0.8 \linewidth}
   \includegraphics[width = 1.1 \linewidth]{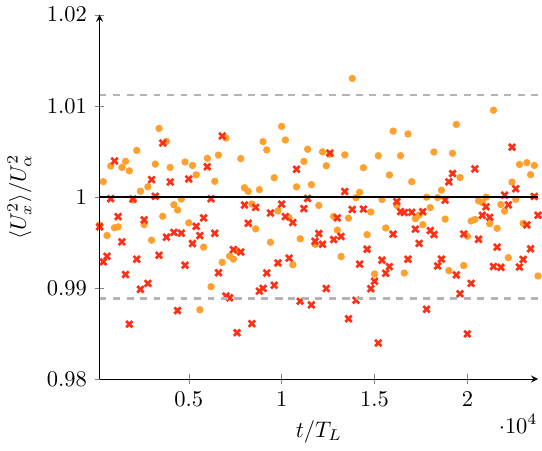}
   \caption{$\langle UU \rangle $ using an anticipating method.}
       \label{fig:disp_long_time_im_uu_naive}
  \end{subfigure}
  \begin{subfigure}{0.8 \linewidth}
   \includegraphics[width = 1.1 \linewidth]{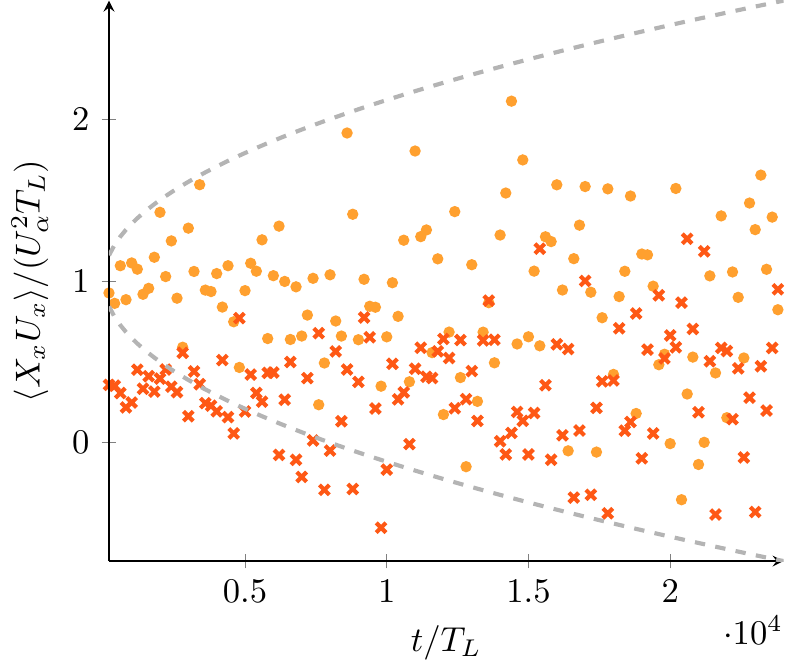}
   \caption{$\langle XU \rangle $ using an anticipating method.}
       \label{fig:disp_long_time_im_ux_naive}
  \end{subfigure}
    \end{minipage}
  \caption{Evolution of $\langle XX \rangle $,  $\langle UU \rangle $, $\langle UX \rangle $ as a function of the dimensionless time in the diffusive limit case with a time step $\Delta t = 200 T_L$. Two spatial refinements are considered: $\times$ (resp. $\bullet$) corresponds to simulations with a cell size $\Delta x = U_\alpha \Delta t / 20 $ (resp. $2.5\  U_\alpha \Delta t$). On the left part, Comparison between the analytical solution (black line) and numerical results obtained with an anticipating method. On the right part, errors obtained with the new algorithm. The dashed lines correspond to the envelop for the \num{99}\% confidence interval (analytical formula).}
  \label{fig:disp_long_time_im}
 \end{figure}

\end{itemize}

At this stage, it is worth emphasizing that present results were obtained in a 1-D dispersion case. However, these results are directly generalized to 3-D dispersion cases since each direction is treated independently. Furthermore, similar behaviors were obtained using 3-D unstructured meshes instead of regular 1-D Cartesian meshes (details are provided in \ref{sec:app:result_unstr_mesh}).

 \subsection{Validation in a non-uniform flow}
 \label{sec:results:Couette}

Having validated the new algorithm in a uniform flow, the idea is to validate the algorithm in the case of non-uniform flows. To that end, a laminar flow is considered so that the stochastic terms in the time-integration part of the algorithm are equal to zero. This ensures that the time-integration part is deterministic, hence allowing to check that the computed trajectory in non-uniform flows is exact. In addition, a cylindrical Couette flow has been chosen since it is one of the simplest non-uniform flows. In that case, each fluid particle is indeed expected to follow a purely circular motion around the center of rotation of the cylinder (see Section~\ref{sec:results:Couette}).

In the following, we start by describing the system considered, including both physical aspects ({e.g.}, geometry of the rotating cylinders, flow characteristics) and numerical aspects ({e.g.}, spatial discretization, time step). Then, the accuracy of the new trajectory algorithm is assessed. This validation is performed in two steps: first, we verify that the motion of a single particle actually follows a purely circular motion and, second, that statistics on particle concentration remain constant throughout the simulation time regardless of the time step used.

\subsubsection{System considered: a laminar cylindrical Couette flow}\label{sec:results:Couette:case}

\paragraph*{Physical parameters} The case studied here is a laminar cylindrical Couette flow. It consists of a fluid flow between two cylinders: the inner cylinder has a radius $r_{in} = \SI{1}{m}$, rotating at a given angular velocity equal to $\omega_\theta(r_{in}) = \SI{1}{s^{-1}}$, while the outer cylinder has a radius $r_{out} = \SI{2}{m}$ and remains at rest. This means that the fluid is contained within the annulus of thickness $\delta r = \SI{1}{m}$ that separates the two cylinders. The kinematic viscosity of the fluid is set to $\SI{1}{m^2 s^{-1}}$ so that the Reynolds number based on those quantities is equal to 1 ({i.e.}, well within the laminar regime).

In this configuration, the fluid is flowing in a cylindrical motion. As a result, the velocity field is unidirectional and, written with the cylindrical coordinate $\mbs{e_\theta}$, is given by the analytical solution:
\begin{equation}
 U = U_\theta(r) = \lra{U_\theta}(r) =  \dfrac{\omega_\theta(r_{in})}{\dfrac{r_{in}}{r_{out}} - \dfrac{r_{out}}{r_{in}}} \left( \dfrac{r}{r_{out}} - \dfrac{r_{out}}{r} \right)
\end{equation}
where $r$ is the distance from the rotating centre.

\paragraph*{Numerical parameters} The simulations are carried out using the CFD software code\_saturne to solve the Navier--Stokes equation on a \ang{360} polyhedral mesh: it comprises 360 cells in the azimuth direction (i.e. the azimuth discretization angle is $\Delta \theta = \ang{1}$) and 21 in the radial one. The simulations are run using a constant time step, whose value is taken between $\SI{0.055}{s}$ and $\SI{200}{s}$. This range of values has been chosen to assess the accuracy of the algorithm with respect to the time step, especially at large time steps where the average particle displacement is much greater than the grid size.

The tracking of fluid particles within this laminar cylindrical Couette flow is obtained by applying the algorithm described in Section~\ref{sec:new_alg}. The Lagrangian integral time $T_L$ has been set to $\SI{0}{s}$ to force a laminar flow ({i.e.}, all stochastic integrals are equal to zero).

\subsubsection{Accuracy of numerical results}
\label{sec:results:Couette:validation}

The new algorithm is here validated by checking that, first, the trajectory of a single particle follows a purely cylindrical motion and, second, that the initial particle concentration along the radial direction is conserved in time. In the following, all results are plotted using a dimensionless time defined as $t^+ = t \  \omega_\theta(r_{in}) / (\Delta \theta)$. This means that the dimensionless time step $\Delta t^+$ measures the ratio between the average particle displacement and the grid size (in the azimuthal direction) near the inner wall.

\paragraph*{Results on a single particle trajectory} The motion of a fluid particle in a laminar cylindrical Couette flow is expected to follow a purely circular trajectory. This means that each particle remains at a constant distance $r$ from the rotating centre throughout the simulation. A simulation has been run for 400 iterations using a time step of \SI{1.024}{s}. This means that the average particle displacement is equal to 55 times the grid size $\Delta \theta  \times r_{in}$ near the inner cylinder and to 0.5 times the grid size $\Delta \theta \times r_{out}$ near the outer one.

\begin{figure}[h]
    \centering
    \includegraphics[width = 0.45 \linewidth]{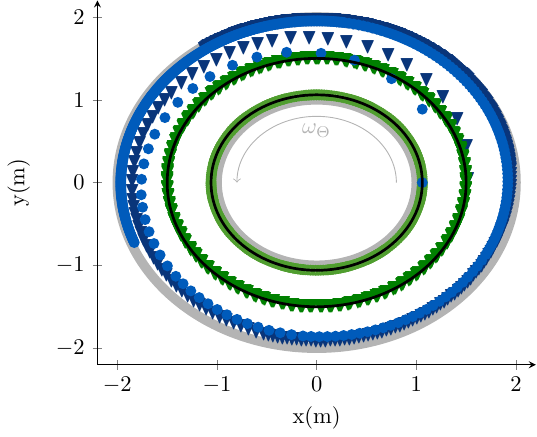}
    \includegraphics[width = 0.45 \linewidth]{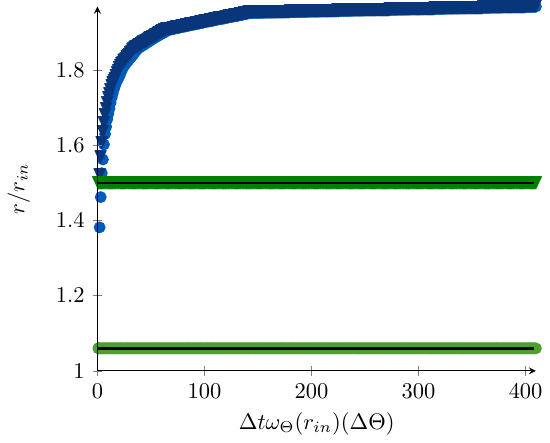}

    \caption{Trajectory of particles (left) and corresponding evolution of the radius $r$ as a function of the dimensionless time $t^+$ (right) for a constant time step $\Delta t^+ = 55$. Two particles are followed: one located initially in the vicinity of the wall $(\bullet)$ and one in the center of the domain $(\blacktriangledown)$. For each particle, the results obtained with the new algorithm (green) are in agreement with the expected trajectory (orange dotted line), while the results obtained with the reference scheme (blue) are flawed by spurious drift effects (leading to accumulation in the region near the outer cylinder).}
    \label{fig:Couette_one_part_traj_im}
\end{figure}

\figurename{}~\ref{fig:Couette_one_part_traj_im} displays the trajectory of two particles: one is initially located very close to the inner cylinder while the second one is initially located in the bulk. The trajectory (left-hand plot) clearly shows that the new algorithm does lead to a circular motion. These results confirm the accuracy of the new algorithm even for large values of the time step. This is further supported by the time-evolution of the radius $r$ (right-hand plot). In fact, the results are not distinguishable from the theoretical expectations, even for the particle initially close to the inner cylinder (which has circled the cylinder close to 60 times by the end of the simulation).

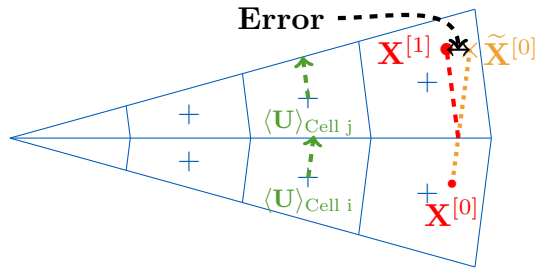
\begin{figure}[h]
 \centering

 \input{tikz/Reference_scheme_error}
 \caption{Scheme highlighting the origin of the numerical error in the algorithm without cell-to-cell integration. This error induces an accumulation of particles near the outer cylinder as time increases without cell-to-cell integration (see \figurename{}~\ref{fig:Couette_conc_im}).}
 \label{fig:Couette_diverg_tikz}
\end{figure}

\figurename{}~\ref{fig:Couette_one_part_traj_im} also displays the results obtained with the reference algorithm, which does not account for cell-to-cell integration. It can be seen that the radius increases with time all the more when the time step increases. This is due to the error made when no cell-to-cell integration is made while using large time steps. The origin of the error is illustrated in \figurename{}~\ref{fig:Couette_diverg_tikz}: the spatial discretization and $P_0$ interpolation mean that each particle experiences a constant velocity within a cell. As a particle crosses a face, if the velocity is not updated immediately, the particle will move for the remainder of the time step with the velocity encountered in the previous cell. Since this previous fluid velocity is slightly shifted with respect to the radial direction in the current new cell, it induces an error without cell-to-cell integration which is directly proportional to the time step.

\paragraph*{Results on particle concentration} We focus now our attention on simulations in which a large number of fluid particles is tracked. In that case, the statistics of interest is the particle concentration along the radial direction. In fact, since each fluid particle follows a circular orbit, the particle concentration should remain constant throughout time. With this result in mind, we verify here that the particle concentration is homogeneous in space provided that fluid particles are homogeneously distributed in space initially.

\begin{figure}[h]
 \centering
 \includegraphics[width=0.6\textwidth]{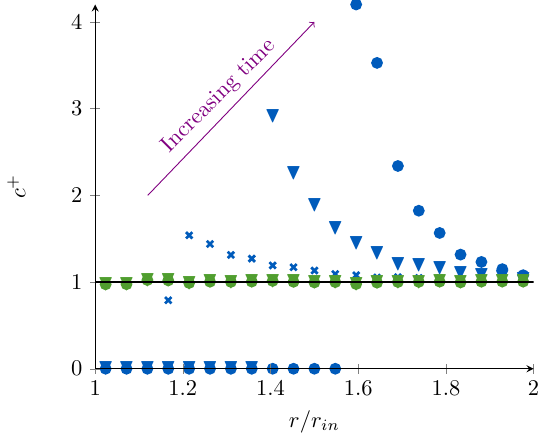}
 \caption{Evolution of the dimensionless concentration $c^+$ (concentration normalized by the initial concentration) as a function of the radius $r$. Results are displayed for various simulation times in the transitional regime:  $t^+$=275 $(\boldsymbol{\times})$, $t^+$=1100 $(\blacktriangledown)$ and $t^+$=2200 $(\bullet)$. The time step is constant such that $\Delta t^+\simeq 5.5$. The results obtained with the new algorithm (green) are in agreement with the analytical results (yellow) while the results obtained with the reference trajectory (blue) lead to an accumulation toward the outer cylinder which increases with time (arrow).}
 \label{fig:Couette_conc_im}
\end{figure}

Numerical simulations have been carried out with 200{,}000 fluid particles initially homogeneously distributed in space. The concentration is then analyzed by computing the particle number concentration in each of the 21 cells along the radial direction. \figurename{}~\ref{fig:Couette_conc_im} displays the evolution of the concentration, which has been normalized with the initial concentration, as a function of the distance $r$. It can be seen that, for a dimensionless time step equal to $\Delta t^+ \simeq 5.5$, the concentration obtained remains constant in all the domain throughout the whole simulated time. Meanwhile, the results obtained with the reference algorithm ({i.e.}, without cell-to-cell integration) show that particles tend to accumulate in the outer region, leading to an increasing concentration in the outer region with time (the stationary state at longer times will consist in having all particles located at the outer cylinder).

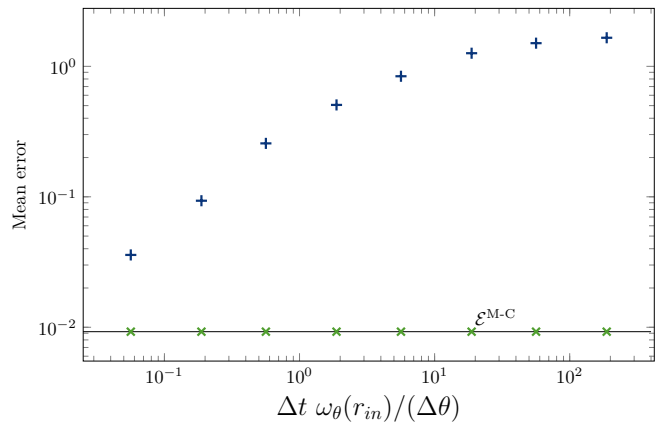
\begin{figure}[h!]
 \centering
 \begin{tikzpicture}[scale = 0.70]
  \def\shift{1e-3}
  \hspace{-2pt}
  \begin{axis}[width= 0.75\textwidth,
               height=0.5\textwidth,
               xmode = log,
               ymode = log,
               ylabel={Mean error},
               xlabel = {\Large $\Delta t \  \omega_\theta(r_{in})/ (\Delta \theta) $},
              ]
   \addplot[bluededf, mark size = 3pt, only marks, mark = +, very thick] table [x index = 1, y index= 2 , col sep=comma]{data/VR_conv_mean_error_frac_vol.csv};
   \addplot[greenedf, mark size = 3pt, only marks , mark = x, very thick] table [x index = 1, y index = 2, col sep=comma]{data/LAGR_REINTEG_conv_mean_error_frac_vol.csv};
   \draw (1e-2 , 0.009250804)  -- (400, 0.009250804) node[near end, above ]{$ \mathcal{E}^{\text{M-C}}$ };
  \end{axis}
 \end{tikzpicture}
 \caption{Convergence of the mean error on particle concentration computed at time $t=$\SI{41}{s} as a function of the time step, using the reference algorithm (\textcolor{blueedf}{\textbf{+}}) and the new cell-to-cell algorithm (\textcolor{greenedf}{\textbf{$\times$}}). These results confirm that the reference algorithm induces a spurious drift (the error increases with increasing time steps) contrary to the new algorithm which remains accurate regardless of the time step (the constant non-zero error is due to the statistical noise inherent to Monte Carlo methods).}
 \label{fig:convergence_couette_tikz}
\end{figure}

To further assess the accuracy of the new algorithm, the error between the numerical value and the theoretical value of the number concentration has been computed within each of the 21 regions. The average error over the whole domain and simulation time (here \SI{41}{s}) is displayed in \figurename{}~\ref{fig:convergence_couette_tikz} as a function of the dimensionless time step $\Delta t^+$. It confirms that the new algorithm provides accurate results regardless of the time step used, i.e., both when the average displacement is smaller or greater than the grid size. The small (constant) error with the new algorithm comes from the statistical noise due to Monte Carlo methods. Meanwhile, the results obtained without cell-to-cell integration clearly show an increasing numerical error with increasing time steps.

\newpage
\section{Conclusions and perspectives}
\label{sec:conclusion}

In this paper, we have developed a new cell-to-cell algorithm for particle tracking in the context of hybrid approaches that couple Lagrangian stochastic methods with mean-field ones. The cell-to-cell integration consists in dividing the time step into a number of sub-iterations, each one corresponding to the motion of a particle within one cell. This means that the motion of a particle is stopped each time it crosses a face and leaves a cell, at which point the mean fields are updated to compute particle displacement over the remainder of the time step. This decomposition allows naturally to account for changes in the mean fields every time a particle enters a new cell. Given the stochastic nature of the model, an additional constraint has to be satisfied: no anticipation should be made when estimating the residence time in each cell. Indeed, a naive formulation would consist in first predicting particle displacements over a time step and then try to deduce the residence times in the cells that have been crossed. Yet, this would yield estimations of these residence times that become functions of the future of the Wiener process driving the diffusion term, in direct violation of the very definition of the stochastic integral in the It\^{o} sense. To avoid this pitfall, and the resulting spurious numerical drift and diffusion values it would entail, careful estimations based only on values at the beginning of each sub-time steps must be made. In the present algorithm, this is achieved by introducing the notion of a virtual partner whose motion is governed by mean conditional increments and which is used to provide free-of-statistical-bias of the residence times in the different cells being crossed during one time step.

Drawing on these two notions (cell-to-cell integration and non-anticipation), the new algorithm is built on a three-step process. The first step consists in computing the deterministic estimation of the particle position at the end of the time step. Then, the trajectory algorithm based on a free-flight assumption is applied to determine if the particle leaves/remains in the cell. If it remains in the cell, the position and velocity of the particle at the end of the time step are computed. Otherwise, the trajectory algorithm provides information on the exit time and exit location. The whole three-step process is repeated again but starting from the last known exit location and for the remainder of the time step. This process is repeated until the particle reaches its final destination at the end of the time step $\Delta t$ ({i.e.} it does remain in the cell).

A first analysis consists in checking that the new algorithm remains exact for constant mean fields. In the more general case of non-uniform flows, since the residence times are now only approximated, the time integration scheme cannot be regarded as exact but the overall scheme is guaranteed to respect the non-anticipating constraint and is shown to provide considerable improvements over the current algorithm for particle tracking in non-homogeneous flows. This is further demonstrated by considering numerical predictions in two test cases: particle dispersion in a uniform mean flow and particle dynamics in a non-uniform flow. These two cases have confirmed the improved accuracy of the new algorithm, even when large time steps are used.

At this point, it is worth emphasizing that, although we have considered fluid particles and a simple Langevin model, the same method can be directly applied to dispersed turbulent two-phase flows and, more generally, to similar stochastic dynamical models which require to track particles in complex geometries and meshes. Note that in the examples presented in Section \ref{sec:results} one concerns a laminar flow and the second one a high-Reynolds-number turbulent flow. However, the present methodology (i.e. the time-splitting algorithm) is not limited to these situations and can be applied to any flows regardless of the Reynolds number, provided that an adequate dynamical model and its corresponding numerical scheme are applied.
Furthermore, this new algorithm paves the way toward the development of refined algorithms for such hybrid approaches that couple Lagrangian methods with mean-field approaches. In this work, we have applied a cell-to-cell integration every time a particle enters a new cell. On the other hand, this time-splitting scheme could be applied only when mean fields differ significantly from one cell to another. Hence, a refined algorithm could be devised using test functions to quantify variations in the mean-field quantities along particle trajectories and apply this time-splitting method only when variations are larger than a given threshold.

\newpage
\appendix
\input{trajecto}

\section*{Acknowledgements}
G. Balvet has received a financial support by ANRT through the EDF-CIFRE contract number 2020/1387. The authors acknowledge the infrastructures at EDF R\&D and the CEREA laboratory for providing access to computational resources.

\bibliographystyle{abbrv}
\bibliography{references}
\end{document}

%% file: tikz/Gaussian_traj.tex
\begin{tikzpicture}[spy using outlines, connect spies]
\def\width{0.5\textwidth}
\def\height{0.4\textwidth}
\def\xpropfact{0.8}
\def\ypropfact{0.75}

    \begin{axis}[
    xmin = 0, xmax = 1,
    ymin = -2, ymax = 2 ,
    width = \width,
    height = \height,
]

\addplot[ greenmf!80, mark size = 0pt,  ] table [x index = 0, y index= 1 , col sep=comma] {data/Gaussian_traj_example.csv};

\addplot[ redmf!80, mark size = 0pt,  ]  table [x index = 0, y index= 2 , col sep=comma] {data/Gaussian_traj_example.csv};

\addplot[ blueedf!80 , mark size = 0pt,  ] table [x index = 0, y index= 3 , col sep=comma] {data/Gaussian_traj_example.csv};
 \addplot[bluededf!80, mark size = 0pt,  ] table [x index = 0, y index= 5 , col sep=comma] {data/Gaussian_traj_example.csv};
 \addplot[orangededf!95, mark size = 0pt,  ] table [x index = 0, y index= 6 , col sep=comma] {data/Gaussian_traj_example.csv};


\addplot[ greendedf!80, mark size = 0pt,  ] table [x index = 0, y index= 8 , col sep=comma] {data/Gaussian_traj_example.csv};
 \addplot[  indigo!80 , mark size = 0pt,  ] table [x index = 0, y index= 9, col sep=comma] {data/Gaussian_traj_example.csv};
\addplot[
    domain = 0:30,
    samples = 200,
    smooth,
    ultra thick,
    gray,
    loosely dashed
] {sqrt(x)};
\addplot[
    domain = 0:30,
    samples = 200,
    smooth,
    ultra thick,
    gray,
    loosely dashed
] {-sqrt(x)};

 \end{axis}
\draw [loosely dashed, ultra thick] (0, 0.5* \ypropfact * \height) node[above left]{$W(t_0)$}  node {$\bullet$} -- (\xpropfact * \width, 0.5* \ypropfact * \height) node[right] {$\langle dW \rangle$ = 0 };
\draw [dashed, ultra thick, <->,gray ] (\xpropfact * \width, 0.75* \ypropfact * \height) -- (\xpropfact * \width, 0.25* \ypropfact * \height) node[right] {\textcolor{black}{$\langle dW^2 \rangle$ = dt} };
            \begin{scope}

        \coordinate (location) at (-0.4 *\xpropfact * \width, 0.82 * \ypropfact * \height);

        \coordinate (tospy) at (0.3*\xpropfact * \width, 0.82 * \ypropfact * \height);

        \spy[magnification=6 , size=2.5cm] on (tospy) in node at (location);

      \end{scope}
    \end{tikzpicture}

%% file: tikz/tikz_hybrid_scheme.tex
 \tikzset{
	connect/.style={ thick, color=#1},
	connect/.default=blueedf,
	light/.style={thin, densely dashed, color=#1},
	light/.default=Dual,
	lab/.style 2 args={scale=1.05, thin, #2, color=#1, fill=#1!8, inner sep=1pt},
	lab/.default={blueedf}{rectangle, rounded corners=3pt},
	focus/.style={very thick,color=orangeedf!65},
	front/.style={fill=blueedf!5, thick,opacity=0.5},
	back/.style={dashed,thin,fill=blueedf!10},
	backline/.style={dashed,thin,draw=blueedf},
	dualvol/.style={color=greenedf, fill=greenedf!20, opacity=0.5},
	subvol/.style={fill=orangeedf!20, thick,opacity=0.5},
	frakvol/.style={fill=black!25, thick,opacity=0.5},
}
\tikzset{->-/.style={decoration={
  markings,
  mark=at position #1 with {\arrow{>}}},postaction={decorate}}}

 \begin{tikzpicture}[scale = 2]
\def\xshift{0}
\def\yshift{0.5}

\coordinate(P0) at (1.8+\xshift,1.6+\yshift);
\coordinate(P1) at (2.1+\xshift,2.2+\yshift);
\coordinate(P2) at (2.75+\xshift,2.35+\yshift);
\coordinate(P3) at (3.2+\xshift,2+\yshift);
\coordinate(P4) at (3.6+\xshift,2.15+\yshift);
\coordinate(P5) at (4.2+\xshift,2.05+\yshift);
\coordinate(P6) at (4.6+\xshift,1.75+\yshift);
\coordinate(P7) at (4.5+\xshift,1.5+\yshift);
\coordinate(P8) at (4.1+\xshift,1.2+\yshift);
\coordinate(P9) at (3.7+\xshift,1.35+\yshift);
\coordinate(P10) at (3.5+\xshift,0.9+\yshift);
\coordinate(P11) at (2.9+\xshift,0.8+\yshift);
\coordinate(P12) at (2.4+\xshift,1.32+\yshift);
\coordinate(P20) at (3.+\xshift,1.6+\yshift);

\coordinate(P30) at (1.65+\xshift,1.6+\yshift);
\coordinate(P312) at (2.3+\xshift,1.15+\yshift);
\coordinate(P311) at (2.8+\xshift,0.7+\yshift);
\coordinate(P310) at (3.6+\xshift,0.8+\yshift);
\coordinate(P38) at (4.1+\xshift,1.05+\yshift);
\coordinate(P37) at (4.6+\xshift,1.25+\yshift);
\coordinate(P36) at (4.8+\xshift,1.9+\yshift);
\coordinate(P35) at (4.2+\xshift,2.2+\yshift);
\coordinate(P34) at (3.6+\xshift,2.3+\yshift);
\coordinate(P32) at (2.75+\xshift,2.5+\yshift);
\coordinate(P31) at (1.95+\xshift,2.3+\yshift);

\draw[blueedf,dashed] (P30) -- (P0);
\draw[blueedf,dashed] (P312) -- (P12);
\draw[blueedf,dashed] (P311) -- (P11);
\draw[blueedf,dashed] (P310) -- (P10);
\draw[blueedf,dashed] (P38) -- (P8);
\draw[blueedf,dashed] (P37) -- (P7);
\draw[blueedf,dashed] (P36) -- (P6);
\draw[blueedf,dashed] (P35) -- (P5);
\draw[blueedf,dashed] (P34) -- (P4);
\draw[blueedf,dashed] (P32) -- (P2);
\draw[blueedf,dashed] (P31) -- (P1);

\draw[] (0.75+\xshift,2+\yshift) node[]{\textbf{a) FV solver}};
\draw[] (0.75+\xshift,1.75+\yshift) node[]{$\xrightarrow{}$ \text{mean flow field}};

\coordinate(C1) at (barycentric cs:P0=0.166,P1=0.166,P2=0.166,P3=0.166,P20=0.166,P12=0.166);
\coordinate(C2) at (barycentric cs:P3=0.125,P4=0.125,P5=0.125,P6=0.125,P7=0.125,P8=0.125,P9=0.125,P20=0.125);
\coordinate(C3) at (barycentric cs:P20=0.2,P9=0.2,P10=0.2,P11=0.2,P12=0.2);

\draw[greenedf,->] (C1) node[scale=0.8,blueedf] {$\times$} node[blueedf,above left] {} node[above , greenedf,scale=1.1]{$\lra{\vect{U}}_{\text{Cell i}}(t)$} -- (2.9,2.15) ;
\draw[greenedf,->] (C2) node[scale=0.8] {$\times$} node[blueedf,below right= 0.5ex and -0.5ex]{} node[above , greenedf,scale=1.1]{$\lra{\vect{U}}_{\text{Cell j}}(t)$}  -- (4.3,2.2);
\draw[greenedf,->] (C3) node[scale=0.8] {$\times$} node[blueedf,below right= 0.5ex and -0.5ex] {} node[above , greenedf,scale=1.1]{$\lra{\vect{U}}_{\text{Cell k}}(t)$}  -- (3.4,1.75);

\draw[ultra thick,connect=blueedf] (P0) -- (P1) -- (P2) -- (P3) -- (P4) -- (P5) -- (P6) -- (P7) --(P8) --(P9) --(P10) -- (P11) -- (P12) -- cycle  ; 
\draw[ultra thick,connect=blueedf] (P3) -- (P20) -- (P9);
\draw[ultra thick,connect=blueedf]  (P20) -- (P12);

\def\xshift{0}
\def\yshift{-1.75}

\coordinate(P0) at (1.8+\xshift,1.6+\yshift);
\coordinate(P1) at (2.1+\xshift,2.2+\yshift);
\coordinate(P2) at (2.75+\xshift,2.35+\yshift);
\coordinate(P3) at (3.2+\xshift,2+\yshift);
\coordinate(P4) at (3.6+\xshift,2.15+\yshift);
\coordinate(P5) at (4.2+\xshift,2.05+\yshift);
\coordinate(P6) at (4.6+\xshift,1.75+\yshift);
\coordinate(P7) at (4.5+\xshift,1.5+\yshift);
\coordinate(P8) at (4.1+\xshift,1.2+\yshift);
\coordinate(P9) at (3.7+\xshift,1.35+\yshift);
\coordinate(P10) at (3.5+\xshift,0.9+\yshift);
\coordinate(P11) at (2.9+\xshift,0.8+\yshift);
\coordinate(P12) at (2.4+\xshift,1.32+\yshift);
\coordinate(P20) at (3.+\xshift,1.6+\yshift);

\coordinate(P30) at (1.65+\xshift,1.6+\yshift);
\coordinate(P312) at (2.3+\xshift,1.15+\yshift);
\coordinate(P311) at (2.8+\xshift,0.7+\yshift);
\coordinate(P310) at (3.6+\xshift,0.8+\yshift);
\coordinate(P38) at (4.1+\xshift,1.05+\yshift);
\coordinate(P37) at (4.6+\xshift,1.25+\yshift);
\coordinate(P36) at (4.8+\xshift,1.9+\yshift);
\coordinate(P35) at (4.2+\xshift,2.2+\yshift);
\coordinate(P34) at (3.6+\xshift,2.3+\yshift);
\coordinate(P32) at (2.75+\xshift,2.5+\yshift);
\coordinate(P31) at (1.95+\xshift,2.3+\yshift);
\coordinate(X1) at (3+\xshift,1.8+\yshift) ;

\draw[blueedf,dashed] (P30) -- (P0);
\draw[blueedf,dashed] (P312) -- (P12);
\draw[blueedf,dashed] (P311) -- (P11);
\draw[blueedf,dashed] (P310) -- (P10);
\draw[blueedf,dashed] (P38) -- (P8);
\draw[blueedf,dashed] (P37) -- (P7);
\draw[blueedf,dashed] (P36) -- (P6);
\draw[blueedf,dashed] (P35) -- (P5);
\draw[blueedf,dashed] (P34) -- (P4);
\draw[blueedf,dashed] (P32) -- (P2);
\draw[blueedf,dashed] (P31) -- (P1);

\draw[] (0.75+\xshift,2+\yshift) node[]{\textbf{b) Tracking  module}};
\draw[] (0.75+\xshift,1.75+\yshift) node[]{$\xrightarrow{}$ \text{particles dynamics}};

\coordinate(C1) at (barycentric cs:P0=0.166,P1=0.166,P2=0.166,P3=0.166,P20=0.166,P12=0.166);
\coordinate(C2) at (barycentric cs:P3=0.125,P4=0.125,P5=0.125,P6=0.125,P7=0.125,P8=0.125,P9=0.125,P20=0.125);
\coordinate(C3) at (barycentric cs:P20=0.2,P9=0.2,P10=0.2,P11=0.2,P12=0.2);

\draw[blueedf] (C1) node[scale=0.8] {$\times$} ;
\draw[blueedf] (C2) node[scale=0.8] {$\times$};
\draw[blueedf] (C3) node[scale=0.8] {$\times$} ;
\draw (X1) node[red,scale=1.75]{$\bullet$} node[above,red ]{$X(t)$};

\draw [red,dotted,ultra thick, ->-=0.5] (C1) to [out=45,in=135] (X1);
\draw [red,dotted,ultra thick, ->-=0.5] (C2) to [out=120,in=45] (X1);
\draw [red,dotted,ultra thick, ->-=0.5] (C3) to [out=180,in=210]  (X1);

\draw[ultra thick,connect=blueedf] (P0) -- (P1) -- (P2) -- (P3) -- (P4) -- (P5) -- (P6) -- (P7) --(P8) --(P9) --(P10) -- (P11) -- (P12) -- cycle  ; 
\draw[ultra thick,connect=blueedf] (P3) -- (P20) -- (P9);
\draw[ultra thick, connect=blueedf]  (P20) -- (P12);
\draw[red] (3+\xshift,0.5+\yshift) node[]{\textbf{ Interpolation of} $\mathbf{\lra{\vect{U}}(t,(X(t))}$\textbf{?}};

\draw[->] (0.75,2.1) to [out=270,in=90] (0.9+\xshift,2.1+\yshift);

 \end{tikzpicture}

%% file: tikz/new_scheme_cell_to_cell.tex
\begin{tikzpicture}[ultra thick,scale=0.70,transform shape]
\def\xshift{0}
\def\yshift{0}

		\draw[->] (-1.4+\xshift,0.66+\yshift) to [out = 270, in= 180] (-1 + \xshift,-2 +\yshift) ;
		\draw (-1+\xshift,1+\yshift) node[scale=1.2]{$\mathbf{[0,t_{out}^{[1]}]}$};

       \draw[blueedf,dashed]   (0+\xshift,0+\yshift)  rectangle (9,2);
       \draw[blueedf,dashed]   (3+\xshift,2+\yshift)  -- (3 ,0);
       \draw[blueedf,dashed]   (6+\xshift,2+\yshift)  -- (6 ,0);

		\draw[greenedf,->,dashed] (1.5+\xshift,1+\yshift) node[black]{$+$} -- (2.3+\xshift,0.8+\yshift) node[midway,below,scale=1.3]{$\lra{\vect{U}}_{\text{Cell i}}$};

       \draw[orangeedf] (2.5+\xshift,1.275+\yshift)node[black!50,scale=1.75]{$\bullet$} node[ above  left,black!50,scale=1.3]{$\{\vect{X}^n, \, \vect{U}^n\}$} -- (3+\xshift,1.15+\yshift) node[scale=1.75]{$\bullet$} node[above right,scale=1.3]{$\vect{X}_{out}^{[1]}$};
    \draw[orangeedf,dotted]  (3+\xshift,1.15+\yshift)   -- (6.4+\xshift,0.3+\yshift) node[orangeedf,scale=1.75]{$\circ$} ;

\def\xshift{1}
\def\yshift{-3}
		\draw[->] (-1.4+\xshift,0.66+\yshift) to [out = 270, in= 180] (-1 + \xshift,-2 +\yshift) ;

		\draw (-1+\xshift,1+\yshift) node[scale=1.2]{$\mathbf{[t_{out}^{[1]}, t_{out}^{[2]}]}$};

       \draw[blueedf,dashed]   (0+\xshift,0+\yshift)  rectangle (9+\xshift,2 +\yshift);
       \draw[blueedf,dashed]   (3+\xshift,2+\yshift)  -- (3 +\xshift,0+\yshift);
       \draw[blueedf,dashed]   (6+\xshift,2+\yshift)  -- (6 +\xshift,0+\yshift);

		\draw[greenedf,->,dashed] (4.5+\xshift,1+\yshift) node[black]{$+$} -- (5.2+\xshift,1.1+\yshift) node[midway,below ,scale=1.3 ]{$\lra{\vect{U}}_{\text{Cell j}}$};

       \draw[orangeedf,dashed] (2.5+\xshift,1.275+\yshift)node[black!50,scale=1.75]{$\bullet$} node[ above  left,black!50,scale=1.3]{$\{\vect{X}^n, \, \vect{U}^n\}$} -- (3+\xshift,1.15+\yshift) node[scale=1.75]{$\bullet$} node[above right,scale=1.3]{$\vect{X}_{out}^{[1]}$};
    \draw[orangeedf]  (3+\xshift,1.15+\yshift)   -- (6 +\xshift,1.58+\yshift) node[scale=1.75]{$\bullet$} node[below right,scale=1.3] {$\vect{X}_{out}^{[2]}$} ;
\draw[orangeedf,dotted]  (6  +\xshift,1.58+\yshift) -- (6.5+\xshift,1.65+\yshift) node[orangeedf,scale=1.75]{$\circ$};

\def\xshift{2}
\def\yshift{-6}

		\draw (-1+\xshift,1+\yshift) node[scale=1.2]{$\mathbf{[t_{out}^{[2]}, t_{out}^{[3]}]}$};

       \draw[blueedf,dashed]   (0+\xshift,0+\yshift)  rectangle (9+\xshift,2 +\yshift);
       \draw[blueedf,dashed]   (3+\xshift,2+\yshift)  -- (3 +\xshift,0+\yshift);
       \draw[blueedf,dashed]   (6+\xshift,2+\yshift)  -- (6 +\xshift,0+\yshift);

		\draw[greenedf,->,dashed] (7.5+\xshift,1+\yshift) node[black]{$+$} -- (8.1+\xshift,1.3+\yshift) node[midway, below =5pt ,scale=1.3]{$\lra{\vect{U}}_{\text{Cell k}}$};

       \draw[orangeedf,dashed] (2.5+\xshift,1.275+\yshift)node[black!50,scale=1.75]{$\bullet$} node[ above  left,black!50,scale=1.3]{$\{\vect{X}^n, \, \vect{U}^n\}$} -- (3+\xshift,1.15+\yshift) node[scale=1.75]{$\bullet$} node[below right,scale=1.3]{$\vect{X}_{out}^{[1]}$};
    \draw[orangeedf,dashed]  (3+\xshift,1.15+\yshift)   -- (6 +\xshift,1.58+\yshift) node[scale=1.75]{$\bullet$} node[below right,scale=1.3] {$\vect{X}_{out}^{[2]}$} ;
\draw[orangeedf]  (6  +\xshift,1.58+\yshift) -- (6.5+\xshift,1.83+\yshift) node[black,scale=1.75]{$\bullet$} node[above, black,scale=1.3]{$\{\vect{X}^{n+1}, \, \vect{U}^{n+1}\}$};


  \end{tikzpicture}

%% file: tikz/summary_new_alg.tex
\begin{tikzpicture}[ultra thick,scale=0.50,transform shape]

\tikzset{->-/.style={decoration={
  markings,
  mark=at position #1 with {\arrow{>}}},postaction={decorate}}}

\def\xshift{0}
\def\yshift{0}

	\draw[->] (-5+\xshift,-5+\yshift) to [out = 270, in= 90] (-3.5 + \xshift,-7 +\yshift) ;

    \draw[decorate,decoration={brace,amplitude=10}] (-3.5+ \xshift,-5.25+\yshift)  -- (-3.5+ \xshift,2.25+\yshift);	
	\draw (-5+ \xshift,-1.75+\yshift) node[rotate = 90]{\huge \textbf{sub-iteration 1}};

\draw[dashed,violet] (13+\xshift,2.5+\yshift) node[above,scale = 1.4]{\Large $\mathbf{t=t^0}$} -- (14.5+\xshift,-5+\yshift) node[below,scale = 1.4]{ \Large $\mathbf{t=t^0 + \theta^{[1]} \Delta t}$};
	
	\draw (-2+\xshift,1.3+\yshift) node{\Large \textbf{$\mathbf{1^{st}}$ integration}};
	\draw (-2+\xshift,0.7+\yshift) node{\Large(deterministic)};

       \draw[blueedf,thick,dashed]   (0+\xshift,0+\yshift)  rectangle (12+ \xshift,2+\yshift);
       \draw[blueedf,thick,dashed]   (3+\xshift,2+\yshift)  -- (3 + \xshift,0+\yshift);
       \draw[blueedf,thick,dashed]   (6+\xshift,2+\yshift)  -- (6 + \xshift,0+\yshift);
       \draw[blueedf,thick,dashed]   (9+\xshift,2+\yshift)  -- (9 + \xshift,0+\yshift);

       \draw[orangeedf,dashed] (1.45+\xshift,1.2+\yshift) node[red,scale=2.5]{$\bullet$} node[greenedf,scale=2.5]{$\times$} node[ below , red,scale=1.75]{$\vect{X}^{[0]}{\color{black}=}{\color{greenedf}\widetilde{\vect{X}^{[0]}}}$};
    \draw[orangeedf]  (10.1+\xshift,0.6+\yshift) node[orangeedf,scale=2.5]{$+$}  node[above,scale=1.75]{$\widehat{\vect{X}^{[1]}}$};


\def\yshift{-2.5}

	\draw (-2+\xshift,1.6+\yshift) node{\Large \textbf{Track  }};
	\draw (-2+\xshift,1+\yshift) node{\Large \textbf{virtual partner}};
	\draw (-2+\xshift,0.4+\yshift) node[violet]{\LARGE $\rightarrow \theta[1]$};

           \draw[blueedf,thick,dashed]   (0+\xshift,0+\yshift)  rectangle (12 +\xshift,2 +\yshift);
       \draw[blueedf,thick,dashed]   (3+\xshift,2+\yshift)  -- (3+\xshift ,0+\yshift);
       \draw[blueedf,thick,dashed]   (6+\xshift,2+\yshift)  -- (6+\xshift ,0+\yshift);
       \draw[blueedf,thick,dashed]   (9+\xshift,2+\yshift)  -- (9+\xshift ,0+\yshift);

 \draw[greenedf] (1.45+\xshift,1.2+\yshift)node[greenedf,scale=2.55]{$\times$} node[ below, greenedf,scale=1.75]{$\widetilde{\vect{X}^{[0]}}$} -- (3+\xshift,1.1+\yshift) node[scale=2.5]{$\times$}  node[ below right=1pt, greenedf,scale=1.75]{$\widetilde{\vect{X}^{[1]}}$} ;
\draw[greenedf,loosely dotted] (3+\xshift,1.1+\yshift)  -- (10.1+\xshift,0.6+\yshift) node[orangeedf,scale=2.5]{$+$}  node[above,scale=1.75,orangeedf]{$\widehat{\vect{X}^{[1]}}$} ;


\def\yshift{-5}

		\draw (-2+\xshift,1.25+\yshift) node{\Large\textbf{$\mathbf{2^{nd}}$ integration}};
	\draw (-1.5+\xshift,0.75+\yshift) node{\Large(stochastic)};

          \draw[blueedf,thick,dashed]   (0+\xshift,0+\yshift)  rectangle (12+\xshift,2+\yshift);
       \draw[blueedf,thick,dashed]   (3+\xshift,2+\yshift)  -- (3+\xshift ,0+\yshift);
       \draw[blueedf,thick,dashed]   (6+\xshift,2+\yshift)  -- (6+\xshift ,0+\yshift);
       \draw[blueedf,thick,dashed]   (9+\xshift,2+\yshift)  -- (9+\xshift ,0+\yshift);
		
\draw[red,dashed,->-=0.7] (2+\xshift,1.2+\yshift) node[red,scale=2.5]{$\bullet$}  node[ below, red,scale=1.75]{$\vect{X}^{[0]}$}  to [out = 40, in= 115] (3.4+\xshift,0.9 + \yshift) node[red,scale=2.5]{$\bullet$}  node[ above right, red,scale=1.75]{$\vect{X}^{[1]}$} ;

\def\xshift{1.5}
\def\yshift{-8.5}
\draw[->] (-5+\xshift,-5+\yshift) to [out = 270, in= 90] (-3.5 + \xshift,-7 +\yshift) ;

    \draw[decorate,decoration={brace,amplitude=10}] (-3.5+ \xshift,-5.25+\yshift)  -- (-3.5+ \xshift,2.25+\yshift);	
	\draw (-5+ \xshift,-1.75+\yshift) node[rotate = 90]{\huge \textbf{sub-iteration 2}};

\draw[dashed,violet] (13+\xshift,2.5+\yshift) -- (14.5+\xshift,-5+\yshift) node[below =-5pt,scale = 1.4]{\Large $\mathbf{t=t^0 + \theta^{[1]} \Delta t}$} node[below =14pt,scale = 1.4]{\Large $\mathbf{+ \theta^{[2]}(1- \theta^{[1]}) \Delta t}$};

	\draw (-2+\xshift,1.3+\yshift) node{\Large \textbf{$\mathbf{1^{st}}$ integration}};
	\draw (-2+\xshift,0.7+\yshift) node{\Large(deterministic)};

       \draw[blueedf,thick,dashed]   (0+\xshift,0+\yshift)  rectangle (12+ \xshift,2+\yshift);
       \draw[blueedf,thick,dashed]   (3+\xshift,2+\yshift)  -- (3 + \xshift,0+\yshift);
       \draw[blueedf,thick,dashed]   (6+\xshift,2+\yshift)  -- (6 + \xshift,0+\yshift);
       \draw[blueedf,thick,dashed]   (9+\xshift,2+\yshift)  -- (9 + \xshift,0+\yshift);

		\draw (3.4+\xshift,0.9 + \yshift) node[red,scale=2.55]{$\bullet$}  node[ above right, red,scale=1.75]{$\vect{X}^{[1]}$};

    \draw[orangeedf]  (8.2+\xshift,1.3+\yshift) node[orangeedf,scale=2.5]{$+$}  node[below=5pt,scale=1.75]{$\widehat{\vect{X}^{[2]}}$};


\def\yshift{-11}

	\draw (-2+\xshift,1.6+\yshift) node{\Large \textbf{Track }};
	\draw (-2+\xshift,1+\yshift) node{\Large \textbf{virtual partner}};
	\draw (-2+\xshift,0.4+\yshift) node[violet]{\LARGE $\rightarrow \theta[2]$};

           \draw[blueedf,thick,dashed]   (0+\xshift,0+\yshift)  rectangle (12 +\xshift,2 +\yshift);
       \draw[blueedf,thick,dashed]   (3+\xshift,2+\yshift)  -- (3+\xshift ,0+\yshift);
       \draw[blueedf,thick,dashed]   (6+\xshift,2+\yshift)  -- (6+\xshift ,0+\yshift);
       \draw[blueedf,thick,dashed]   (9+\xshift,2+\yshift)  -- (9+\xshift ,0+\yshift);

\draw[greenedf] (3+\xshift,1.1+\yshift) node[scale=2.5]{$\times$}  node[below right, greenedf,scale=1.75]{$\widetilde{\vect{X}^{[1]}}$} -- (6+\xshift,1.22 + \yshift) node[scale=2.5]{$\times$} node[ below right, greenedf,scale=1.75]{$\widetilde{\vect{X}^{[2]}}$};

\draw[greenedf, loosely dotted] (6+\xshift,1.22 + \yshift) -- (8.2+\xshift,1.3+\yshift) node[orangeedf,scale=2.55]{$+$}  node[below = 5pt,orangeedf,scale=1.75]{$\widehat{\vect{X}^{[2]}}$};


\def\yshift{-13.5}

		\draw (-2+\xshift,1.25+\yshift) node{\Large\textbf{$\mathbf{2^{nd}}$ integration}};
	\draw (-1.5+\xshift,0.75+\yshift) node{\Large(stochastic)};

          \draw[blueedf,thick,dashed]   (0+\xshift,0+\yshift)  rectangle (12+\xshift,2+\yshift);
       \draw[blueedf,thick,dashed]   (3+\xshift,2+\yshift)  -- (3+\xshift ,0+\yshift);
       \draw[blueedf,thick,dashed]   (6+\xshift,2+\yshift)  -- (6+\xshift ,0+\yshift);
       \draw[blueedf,thick,dashed]   (9+\xshift,2+\yshift)  -- (9+\xshift ,0+\yshift);
		
\draw[red,dashed,->-=0.7] (3.4+\xshift,0.9 + \yshift) node[red,scale=2.5]{$\bullet$}  node[ above right, red,scale=1.75]{$\vect{X}^{[1]}$}  to [out = 330, in= 230] (6.6+\xshift,0.8+\yshift) node[red,scale=2.55]{$\bullet$}  node[ above , red,scale=1.75]{$\vect{X}^{[2]}$}  ;

\def\xshift{3}
\def\yshift{-17}
\draw[->] (-5+\xshift,-5+\yshift) to [out = 270, in= 90] (-5 + \xshift,-6.3 +\yshift) ;

    \draw[decorate,decoration={brace,amplitude=10}] (-3.5+ \xshift,-5.25+\yshift)  -- (-3.5+ \xshift,2.25+\yshift);	
	\draw (-5+ \xshift,-1.75+\yshift) node[rotate = 90]{\huge \textbf{sub-iteration 3}};

\draw[dashed,violet] (13+\xshift,2+\yshift) -- (14.5+\xshift,-5.5+\yshift) node[below,scale = 1.4]{\Large $\mathbf{t=t^0 + \Delta t}$} ;

	\draw (-2+\xshift,1.3+\yshift) node{\Large \textbf{$\mathbf{1^{st}}$ integration}};
	\draw (-2+\xshift,0.7+\yshift) node{\Large(deterministic)};

       \draw[blueedf,thick,dashed]   (0+\xshift,0+\yshift)  rectangle (12+ \xshift,2+\yshift);
       \draw[blueedf,thick,dashed]   (3+\xshift,2+\yshift)  -- (3 + \xshift,0+\yshift);
       \draw[blueedf,thick,dashed]   (6+\xshift,2+\yshift)  -- (6 + \xshift,0+\yshift);
       \draw[blueedf,thick,dashed]   (9+\xshift,2+\yshift)  -- (9 + \xshift,0+\yshift);

\draw (6.6+\xshift,0.8+\yshift) node[red,scale=2.5]{$\bullet$}  node[ above, red,scale=1.75]{$\vect{X}^{[2]}$}  ;

    \draw[orangeedf]  (8.4+\xshift,0.6+\yshift) node[orangeedf,scale=2.5]{$+$}  node[above,scale=1.75]{$\widehat{\vect{X}^{[3]}}$};


\def\yshift{-19.5}

	\draw (-2+\xshift,1.6+\yshift) node{\Large \textbf{Track }};
	\draw (-2+\xshift,1+\yshift) node{\Large \textbf{virtual partner}};
	\draw (-2+\xshift,0.4+\yshift) node[violet]{\LARGE $\rightarrow \theta[3]$};

      \draw[blueedf,thick,dashed]   (0+\xshift,0+\yshift)  rectangle (12 +\xshift,2 +\yshift);
       \draw[blueedf,thick,dashed]   (3+\xshift,2+\yshift)  -- (3+\xshift ,0+\yshift);
       \draw[blueedf,thick,dashed]   (6+\xshift,2+\yshift)  -- (6+\xshift ,0+\yshift);
       \draw[blueedf,thick,dashed]   (9+\xshift,2+\yshift)  -- (9+\xshift ,0+\yshift);

\draw[greenedf] (6+\xshift,1.22 + \yshift) node[scale=2.5]{$\times$} node[ left, greenedf,scale=1.75]{$\widetilde{\vect{X}^{[2]}}$} --  (8.4+\xshift,0.6+\yshift) node[orangeedf,scale=2.5]{$+$}  node[above right = 0.3 and -0.9,scale=1.75,orangeedf]{$\widehat{\vect{X}^{[3]}} {\color{black}=} {\color{greenedf} \widetilde{\vect{X}^{[3]}} }$};


\def\yshift{-22}

		\draw (-2+\xshift,1.25+\yshift) node{\Large\textbf{$\mathbf{2^{nd}}$ integration}};
	\draw (-1.5+\xshift,0.75+\yshift) node{\Large(stochastic)};

          \draw[blueedf,thick,dashed]   (0+\xshift,0+\yshift)  rectangle (12+\xshift,2+\yshift);
       \draw[blueedf,thick,dashed]   (3+\xshift,2+\yshift)  -- (3+\xshift ,0+\yshift);
       \draw[blueedf,thick,dashed]   (6+\xshift,2+\yshift)  -- (6+\xshift ,0+\yshift);
       \draw[blueedf,thick,dashed]   (9+\xshift,2+\yshift)  -- (9+\xshift ,0+\yshift);
		
\draw[red,dashed,->-=0.7] (6.6+\xshift,0.8+\yshift) node[red,scale=2.5]{$\bullet$}  node[ above, red,scale=1.75]{$\vect{X}^{[2]}$} to [in=165, out = 20]   (9.4+\xshift,0.6+\yshift) node[red,scale=2.5]{$\bullet$}  node[ above right, red,scale=1.75]{$\vect{X}^{[3]}$};

\def\xshift{3}
\def\yshift{-26}

    \draw[decorate,decoration={brace,amplitude=10}] (-3.5+ \xshift,-0.25+\yshift)  -- (-3.5+ \xshift,2.25+\yshift);	
	\draw (-5+ \xshift,1+\yshift) node[rotate = 90]{\huge \textbf{Finalise}};

	\draw (-2+\xshift,1.3+\yshift) node{\Large \textbf{Track  }};
	\draw (-2+\xshift,0.7+\yshift) node{\Large final position};
	
          \draw[blueedf,thick,dashed]   (0+\xshift,0+\yshift)  rectangle (12+\xshift,2+\yshift);
       \draw[blueedf,thick,dashed]   (3+\xshift,2+\yshift)  -- (3+\xshift ,0+\yshift);
       \draw[blueedf,thick,dashed]   (6+\xshift,2+\yshift)  -- (6+\xshift ,0+\yshift);
       \draw[blueedf,thick,dashed]   (9+\xshift,2+\yshift)  -- (9+\xshift ,0+\yshift);
       
  \draw[red]  (8.4+\xshift,0.6+\yshift) node[orangeedf,scale=2.55]{$+$}  node[above = 5pt,scale=1.75,orangeedf]{$\widehat{\vect{X}^{[3]}}$} --   	 (9.4+\xshift,0.6+\yshift) node[red,scale=2.5]{$\bullet$}  node[ above right, red,scale=1.75]{$\vect{X}^{[3]}$};

	       \end{tikzpicture}    

%% file: tikz/Reference_scheme_error.tex
 
\begin{tikzpicture}[scale=0.8, transform shape]
:
\def\a{15}
\def\tana{0.267949192}
\def\cosa{0.965925826}
\def\htana{0.131652498}
\def\hcosa{0.991444861}
\def\d{0.7}
\def\xcros{7.45}
\def\negc{1.1}
\def\posc{2.1}

\draw[blueedf] (0,0) -- (8 ,0);
\draw[blueedf] (0,0) -- (8* \cosa,8 * \tana);
\draw[blueedf] (0,0) -- (8* \cosa,-8 * \tana);

\draw[blueedf] (2,0) -- (2* \cosa, 2*\tana);\draw[blueedf] (2,0) -- (2* \cosa,- 2*\tana);   
\foreach \i in {4,6,...,8}
{
   \draw[blueedf] (\i,0) -- (\i* \cosa, \i*\tana);
   \draw[blueedf] (\i,0) -- (\i* \cosa,- \i*\tana);
   \draw[blueedf] (\i*\hcosa-\hcosa, \i*\htana -\htana) node{\Large $+$};   
   \draw[blueedf] (\i*\hcosa-\hcosa,- \i*\htana +\htana) node{\Large $+$};   
}

\draw[greenedf,->,dashed,ultra thick]  (6*\hcosa-\hcosa,- 6*\htana +\htana) node[below,scale=1.2]{$\lra{\vect{U}}_{\text{Cell i}}$} -- ++(\d*\htana,\d);

\draw[greenedf,->,dashed,ultra thick]  (6*\hcosa-\hcosa, 6*\htana -\htana)node[below,scale=1.2]{$\lra{\vect{U}}_{\text{Cell j}}$}-- ++(-\d*\htana,\d) ;

\draw[orangeedf, dotted, ultra thick] (\xcros - \negc* \d * \htana , - \negc *\d) node[red]{$\bullet$}node[below, red,scale=1.5]{$\vect{X}^{[0]}$} -- (\xcros + \posc * \d * \htana ,\posc * \d ) node[orangeedf,scale=1.5]{$\times$} node[right , orangeedf,scale=1.5]{$\widetilde{\vect{X}}^{[0]}$};
\draw[red, dashed, ultra thick] (\xcros,0) -- (\xcros - \d * \posc * \htana, \d * \posc) node[red,scale=1.5]{$\bullet$}node[left  , red,scale=1.5]{$\vect{X}^{[1]}$};

\draw[thick,<->] (\xcros - \d * \posc * \htana, \d * \posc) -- (\xcros + \d * \posc * \htana, \d * \posc);

\draw[dashed,->, ultra thick] (\xcros - 2, \xcros *\tana) node[left,scale=1.5]{\textbf{Error}} to [out =0, in = 90]  (\xcros , \d * \posc);

\end{tikzpicture}

%% file: trajecto.tex
\section{Details of the particle tracking algorithm for 3-D unstructured mesh}
\label{sec:app:neighbor_searches}

This appendix presents the trajectory algorithm that is used in the present study. This algorithm is able to track the motion of particles even in 3-D fully unstructured meshes with warped faces. 
The tracking algorithm is first described in \ref{sec:app:numerical_methods_face_det}. It includes the description of the original tracking algorithm with the detection of face-crossing events in Section \ref{eq:face-crossing-events}. 
Then, the algorithm is extended in \ref{sec:app:neighbor_searches:extended} to compute the location and exit time when a particle crosses a face. Last, the algorithm is validated in \ref{sec:app:result_unstr_mesh} by comparing the numerical results obtained using various 3-D unstructured meshes in a simple non-uniform flow.

\subsection{Principle of the neighbor search algorithm}
\label{sec:app:numerical_methods_face_det}

The algorithm is based on a successive neighbor search. This means that the cell inside which a particle currently resides is determined by browsing through the neighboring cells. Such algorithms require three pieces of information: (a) the origin particle location $\vect{X}_O = \vect{X}^n$, (b) the corresponding cell inside which it was initially and (c) the destination particle location $\vect{X}_D= \vect{X}^{n+1}$ (as depicted in \figurename{}~\ref{fig:numerical_scheme:tracking}). 
The principle is then to determine if the particle leaves the current cell assuming a free-flight motion between point $\vect{X}_O$ and $\vect{X}_D$. This is performed by:
\begin{enumerate}
    \item computing which faces of the current cell are intersected by the line $(\vect{X}_O\vect{X}_D)$;
    \item checking if the intersection belongs to the straight-line vector $\vect{X}_O\vect{X}_D= \vect{X}_D - \vect{X}_O$. 
\end{enumerate}
The key issue is then to have a robust method to detect the intersection between a displacement vector and any face. 
It is of prime importance to prevent any particle from being permanently lost in the computational domain. 
For that reason, the method uses Boolean elementary tests which are reproducible from one cell to another so that it can handle pathological cases such as when the vector $\vect{X}_O\vect{X}_D$ crosses a face through one of its edges (to the machine precision).

\subsubsection{Method to detect face-crossing events through warped-faces}\label{eq:face-crossing-events}

The method to detect face-crossing events is based on the decomposition of each face into a set of triangular sub-faces (see also \figurename{}~\ref{tikz_multiple_face_crossing}). 
Each triangular sub-face is built using one of the oriented edges of the face (formed by two consecutive vertices $\vect{X}_i$ and $\vect{X}_j$) and the center of gravity of the face $\vect{X}_f $. This decomposition of faces ensures that each triangular sub-face treated is planar, hence making the method tractable even for warped meshes ({i.e.} with faces whose vertices do not belong to the same plane).

Once a face is decomposed into a set of triangular sub-faces, the intersection between a vector and each planar sub-face is detected using a kind of Möller--Trumblemore algorithm \cite{moller1997fast}. 
The principle is to project the vertices of each sub-face in the oriented plane orthogonal to the displacement vector $\vect{X}_O\vect{X}_D$ passing through $\vect{X}_O$ noted $(\vect{X}_O ,\, \vect{X}_O\vect{X}_D^\perp)$. 
Then, to compute if line $(\vect{X}_O\vect{X}_D)$ crosses a sub-face, we simply need to check if the point $\vect{X}_O$ belongs to the triangle formed by the projection of the sub-face on this plane (as displayed in \figurename{}~\ref{tikz_face_crossing}, where the superscript $(.)^\dagger$ corresponds to the projection in the plane perpendicular to the displacement). For that purpose, we resort here to a series of three elementary Boolean tests, each one allowing to verify if the point $\vect{X}_O$ is located on the "proper side" of a projected edge. 

\paragraph*{Elementary Boolean tests} For each of the three edges forming a projected sub-face (namely $\vect{X}_f^\dagger \vect{X}_i^\dagger$, $\vect{X}_f^\dagger \vect{X}_j^\dagger$ and $\vect{X}_i^\dagger \vect{X}_j^\dagger$), we have to verify whether the point $\vect{X}_O $ is on the proper side of the projected edge. To that extent, we resort here to simple logical tests. For the sake of clarity, let's consider the case of an oriented edge connecting two points $ \vect{X}_{\alpha}\vect{X}_\beta$ (where $_\alpha$ and $_\beta$ are the indexes of two vertexes of a sub-face) on the projection plane. In that case, the logical test $\mathcal{L^\text{edge}_{\alpha,\beta}}$ reads:
\begin{equation}
   \mathcal{L}^\text{edge}_{\alpha,\beta} =
    \begin{cases}
    &    true  \hspace{0.1 \linewidth} \text{if }   (\vect{X}_{\alpha} \vect{X}_{\beta}\wedge  \vect{X}_{\alpha} \vect{X}_O ) \cdot \vect{X}_O\vect{X}_D > 0. \\
        &  false  \hspace{0.1 \linewidth} \text{otherwise.}
    \end{cases} 
    \label{condition_vol}
\end{equation}

As displayed in \figurename{}~\ref{tikz_test_edge}, this elementary Boolean test provides information on whether a point $\vect{X}_O $ is in the half plane on the left (true) or on the right (false) of the projected line $(\vect{X}_{\alpha}^\dagger \vect{X}_{\beta}^\dagger)$. It is worth noticing that the inequality in this test being strict, the behavior on the oriented line is asymmetric. Indeed, if point $\vect{X}_O$ belongs to this line at the machine precision, test $\mathcal{L}^\text{edge}_{\alpha,\beta}$  returns $false$. In other words, we arbitrarily consider that the line $(\vect{X}_{\alpha}^\dagger \vect{X}_{\beta}^\dagger)$ belongs to the closed right half plane and not to the open left one. As a result, the Boolean test $\mathcal{L}^\text{edge}_{\alpha,\beta} \bigcup \text{not}( \mathcal{L}^\text{edge}_{\alpha,\beta})  $ is a partition of the domain. This means that the computation of $\mathcal{L}^\text{edge}_{\alpha,\beta}$ is reproducible for a given displacement vector $ \vect{X}_O \vect{X}_D$ for two faces sharing the same edge\footnote{This is the case for instance for a sub-face seen from one cell or from the neighboring one.} provided that the ordering 
of the vertices $\alpha$ and $\beta$ is fixed. In the present work, we have imposed to sort the vertices in the following order: first $\vect{X}_f $, then $\vect{X}_i $ and finally $\vect{X}_j $ (with $i < j$).

\input{tikz/tikz_trajecto_test_edge}

\figurename{}~\ref{tikz_test_edge} also shows that this elementary Boolean test is not enough to determine if a point $\vect{X}_O$ belongs to the projected triangle sub-face. In fact, a point can be located either on the right side or on the left side of the line depending on the orientation of the vector $\vect{X}_i^\dagger \vect{X}_j^\dagger$ but also on whether the point is above or below the plane. While the orientation of the vector is now fixed thanks to the sorted vertices introduced in the previous paragraph, we have to introduce the notion of plane orientation to deal with the second issue. For that purpose, we rely on the orientation of the sub-face and we introduce an additional logical test to know if the displacement vector $\vect{X}_O \vect{X}_D$ is aligned with the oriented sub-face. It returns true if they are aligned and false if they are not. This condition reads:
\begin{equation}
   \mathcal{A}^\text{face}_{f,\, i, \, j} = 
    \begin{cases}
    &   true  \hspace{0.1 \linewidth} \text{if }    (\vect{X}_f \vect{X}_i \wedge \vect{X}_f \vect{X}_j) \cdot \vect{X}_O\vect{X}_D  > 0. \\
        &  false  \hspace{0.1 \linewidth} \text{otherwise.}
    \end{cases} 
    \label{condition_Orient}
\end{equation}
By combining both logical tests $\left( \mathcal{L}^\text{edge}_{i, \, j} \equiv \mathcal{A}^\text{face}_{f,\, i, \, j} \right)$ (where $\equiv$ means that both Boolean tests have the same value), we are able to determine on which side of the line $\vect{X}_i^\dagger \vect{X}_j^\dagger$ the particle lies with respect to the face orientation.

\paragraph*{Combined elementary Boolean tests} To determine if the intersection point $\vect{X}_i $ is within the projected face, we have then to combine these elementary logical tests together. As displayed in \figurename{}~\ref{tikz_test_face}, the intersection point belongs to the projected face if three conditions are met. These conditions depend on the orientation of the sub-face with respect to the displacement vector:
\begin{itemize}
 \item When the displacement vector is aligned with the sub-face orientation (i.e., $\mathcal{A}^\text{face}_{f,\, i, \, j} = true$), the point is located to the left of each of the projected edges provided that we follow the sorted vertices (namely $\vect{X}_f^\dagger \vect{X}_i^\dagger$, $\vect{X}_i^\dagger \vect{X}_j^\dagger$ and $\vect{X}_j^\dagger \vect{X}_f^\dagger$). This means that the following three conditions have to be met: first, $\mathcal{L}^\text{edge}_{f, \, i} = true$; second, $\mathcal{L}^\text{edge}_{f, \, j} = false$ (due to its reverse orientation); third, $\mathcal{L}^\text{edge}_{i, \, j} = true$. This case is displayed in the LHS of \figurename{}~\ref{tikz_test_face}.
 \item When the displacement vector is not aligned with the sub-face orientation (i.e., $\mathcal{A}^\text{face}_{f,\, i, \, j} = false$), the point is located to the right of each of the projected edges provided that we follow the sorted vertices (namely $\vect{X}_f^\dagger \vect{X}_i^\dagger$, $\vect{X}_i^\dagger \vect{X}_j^\dagger$ and $\vect{X}_j^\dagger \vect{X}_f^\dagger$). This means that the following three conditions have to be met: first, $\mathcal{L}^\text{edge}_{f, \, i} = false$; second, $\mathcal{L}^\text{edge}_{f, \, j} = true$ (due to its reverse orientation); third, $\mathcal{L}^\text{edge}_{i, \, j} = false$. This case is displayed in the RHS of \figurename{}~\ref{tikz_test_face}.
\end{itemize}

\input{tikz/tikz_trajecto_test_face_Orient}

To sum it up, the intersection point is inside a given sub-face if the following condition is respected.
\begin{equation}
    \label{condition}   
\begin{array}{rl}
             & \left( \mathcal{L}^\text{edge}_{i, \, j} \equiv \mathcal{A}^\text{face}_{f,\, i, \, j} \right) 
    \\ 
 \text{and}  & \left( \mathcal{L}^\text{edge}_{f, \,  i}  \equiv \mathcal{A}^\text{face}_{f,\, i, \, j} \right) \\
     \text{and}  & \left( \text{not}( \mathcal{L}^\text{edge}_{f, \, j}) \equiv \mathcal{A}^\text{face}_{f,\, i, \, j} \right) = true.
\end{array}
\end{equation}
where the symbol $\equiv$ corresponds to the operator ``equivalent'' ({i.e.}, it is true if the two Boolean variables have the same value). This test is made for each sub-face. The algorithm is illustrated in \figurename{}~\ref{tikz_face_crossing} (where the intersection point is on the right hand face).

\input{tikz/tikz_trajecto}

In the case of warped faces, it is possible for a line $(\vect{X}_O\vect{X}_D)$ to cross the same face several times. This is depicted in \figurename{}~\ref{tikz_multiple_face_crossing}. In that case, the algorithm monitors the number of times that the line $(\vect{X}_O\vect{X}_D)$ crossed the face. If this number is even, it means that the particle remains in the current cell (it has left and reentered the cell). If this number is odd, it means that the particle leaves the cell through this face.

\input{tikz/tikz_trajecto_warped}

\subsubsection{Method to estimate the intersection time and position}
 \label{sec:app:neighbor_searches:extended}

As mentioned in Section~\ref{sec:new_alg:description:overview}, the new algorithm not only requires information on the cell containing the particle but also on the intersection time and location. This means that the trajectory algorithm described previously has to be extended to provide this information. 

Having determined that the line 
$(\vect{X}_O\vect{X}_D)$ does cross a sub-face, the relative time $\theta = t_{\text{cross}}/\Delta t$ necessary to reach this sub-face can be estimated using the free-flight assumption. It gives:
\begin{equation}
 \theta =  \frac{\vect{X}_O\vect{X}_f \cdot(\vect{X}_f \vect{X}_i \wedge\vect{X}_f\vect{X}_j )}{\vect{X}_O\vect{X}_D\cdot(\vect{X}_f \vect{X}_i \wedge\vect{X}_f \vect{X}_j)}.
\label{eq:lagrangian:trajecto_3}
\end{equation}

This equation is well-posed since we apply it only when we have previously determined that the line actually crosses the face. In fact, the value of $\theta$ actually provides additional information. When $\theta$ is negative, it means that the intersection point is an entrance point for the oriented axis $(\vect{X}_O \vect{X}_D)$. When $\theta$ is positive, it means that the intersection point is an exit point. The number of sub-faces through which the oriented axis $(\vect{X}_O\vect{X}_D)$ enters $(n_{in})$ and leaves $(n_{out})$ is then counted. We then check if the particle is indeed in the correct cell thanks to these numbers. The particle is in the correct cell if the number of sub-faces through which the line $(\vect{X}_O\vect{X}_D)$ enters in the cell equals to the number of sub-faces through which it exits the cell and if both of these numbers are not zero ({i.e.}, $ n_{in}= n_{out} >0 $). 

If the value of $\theta$ defined in Eq.~\eqref{eq:lagrangian:trajecto_3} is in the interval $ \theta \in [ 0,1[$, the particle does actually cross a face during the time step. The position of the particle at the intersection is then simply given  using a simple linear interpolation (this linear interpolation is coherent with the assumption of free-flight motion):

\begin{equation}
 \vect{X}_I = \vect{X}_O  + \theta \times \vect{X}_O\vect{X}_D.
\label{eq:lagrangian:trajecto_2}
\end{equation}

When a particle leaves a cell after crossing one face several times (as in wrapped faces), the exit time is considered equal to the largest value in the range $[0,1[$ ({i.e.}, the last exit time).

\subsection{Validation on 3-D unstructured meshes}
\label{sec:app:result_unstr_mesh}
 
\begin{figure}[h!]
    \begin{subfigure}{0.24\textwidth}
      \includegraphics[width=\textwidth]{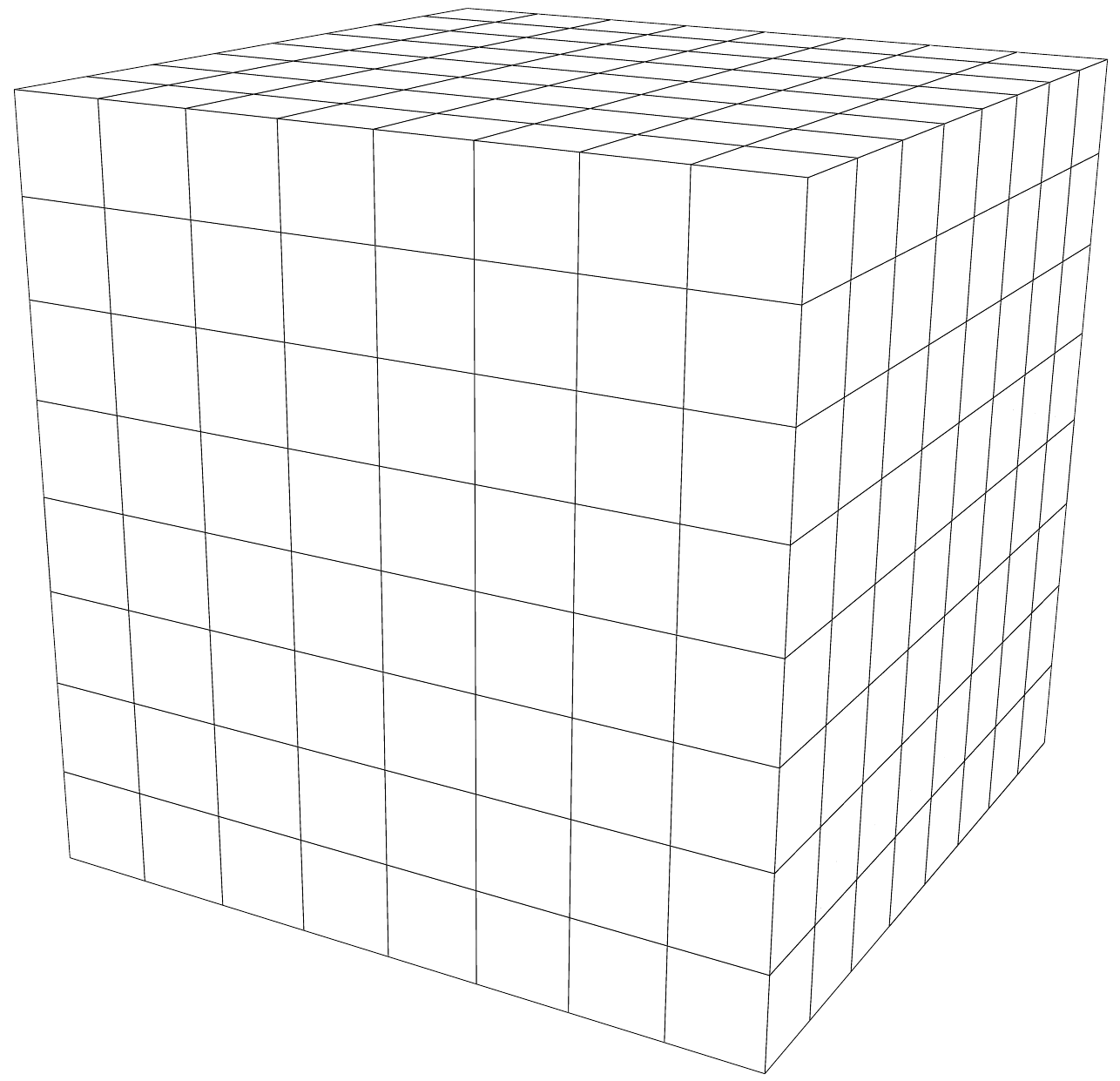}
      \caption{Hexa mesh.}
    \end{subfigure}
        \begin{subfigure}{0.24\textwidth}
      \includegraphics[width=\textwidth]{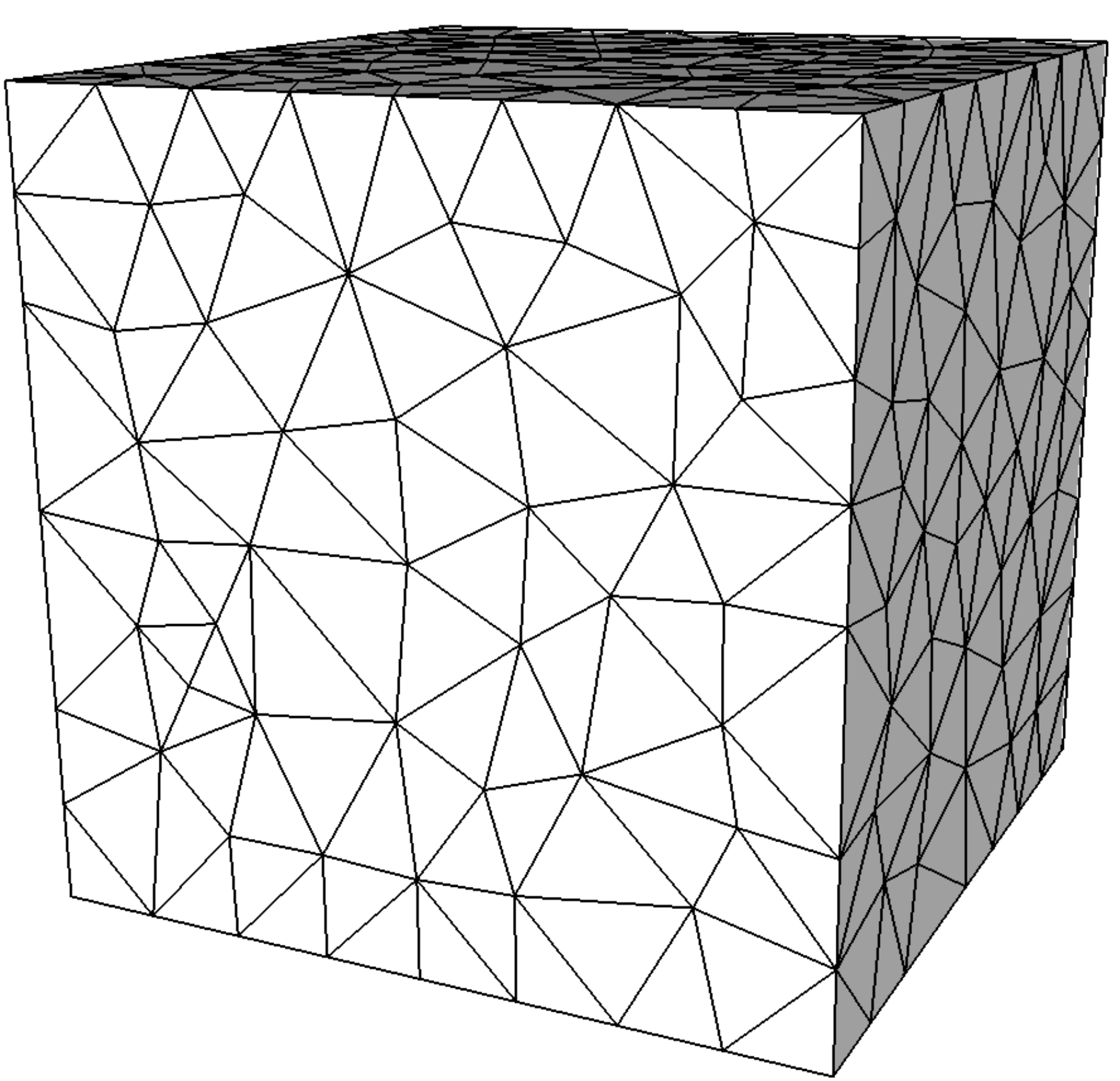}
      \caption{Tetra mesh.}
    \end{subfigure}
        \begin{subfigure}{0.24\textwidth}
      \includegraphics[width=\textwidth]{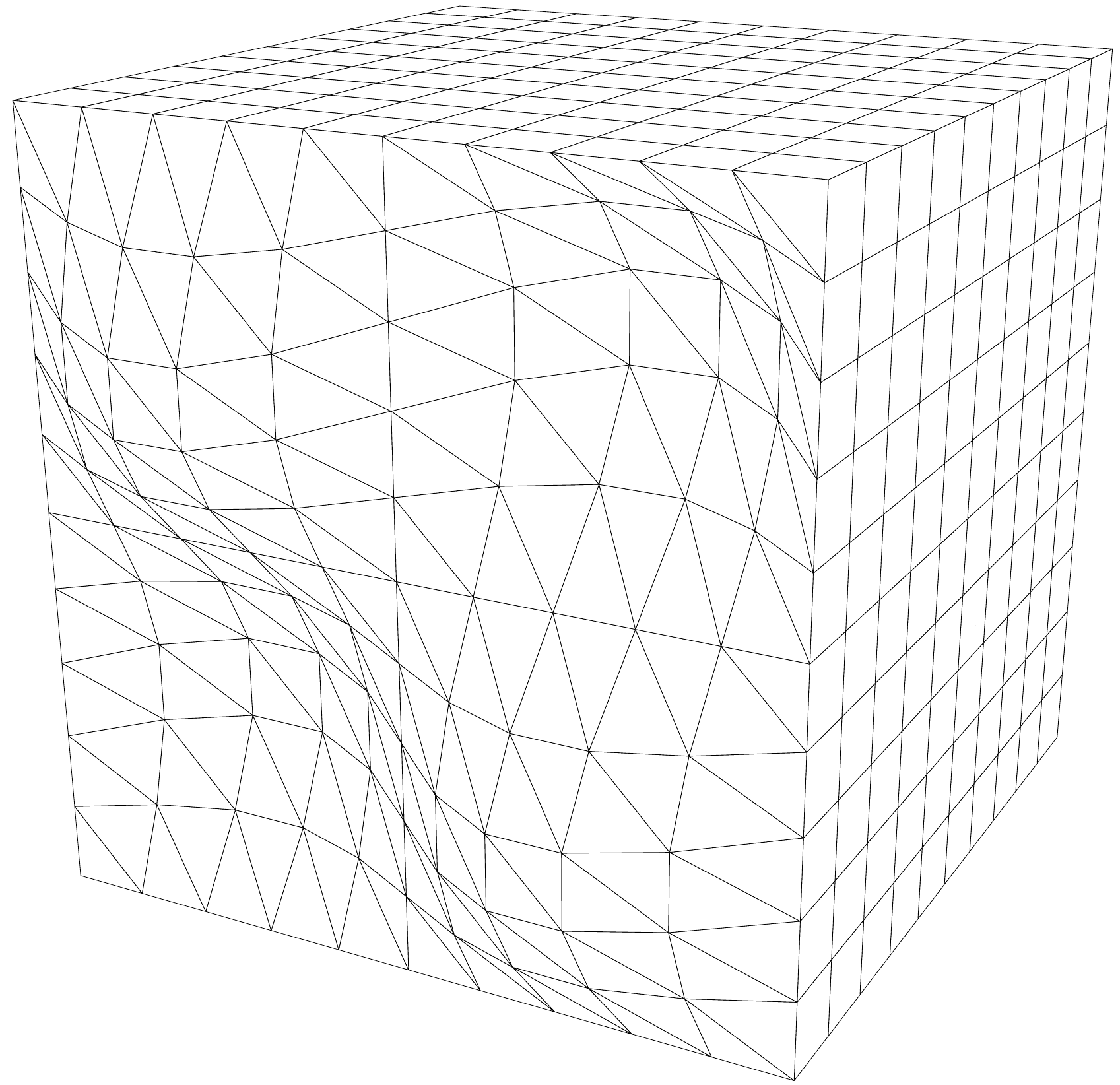}
           \caption{PrT mesh.}
    \end{subfigure}
        \begin{subfigure}{0.24\textwidth}
      \includegraphics[width=\textwidth]{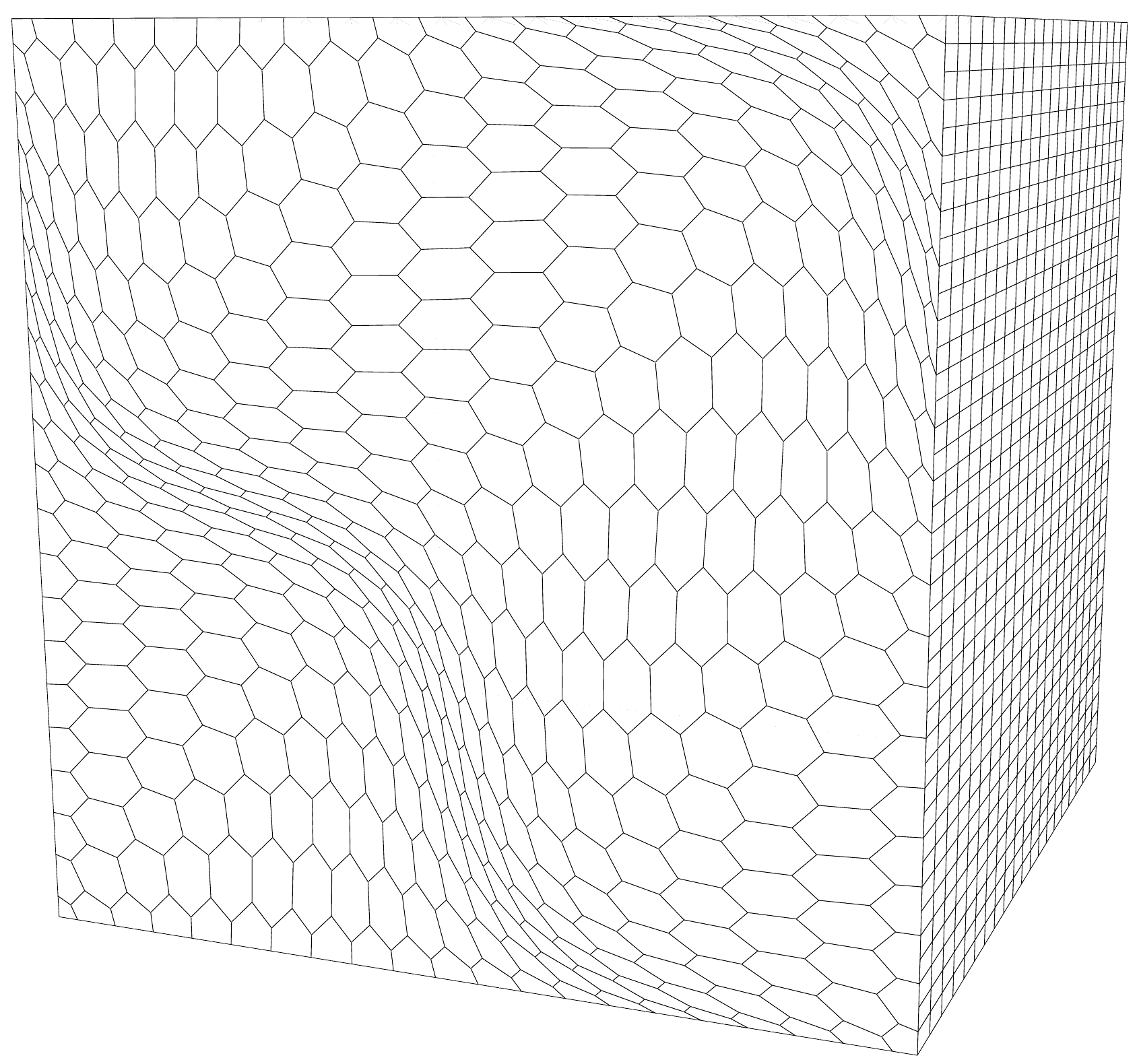}
               \caption{PrG mesh.}
    \end{subfigure}
  \caption{Type of mesh used in the present case, which span a range of possible configurations (from simple Cartesian mesh without wrapped faces to highly distorted meshes with wrapped faces.\label{fig:meshes} }
\end{figure}

The trajectory algorithm has been tested using various meshes obtained from the FVCA6 benchmark test cases \cite{eymard20113d}. These meshes were selected to be representative of a range of different meshes, going from a regular Cartesian mesh to a highly distorted mesh with different refinements. The four meshes used here are displayed in \figurename{}~\ref{fig:meshes}.

The case considered for validation actually corresponds to the uniform flow described in Section~\ref{sec:results:point_source:case}. It consists in a point source dispersion within homogeneous isotropic turbulence. To ensure that no particle is lost, the distance between the particles and the cell center (point source) is tracked. In order to have a representative quantity independent of the mesh, the quantity followed is the dimensionless distance $d^*$. It is defined is defined as $d^*=\tfrac{\qquad \left\Vert \vect{X}_c - \vect{X}_p \right\Vert }{\underset{X \in \Omega_c}{\max} \left\Vert \vect{X}_c - \vect{X} \right\Vert}$ with $\vect{X}_c$ the center of gravity of the cell, $\vect{X}_p$ the position of the particle and $ \Omega_c$ the domain defined by the cell $c$. When particles are properly tracked, this distance is always smaller than 1. However, if one of the face crossed by a particle would be missed, the particle would be permanently lost since it could continue its motion without bound. \footnote{If the current cell associated to a particle is not correct, the algorithm would not detect other face-crossing events.} In such cases, the distance from the cell center could diverge and become much greater than 1. 

\begin{figure}
    \centering
    \includegraphics[width = 0.5 \linewidth]{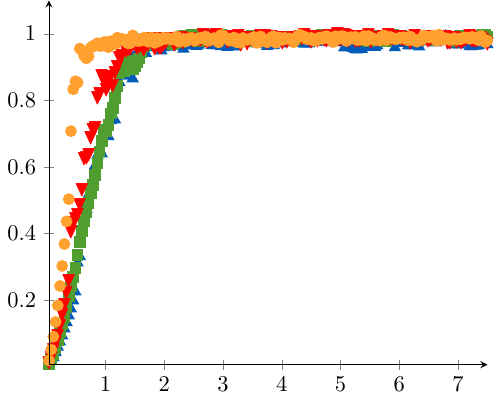}
    \caption{Maximum dimensionless distance  between the particles and their center of gravity using the Hexa mesh  (\textcolor{greenedf}{$\blacksquare$}), the Tetra mesh (\textcolor{blueedf}{$\blacktriangle$}), the PrT mesh (\textcolor{red}{$\blacktriangledown$}) and the PrG mesh (\textcolor{orangeedf}{$\bullet$}. All maximum dimensionless distances are smaller or equal to 1, meaning that the particles stay within the physical domain.)}
    \label{fig:app:distance_center}
\end{figure}

Results obtained with various meshes are compared using a time scale made dimensionless using the Lagrangian time scale $t^*=t/T_L$ (which is constant for all meshes considered). The results are displayed in \figurename{}~\ref{fig:app:distance_center}: we can see that, with \num{100000} particles dispersed initially from the point source, the maximum distance to the cell center converges toward 1 but remains always smaller than unity. This proves that the current algorithm is tractable even for 3-D unstructured meshes.

\begin{figure}[h!]

    \begin{minipage}{0.33 \linewidth}
        \includegraphics[width = 1. \linewidth]{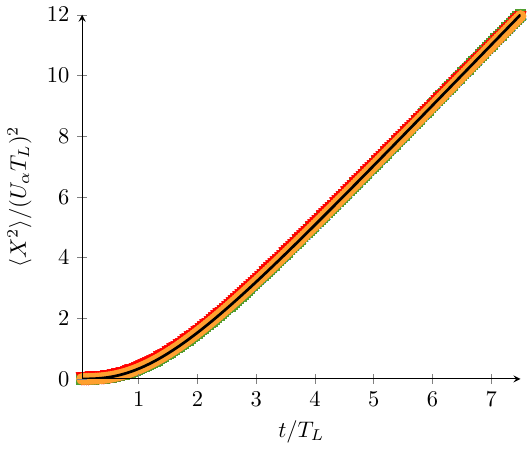}
    \end{minipage}
\begin{minipage}{0.33 \linewidth}
        \includegraphics[width = 1.\linewidth]{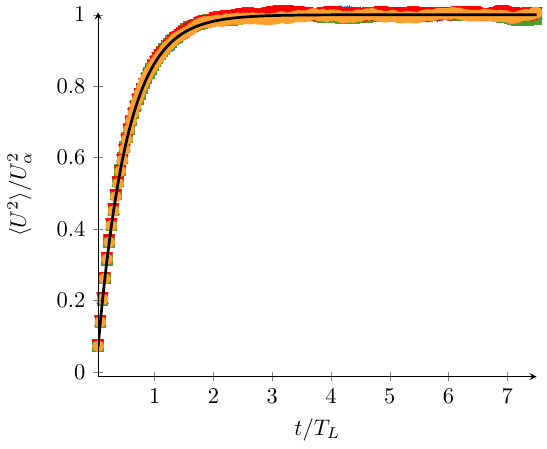}
\end{minipage} 
\begin{minipage}{0.33 \linewidth}
        \includegraphics[width = 1. \linewidth]{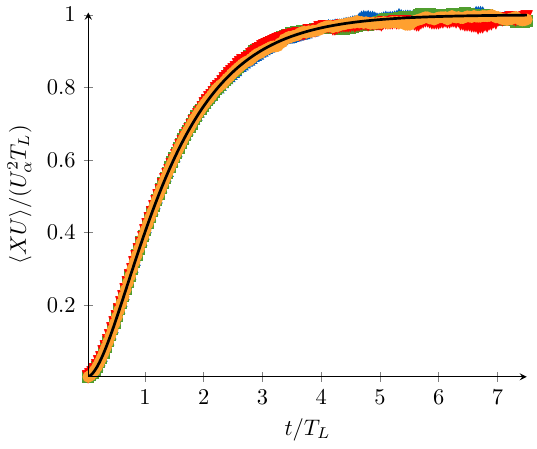}

\end{minipage}  

\caption{Evolution of $\langle XX \rangle $, $\langle UU \rangle $ ,$\langle XU \rangle $ for the point source dispersion in the ballistic limit case. Comparison between the analytical solution (black line) and numerical results obtained with the new algorithm for the various meshes. The numerical results are all in agreement with the analytical solution regardless of the mesh used.}    \label{disp_unstruc_mesh}
\end{figure}

As in Section~\ref{sec:results:point_source}, we can also analyze the results obtained for the different second order moments. We focus here on verifying that the results are consistent regardless of the mesh used using a given time step (the role of the time step has been detailed in Section~\ref{sec:results:point_source}). The results are displayed in \figurename{}~\ref{disp_unstruc_mesh}: it can be seen that all numerical results match the analytical values regardless of the mesh used in such cases. This confirms the accuracy of the algorithm even when 3-D unstructured meshes are used. At this stage, it is also worth noting that 1-D simulation provide the same results as 3-D ones since each of the three directions can be treated independently of the other ones (this is actually a typical characteristic of homogeneous isotropic turbulence).

%% file: tikz/tikz_trajecto_test_edge.tex
\begin{figure}[h]
    \centering
    \tikzset{
	connect/.style={ thick, color=#1},
	connect/.default=blueedf,
	light/.style={thin, densely dashed, color=#1},
	light/.default=Dual,
	lab/.style 2 args={scale=1.05, thin, #2, color=#1, fill=#1!8, inner sep=1pt},
	lab/.default={blueedf}{rectangle, rounded corners=3pt},
	focus/.style={very thick,color=orangeedf!65},
	front/.style={fill=blueedf!5, thick,opacity=0.5},
	back/.style={dashed,thin,fill=blueedf!10},
	backline/.style={dashed,thin,draw=blueedf},
	dualvol/.style={color=greenedf, fill=greenedf!20, opacity=0.5},
	subvol/.style={fill=orangeedf!20, thick,opacity=0.5},
	frakvol/.style={fill=black!25, thick,opacity=0.5},
}
    \tikzset{->-/.style={decoration={
  markings,
  mark=at position #1 with {\arrow{>}}},postaction={decorate}}}

    \def\x{2}
    \def\y{1}
    \def\shifty{0.5}

    \begin{tikzpicture}
    \coordinate(I1) at (-\x+1,-0.4);
    \coordinate(V01) at (-\x,-\y-\shifty);
    \coordinate(V11) at (-\x,-\y);
    \coordinate(V21) at (-\x,\y);
    \coordinate(V31) at (-\x,\y+\shifty);

\path[pattern=north west lines, pattern color=greenedf!50] (-3*\x,\y+\shifty) rectangle (-\x,-\y-\shifty);
\path[pattern= north east lines, pattern color=red!50] (-\x,-\y-\shifty) rectangle (\x,\y+\shifty);
    \draw[black] (I1) node[scale=1.2] {$\odot$} node[black,above=0.3ex] {$\vect{X}_O$};
    \draw[dashed, very thick, redmf!75] (V01) -- (V11);
    \draw[->-=0.5, very thick, red] (V11) node[scale=1.2, blueedf]{$\bullet$} node[right, blueedf]{${\vect{X}_{\alpha}^\dagger}$}   -- (V21) node[right, blueedf]{$\vect{X}_{\beta}^\dagger$}  node[scale=1.2, blueedf]{$\bullet$};
    \draw[dashed, very thick, redmf!75] (V21) -- (V31)  ;
    \draw (-2* \x,\y+\shifty) node[black,above ]{$\mathcal{L}^\text{edge}_{\alpha,\beta} = true$};
    \draw (0,\y+\shifty) node[black,above ]{$\mathcal{L}^\text{edge}_{\alpha,\beta} = false$};


    \end{tikzpicture}
    \caption{Sketch illustrating how the elementary boolean test $\mathcal{L}^\text{edge}_{\alpha,\beta}$ works: the point $(\vect{X}_0$ can either be located on the left-hand side or on the right-hand side of the oriented line $(\vect{X}_\alpha^\dagger \vect{X}_\beta^\dagger)$. Here, the figure is seen from above the plane $(\vect{X}_0, \, \vect{X}_O \vect{X}_D^\perp)$, meaning that the elementary Boolean test is $true$ on the left-hand side of the figure. Note that the line belongs to the closed half-plane (i.e., $false$ in red color).}
        \label{tikz_test_edge}
\end{figure}

%% file: tikz/tikz_trajecto_test_face_orient.tex
\begin{figure}[h]
    \centering
    \tikzset{
	connect/.style={ thick, color=#1},
	connect/.default=blueedf,
	light/.style={thin, densely dashed, color=#1},
	light/.default=Dual,
	lab/.style 2 args={scale=1.05, thin, #2, color=#1, fill=#1!8, inner sep=1pt},
	lab/.default={blueedf}{rectangle, rounded corners=3pt},
	focus/.style={very thick,color=orangeedf!65},
	front/.style={fill=blueedf!5, thick,opacity=0.5},
	back/.style={dashed,thin,fill=blueedf!10},
	backline/.style={dashed,thin,draw=blueedf},
	dualvol/.style={color=greenedf, fill=greenedf!20, opacity=0.5},
	subvol/.style={fill=orangeedf!20, thick,opacity=0.5},
	frakvol/.style={fill=black!25, thick,opacity=0.5},
}
    \tikzset{->-/.style={decoration={
  markings,
  mark=at position #1 with {\arrow{>}}},postaction={decorate}}}

    \def\xf{-3}
    \def\yf{0.3}
    \def\xi{-1.3}
    \def\yi{-1}
    \def\xj{-1}
    \def\yj{1.75}

    \def\shifty{0.5}

    \begin{tikzpicture}
    \coordinate(I1) at (-1.5,0.5);
    \coordinate(Vf1) at (\xf,\yf);
    \coordinate(Vi1) at (\xi,\yi);
    \coordinate(Vj1) at (\xj,\yj);
    \coordinate(Vf2) at (\xf + 4,\yf); 
    \coordinate(Vi2) at (\xj + 4,\yj);
    \coordinate(Vj2) at (\xi + 4,\yi);
    \coordinate(I2) at (-1.5 + 4,0.5);
    \coordinate(M1) at (0,-2 );
    \coordinate(M2) at (0, 2 );
    \coordinate (T1) at (-2,-1.9);
    \coordinate (T2) at (2,-1.9);
    \coordinate (L1) at (-1.65,-2.5);
    \coordinate (L2) at (-1.65,-3);
    \coordinate (L12) at (-1.15,-2.5);
    \coordinate (L22) at (-1.15,-3);

\draw[draw,fill=greenedf!15] (Vf1)--(Vi1)--(Vj1)--(Vf1) ;
\draw [ pattern = vertical lines, pattern color=black!50] (Vf1)--(Vi1)--(Vj1)--(Vf1);
\draw[draw,fill=greenedf!15] (Vf2)--(Vi2)--(Vj2)--(Vf2) ;
\draw[thick, pattern = horizontal lines, pattern color=black!50] (Vf2)--(Vi2)--(Vj2)--(Vf2) ;

    \draw[black] (I1) node[scale=1.2] {$\odot$} node[black,above=0.3ex] {$\vect{X}_O$};
    \draw[dotted] (M1) -- (M2);
    \draw[->-=0.5, very thick, greenedf] (Vi1) -- (Vj1);
    \draw[->-=0.5, very thick, greenedf] (Vf1) node[scale=1.2, blueedf]{$\bullet$} node[right, blueedf]{$\vect{X}_f^\dagger$}   -- (Vi1) node[right, blueedf]{$\vect{X}_{i}^\dagger$}  node[scale=1.2, blueedf]{$\bullet$};
    \draw[->-=0.5, very thick, red] (Vf1)   -- (Vj1) node[right, blueedf]{$\vect{X}_{j}^\dagger$}  node[scale=1.2, blueedf]{$\bullet$};

    \draw (T1) node{$\mathcal{A}^\text{face}_{f,\, i, \, j} = true$};

    \draw[black] (I2) node[scale=1.2] {$\odot$} node[black,above=0.3ex] {$\vect{X}_O$};
    
    \draw[->-=0.5, very thick, red] (Vi2) -- (Vj2);
    \draw[->-=0.5, very thick, red] (Vf2) node[scale=1.2, blueedf]{$\bullet$} node[right, blueedf]{$\vect{X}_f^\dagger$}   -- (Vi2) node[right, blueedf]{$\vect{X}_{i}^\dagger$}  node[scale=1.2, blueedf]{$\bullet$};
    \draw[->-=0.5, very thick, greenedf] (Vf2)   -- (Vj2) node[right, blueedf]{$\vect{X}_{j}^\dagger$}  node[scale=1.2, blueedf]{$\bullet$};
    \draw (T2) node{$\mathcal{A}^\text{face}_{f,\, i, \, j} = false$};

    \draw[->-=0.5, thick, greenedf] (L1) -- (L12) node[right,black]{$\mathcal{L}^\text{edge}_{\alpha , \, \beta} = true$};
      \draw[->-=0.5, thick, red] (L2) -- (L22) node[right,black]{$\mathcal{L}^\text{edge}_{\alpha , \, \beta} = false$};
    
    \end{tikzpicture}
    \caption{Sketch of the Boolean test for alignment $\mathcal{A}^\text{face}_{f,\, i, \, j} $. It determines if the displacement vector $\vect{X}_O \vect{X}_D$ is aligned with the oriented sub-face $\vect{X}_f \vect{X}_{i} \vect{X}_{j}$ (and returns $true$ in that case). The sketch shows the two possible cases: on the left-hand side, the displacement vector is aligned with the sub-face orientation; on the right-hand side, they are not aligned. In each case, the point $\vect{X}_O$ lies within the projected sub-face if it is located on the proper side of all oriented edges (namely $\vect{X}_f \vect{X}_i $, $\vect{X}_f \vect{X}_j $ and $\vect{X}_i \vect{X}_j $, with $i<j$). The logical tests $\mathcal{L}^\text{edge}_{\alpha, \, \beta} = true$ are displayed according to their result (green = $true$ and red=$false$).}
        \label{tikz_test_face}
\end{figure}

%% file: tikz/tikz_trajecto.tex
\newcommand{\mb}[1]{\mathbf{#1}}
\newcommand{\bds}[1]{\boldsymbol{#1}}
\newcommand{\mc}[1]{\mathcal{#1}}

\tikzset{
	connect/.style={ thick, color=#1},
	connect/.default=blueedf,
	light/.style={thin, densely dashed, color=#1},
	light/.default=Dual,
	lab/.style 2 args={scale=1.05, thin, #2, color=#1, fill=#1!8, inner sep=1pt},
	lab/.default={blueedf}{rectangle, rounded corners=3pt},
	focus/.style={very thick,color=orangeedf!65},
	front/.style={fill=blueedf!5, thick,opacity=0.5},
	back/.style={dashed,thin,fill=blueedf!10},
	backline/.style={dashed,thin,draw=blueedf},
	dualvol/.style={color=greenedf, fill=greenedf!20, opacity=0.5},
	subvol/.style={fill=orangeedf!20, thick,opacity=0.5},
	frakvol/.style={fill=black!25, thick,opacity=0.5},
}
\tikzset{->-/.style={decoration={
  markings,
  mark=at position #1 with {\arrow{>}}},postaction={decorate}}}

\newcommand{\defprg}{%
\def\vp{-0.3} 
\def\hp{-0.5} 

\coordinate(P0) at (0,0,0);
\coordinate(P1) at (4,0,0);
\coordinate(P2) at (4,2.5,0);
\coordinate(P3) at (0,2.5,0);
\coordinate(P4) at (2,3.3,0);
\coordinate(P5) at (2,-0.8,0);
\coordinate(P6) at (0-\hp,0-\vp,-4);
\coordinate(P7) at (4-\hp,0-\vp,-4);
\coordinate(P8) at (4-\hp,2.5-\vp,-4);
\coordinate(P9) at (0-\hp,2.5-\vp,-4);
\coordinate(P10) at (2-\hp,3.3-\vp,-4);
\coordinate(P11) at (2-\hp,-0.8-\vp,-4);
\coordinate(leg1_s) at (7,4.8,0);
\coordinate(leg1_e) at (8,4.8,0);
\coordinate(leg2_s) at (7,4,0);
\coordinate(leg2_e) at (8,4,0);
\coordinate(leg3_s) at (7,2.9,0);
\coordinate(leg3_e) at (8,2.9,0);
\coordinate(leg3_m) at (7.5,3.5,0);
\coordinate(leg3_t) at (8,3.2,0);
\coordinate(leg4_s) at (7,2.1,0);
\coordinate(leg4_e) at (8,2.1,0);
\coordinate(leg4_m) at (7.5,2.7,0);
\coordinate(leg4_t) at (8,2.4,0);

\coordinate(leg5_s) at (7,1.3,0);
\coordinate(leg5_e) at (8,1.3,0);
\coordinate(leg5_m) at (7.5,1.9,0);
\coordinate(leg5_t) at (8,1.6,0);

\coordinate(leg6_s) at (7,0.5,0);
\coordinate(leg6_e) at (8,0.5,0);
\coordinate(leg6_m) at (7.5,1.1,0);
\coordinate(leg6_t) at (8,0.8,0);

\coordinate(D0) at (3.5,1,-2);
\coordinate(DI) at (3.885,1.385,-1);
\coordinate(D1) at (4.655,2.155,1);

\coordinate(F1278) at (barycentric cs:P1=0.25,P2=0.25,P7=0.25,P8=0.25);
\coordinate(F24810) at (barycentric cs:P2=0.25,P4=0.25,P8=0.25,P10=0.25);
\coordinate(F012345) at (barycentric cs:P0=0.166,P1=0.166,P2=0.166,P3=0.166,P4=0.166,P5=0.166);

}

\begin{figure}[h]
\centering
\begin{tikzpicture}

\defprg
\path[front] (P3) -- (P4) -- (P2) -- (P1) -- (P5) -- (P0) -- cycle; 
\path[front] (P3) -- (P9) -- (P10) -- (P8) -- (P2) -- (P4) -- cycle; 
\path[front] (P1) -- (P7) -- (P8) -- (P2) -- cycle; 

\draw[draw,fill=greenedf!15] (P1)--(F1278)--(P2) ;
\draw[ pattern = horizontal lines, pattern color=black!50] (P1)--(F1278)--(P2) ;
\draw[draw,fill=red!15] (P2)--(F1278)--(P8) ;
\draw[ pattern = horizontal lines, pattern color=black!50] (P2)--(F1278)--(P8) ;
\draw[draw,fill=red!15] (P8)--(F1278)--(P7) ;
\draw[ pattern = horizontal lines, pattern color=black!50](P8)--(F1278)--(P7) ;
\draw[draw,fill=red!15] (P1)--(F1278)--(P7) ;
\draw[ pattern = vertical lines, pattern color=black!50] (P1)--(F1278)--(P7) ;


\draw[red,->-=.5, very thick]  (F1278) -- (P1) node[blueedf,scale=1.2] {$\bullet$} node[blueedf,below right= 0.5ex and 0.1ex] {$\vect{X}_{v_1}^\dagger$}; 
\draw[greenedf, ->-=0.5, very thick]  (F1278) -- (P2)   node[blueedf,scale=1.2] {$\bullet$} node[blueedf,above =1mm] {$\vect{X}_{v_2}^\dagger$} ;
\draw[red,->-=.5, very thick]  (F1278) -- (P7)node[blueedf,scale=1.2] {$\bullet$} node[blueedf,right=0.1ex] {$\vect{X}_{v_4}^\dagger$};
\draw[greenedf, ->-=0.5, very thick]  (F1278) -- (P8)  node[blueedf,scale=1.2] {$\bullet$} node[blueedf,above right=0.3ex and 0.1ex] {$\vect{X}_{v_3}^\dagger$} ; 
\draw[blueedf] (F1278) node[scale=1.2] {$\times$} node[blueedf,below right= 1.5ex and -0.5ex] {$\vect{X}_{f}^\dagger$};

\draw[black] (DI) node[scale=1.2] {$\odot$} node[black,above=0.3ex] {$\vect{X}_O$};

\draw[connect=blueedf] (P3) -- (P4) -- (P2) -- (P1) -- (P5) -- (P0) -- cycle;
\draw[connect=blueedf] (P9) -- (P10) -- (P8) -- (P7);
\draw[connect=blueedf, ->-=0.5 ] (P3) -- (P9) (P4) -- (P10) (P2) -- (P8) (P1) -- (P7);
\draw[->-=0.5, red, very thick] (P1) -- (P2);
\draw[->-=0.5, red, very thick] (P2) -- (P8);
\draw[->-=0.5, red,very thick] (P8) -- (P7);
\draw[greenedf, ->-=0.5,very thick] (P1) -- (P7);


\draw[backline] (P7) -- (P11) -- (P6) -- (P9);
\draw[backline] (P0) -- (P6) (P5) -- (P11);
\draw[->-=0.5, red,very thick] (leg1_s) -- (leg1_e) node[right,black]{$\mathcal{L}^\text{edge}_{\alpha , \, \beta} = false$};
\draw[greenedf, ->-=0.5,very thick] (leg2_s) -- (leg2_e) node[right,black]{$\mathcal{L}^\text{edge}_{\alpha , \, \beta} = true$};
\draw[ pattern = horizontal lines, pattern color=black!50] (leg3_e) -- (leg3_s)--(leg3_m)--(leg3_e);
\draw[] (leg3_t)node[right,black]{$\mathcal{A}^\text{face}_{f,\, i, \, j} = false$};
\draw[ pattern = vertical lines, pattern color=black!50] (leg4_e)--(leg4_s)--(leg4_m)--(leg4_e) ;
\draw[] (leg4_t)node[right,black]{$\mathcal{A}^\text{face}_{f,\, i, \, j} = true$};
\draw[draw,fill=red!15] (leg5_e) -- (leg5_s)--(leg5_m)--(leg5_e);
\draw[] (leg5_t) node[right,black]{face not crossed};
\draw[draw,fill=greenedf!15] (leg6_e)--(leg6_s)--(leg6_m)--(leg6_e) ;
\draw[] (leg6_t)node[right,black]{face crossed};

\end{tikzpicture}
\caption{Sketch illustrating how the tracking algorithm determines if a particle displacement from $\vect{X}_O$ to $\vect{X}_D$ crosses a face. Here, the particle exits the cell through the face on the right. As a result, when the edges of this face are projected on the plane normal to the displacement vector (here with superscripts $\dagger$), the various logical tests $\mathcal{L}^\text{edge}_{\alpha , \, \beta}$ and $\mathcal{A}^\text{face}_{f,\, i, \, j}$ confirm that one of the triangular sub-face is detected as an exit face (green color) while the three other sub-faces are not (red color). \label{tikz_face_crossing}}
\end{figure}

%% file: tikz/tikz_trajecto_warped.tex

\tikzset{
	connect/.style={ thick, color=#1},
	connect/.default=blueedf,
	light/.style={thin, densely dashed, color=#1},
	light/.default=Dual,
	lab/.style 2 args={scale=1.05, thin, #2, color=#1, fill=#1!8, inner sep=1pt},
	lab/.default={blueedf}{rectangle, rounded corners=3pt},
	focus/.style={very thick,color=orangeedf!65},
	front/.style={fill=blueedf!5, thick,opacity=0.5},
	back/.style={dashed,thin,fill=blueedf!10},
	backline/.style={dashed,thin,draw=blueedf},
	dualvol/.style={color=greenedf, fill=greenedf!20, opacity=0.5},
	subvol/.style={fill=orangeedf!20, thick,opacity=0.5},
	frakvol/.style={fill=black!25, thick,opacity=0.5},
}

\newcommand{\defwarped}{

\coordinate(P1) at (0,0,0);
\coordinate(P2) at (4,0,0);
\coordinate(P3) at (4.4,0,-4);
\coordinate(P4) at (-0.3,0.3,-4.5);
\coordinate(P5) at (2.15,1.0,-4.8);
\coordinate(C) at (2.15,3.75,-2);
\coordinate(projP4) at (-0.3,0,-4.5);
\coordinate(projP5) at (2.15,0,-4.8);

\coordinate(D0) at (0.1,1.2,-1);
\coordinate(D1) at (5.5,0.68,-1);
\coordinate(DI2) at (4,0.83,-1);
\coordinate(DI1) at (1.1,1.105,-1);

\coordinate(F12345) at (barycentric cs:P1=0.2,P2=0.2,P3=0.2,P4=0.2,P5=0.2);
}

\begin{figure}[h]
\centering
\begin{tikzpicture}
\defwarped
\draw[blueedf, thick] (P4) -- (P1) -- (P2) -- (P3) -- (P5) --(P4); 

\draw[connect = orangededf,thin] (F12345) -- (P1)  (F12345) -- (P2) (F12345) -- (P3) (F12345) -- (P4) (F12345) -- (P5) ;

\draw[bluededf, dashed] (-1.5,0,0.5) node[above left]{$(\vect{X}_{v_1} \vect{X}_{v_2} \vect{X}_{v_3})$} -- (4.5,0,0.5) -- (5.05,0, -5) -- (-0.95,0, -5) -- (-1.55,0,0.5) ;

\draw[orangededf] (P1) node[blueedf,scale=1.2] {$\bullet$} node[blueedf,below left= 0.5ex and 0.1ex] {$\vect{X}_{v_1}$};
\draw[orangededf] (P2) node[blueedf,scale=1.2] {$\bullet$} node[blueedf,below =1mm] {$\vect{X}_{v_2}$};
\draw[orangededf] (P3) node[blueedf,scale=1.2] {$\bullet$} node[blueedf,right=0.1ex] {$\vect{X}_{v_3}$};
\draw[orangededf] (P4) node[blueedf,scale=1.2] {$\bullet$} node[blueedf,left=0.3ex] {$\vect{X}_{v_4}$};
\draw[orangededf, thick, dotted] (P4) -- (projP4) node[scale=0.5] {$\bullet$};
\draw[orangededf] (P5) node[blueedf,scale=1.2] {$\bullet$} node[blueedf,above=0.3ex] {$\vect{X}_{v_5}$};
\draw[orangededf, thick, dotted] (P5) -- (projP5) node[scale=0.5] {$\bullet$};

\draw[orangededf] (F12345) node[scale=1.2] {$\times$} node[blueedf,below right= 0.5ex and -0.5ex] {$\vect{X}_{f}$};

\draw[black] (D0) node[scale=1.2] {$\times$} node[black, left=0.3ex ] {$\vect{X}_{O}$};
\draw[black] (D1) node[scale=1.2] {$\times$} node[black, right= 0.1ex] {$\vect{X}_{D}$};
\draw[black] (DI1) node[scale=1.2] {$\times$} node[black,below right=0.3ex and 0.05ex] {$\vect{X}_{I_1}$};
\draw[black] (DI2) node[scale=1.2] {$\times$} node[black,above=0.3ex] {$\vect{X}_{I_2}$};
\draw[black,very thick] (D0) -- (DI1);
\draw[dashed,black] (DI1) -- (DI2);
\draw[->,black,very thick] (DI2) -- (D1) ;
\end{tikzpicture}
\caption{Sketch showing a particle displacement from $\vect{X}_O$ to $\vect{X}_D$ going through a warped face: it can be seen that the particle exits the cell through $\vect{X}_{I_1}$ (which belongs to the sub-face ($\vect{X}_{f}$, $\vect{X}_{v_1}$, $\vect{X}_{v_4}$) and re-enters the cell through $\vect{X}_{I_2}$ (which belongs to the sub-face  ($\vect{X}_{f}$, $\vect{X}_{v_2}$, $\vect{X}_{v_3}$). This is naturally handled by the present algorithm which browse through all sub-faces and then counts the number of times a sub-face is crossed: a pair number means that it stays in the current cell while an odd number means that it exits the current cell.
\label{tikz_multiple_face_crossing}}
\end{figure}